\documentclass[prd,aps,showpacs,superscriptaddress,nofootinbib,twocolumn,floatfix]{revtex4-1}
\usepackage[intlimits]{amsmath}
\usepackage{amssymb}
\usepackage{graphicx}
\usepackage{booktabs}
\usepackage{mathrsfs,slashed}
\usepackage{hyperref}
\usepackage{epsfig}
\usepackage{color}
\usepackage{float}
\usepackage{ulem} 

\newcommand{\bea}{\begin{eqnarray}}
\newcommand{\eea}{\end{eqnarray}}

\begin{document}
\title{Sexaquark dilemma in neutron stars and its solution by quark deconfinement
}
\author{M.~Shahrbaf}
\email{m.shahrbaf46@gmail.com }
\affiliation{Institute of Theoretical Physics,
    University of Wroclaw,
    50-204 Wroclaw, Poland}

\author{D.~Blaschke}
\email{david.blaschke@uwr.edu.pl}
\affiliation{Institute of Theoretical Physics,
    University of Wroclaw,
    50-204 Wroclaw, Poland}
\affiliation{Bogoliubov Laboratory of Theoretical Physics,
    Joint Institute for Nuclear Research,
    141980 Dubna, Russia}
\affiliation{National Research Nuclear University (MEPhI),
    115409 Moscow, Russia}

\author{S.~Typel}
\email{stypel@ikp.tu-darmstadt.de}
\affiliation{Technische Universit\"{a}t Darmstadt, Fachbereich Physik, Institut f\"{u}r Kernphysik,
   Schlossgartenstra\ss{}e 9, D-64289 Darmstadt, Germany}
\affiliation{GSI Helmholtzzentrum f\"{u}r Schwerionenforschung GmbH,
Theorie, Planckstra\ss{}e 1, D-64291 Darmstadt, Germany}

\author{G.~R.~Farrar}
\email{gf25@nyu.edu}
\affiliation{Center for Cosmology and Particle Physics, 
Department of Physics, New York University, NY, 
NY 10003, USA}

\author{D.~E. Alvarez-Castillo}
\email{alvarez@theor.jinr.ru}
\affiliation{Bogoliubov Laboratory of Theoretical Physics,
    Joint Institute for Nuclear Research,
    141980 Dubna, Russia}
 \affiliation{Henryk Niewodnicza{\'n}ski Institute of Nuclear Physics, 
 Polish Academy of Sciences, 
 31-342 Cracow, Poland}

\date{\today}
\begin{abstract}
Following the idea that a stable sexaquark state with quark content (uuddss) would have gone unnoticed by experiment so far and that such a particle would be a good dark matter candidate, we investigate the possible role of a stable sexaquark in the physics of compact stars given the stringent constraints on the equation of state that stem from observations of high mass pulsars and GW170817 bounds on the compactness of intermediate mass stars. We find that there is a ``sexaquark dilemma" (analogous to the hyperon dilemma) for which the dissociation of the sexaquark in quark matter is a viable solution fulfilling all present constraints from multi-messenger astronomy. The parameters needed to model the hybrid star including sexaquarks are in line with parameters of pre-existing quark- and hadronic-matter models.  We find that current constraints -- tidal deformability in accordance with GW170817 and maximum mass above the lower limit from PSR J0740+6620 -- can be satisfied two ways: with early quark deconfinement such that neither sexaquarks nor hyperons are present in any NS interiors, or with later deconfinement such that a neutron-sexaquark shell surrounds the inner quark matter core.
\end{abstract}

\maketitle

\section{Introduction}
The present work is devoted to consideration of the consequences for neutron star (NS) phenomenology that would follow from the existence of a  possible stable sexaquark (S) state with the quark content $uuddss$.

\begin{figure}
    \centering
   \includegraphics[width=0.5\textwidth]{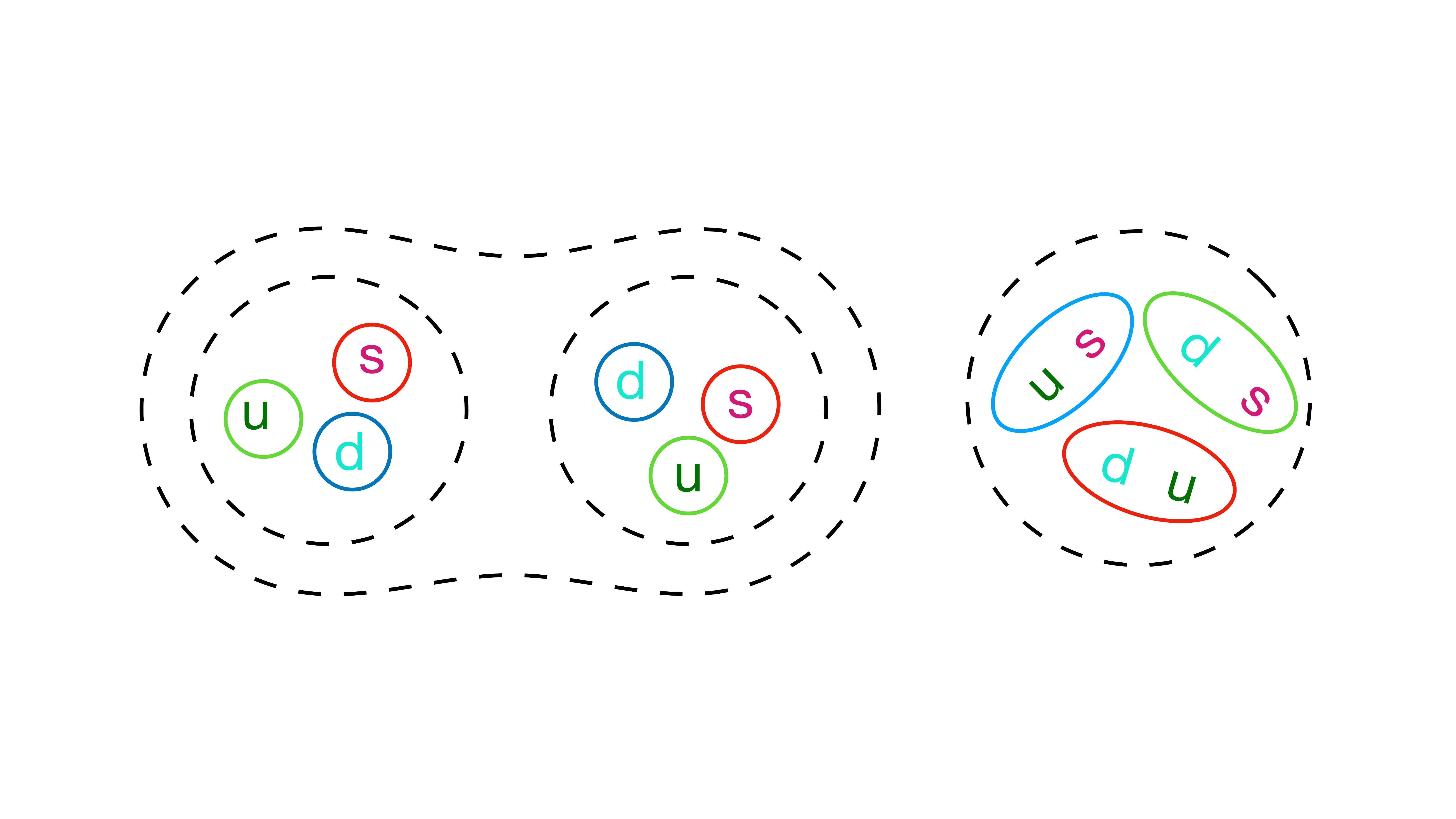}
    \caption{Comparison of two realisations of a color singlet hadron with the quark content (uuddss). 
    Left: hadronic molecule of two $\Lambda$(uds) hyperons, corresponding to the H-dibaryon. Right: compact bound state of three diquarks, bound by color forces, which corresponds to a possible structure of the sexaquark.}
    \label{fig:H-S}
\end{figure}

The S is an electrically neutral spin-less boson with baryon number $B_S=2$ and strangeness $S_S=-2$ in a flavor-singlet state.  If it is light enough to be stable against weak decay ($m_S < m_p + m_\Lambda  + m_e = 2054$ MeV), the S is a good dark matter candidate~\cite{Farrar:2002ic,Farrar:2018hac} and experiments to date would not have been sensitive to it~\cite{Farrar:2022mih}.
The sexaquark has to be distinguished from the H-dibaryon (H) with the same quark content, which was introduced by Jaffe \cite{Jaffe:1976yi} and estimated using the MIT Bag Model to have a mass of 2150 MeV -- greater than $m_\Lambda + m_p + m_e$ and hence weak-interaction unstable with a lifetime $\mathcal{O}(10^{-10}s$).   As illustrated in Fig.~\ref{fig:H-S},  both a molecule of two $\Lambda$ hyperons as well as a more tightly bound state of three diquarks could exist.

Our specific goal is to investigate whether the existence of neutron stars with masses above $2 M_\odot$ and the evidence that NS radii vary slowly with mass above $\sim 1.4 M_\odot$, excludes the existence of a scalar strongly-interacting dibaryon with mass in the range where it is a potential dark matter candidate,  $\approx 2 m_N$ to 2054 MeV.  To this end, we use a pre-existing formalism for the hadronic equation of state and extend it to include an S along with a 2-flavor constant-speed-of-sound equation of state (EoS) for deconfined quark matter, and we consider two methods for interpolating between them.  With a non-exhaustive exploration of model parameters (without adjusting beyond the normal range), we find that a stable S with mass above $\approx 1885$ MeV is compatible with current neutron star observations.  We find solutions of two types: i) with early deconfinement and neither S nor hyperons present in neutron stars of any mass and ii) with a layer containing both S and nucleons outside a quark matter core.  When our understanding of the transition between hadronic and quark matter is improved,  knowledge of neutron stars may provide constraints on the allowed mass of a sexaquark, or even point to its existence.  Indeed, we find that an S naturally explains the soft EoS at low densities implied by observations of GW170817. 

\section{Overview}
One of the puzzling questions in the physics of hadrons concerns the structure of recently discovered multiquark clusters such as pentaquarks and tetraquarks. 
Are these states molecules of color-neutral (hadronic) subclusters analogous to the hydrogen molecule of atomic physics?
Or are they multiquark states bound by confining forces between color charged quarks and diquarks?  Or some of each?
As an example, the recently discovered all-charm tetraquark 
X(6900)~\cite{LHCb:2020bwg} can a priori be viewed as a molecule of two J/$\psi$ mesons or a diquark-antidiquark bound state. 
Similarly, a pentaquark could be a baryon-meson molecule or a bound state of two diquarks and an antiquark. 
For more details, see the textbook \cite{Ali:2019roi}.

While the H state has been investigated in lattice QCD calculations \cite{Green:2021qol,Francis:2018qch,NPLQCD:2012mex,Inoue:2010es,NPLQCD:2010ocs}
and a binding energy of a few to 80 MeV has been found at still unphysically large quark mass, the much lighter and possibly more compact S state is not yet accessible to lattice QCD calculations \cite{Detmold:2021}. 

In a recent calculation within a constituent quark model, Buccella obtained the mass of the S as a bound state of three diquarks to be $m_S=1883$ MeV \cite{Buccella:2020mxi} while Azizi et al. have obtained it to be about $m_S=1180$ MeV using QCD sum rules \cite{Azizi:2019xla}.
Buccella showed also that this estimate does not depend on the choice of the parameters for the constituent quark masses by replacing the latter with an expression using the heavier nonet mesons,
$m_S= 2M_{K(1430)} - (1/2) M_{f^0(1370)} + 40~{\rm MeV} 
+ (1/2) [M_N - M_\Delta] +M_\Sigma - M_{Y^*} = 1876.5$ MeV,
where the first three terms correspond to the constituent mass and the other terms to the contribution of the chromomagnetic interaction \cite{Buccella:2020mxi}.
An S in this mass range is sufficiently heavy to not induce instability in the deuteron (D) and other nuclei~\cite{fw22a,Farrar:2022mih} 

For $m_S \leq m_D  = 1876.122$~MeV,  the S is absolutely stable due to baryon number conservation, while as long as 
$m_S \leq m_p + m_e + m_\Lambda = 2054.466$~MeV its decay requires $\Delta S = 2$ and is hence doubly-weak, and its lifetime exceeds the age of the universe \cite{Farrar:2003qy,fw22a}. 
In this study we suppose that the S may be a deeply bound state with low
enough mass to be absolutely or effectively stable; in that case it is an attractive dark matter (DM) candidate~\cite{Farrar:2018hac,Farrar:2022mih}.
The observed dark matter to baryon ratio is $\Omega_{DM}/\Omega_B = 5.3\pm 0.1$ \cite{Planck:2015fie, ParticleDataGroup:2018ovx} and a successful model for dark matter should account for this value.  In fact, an abundance of S dark matter (SDM)
in agreement with this observation follows~\cite{Farrar:2018hac} from statistical mechanics using quark masses as determined from lattice QCD, given the effective freezeout temperature $T_{\rm eff}\approx 150$ MeV for $m_S$ in the expected range.\footnote{The predicted $\Omega_{DM}/\Omega_B$ is very {\it insensitive} to the assumed freezeout temperature; the relevant range is motivated by the recent result of $T_c=156.5 \pm 1.5$ MeV for the pseudocritical temperature obtained in lattice QCD simulations \cite{HotQCD:2018pds}. Note a plotting error in~\cite{Farrar:2018hac} is corrected in v2 of ~\cite{Farrar:2020zeo}; the formulae given in~\cite{Farrar:2018hac} are correct.}

Another environment for the production of the S, apart from the hadronization transition in the early Universe, is the ``Little Bang" in ultrarelativistic heavy-ion collisions at the LHC, where not only traditional hadrons but also light nuclei and antinuclei are abundantly produced. 
Recently, the ratio of light sexaquarks relative to the deuteron 
to be expected under these conditions has been estimated within a thermal statistical model \cite{Blaschke:2021tul} and found to be sizeable, of the order one.   
Abundant production of sexaquarks is not sufficient for their discovery -- there remains the problem of the detection of the S and discriminating it from the far more abundant neutrons; see~\cite{Farrar:2022mih} for an analysis of detection strategies.

In the present work, we consider S in the mass range of
$1885~\mbox{MeV} < m_S < 2054$~MeV.\footnote{The lower end of the range is sufficient that condensation of S below saturation density 
can be avoided and above the upper end of the range the S decays in $\mathcal{O}(10^{-10}s)$ and is not a Dark Matter candidate.}
We study the possible relevance of the S for the properties of neutron stars in light of recent multi-messenger observations that constrain the mass-radius diagram and tidal deformability.
When the S is present in cold dense baryonic matter in neutron stars, it
forms a Bose condensate as soon as the baryo-chemical potential in the center of the star fulfills $\mu_b=m_S/2$, unless a transition to non-hadronic degrees of freedom occurs first. 
For a density-independent mass, for instance $m_S=1941 \,(2054)$ MeV, this occurs for a star with $M=0.21~M_\odot\,(0.7~M_\odot)$ respectively. The mass $m_S = 1941$ MeV is chosen as an illustrative case, because for a density-independent mass and $m_S = 1941$ MeV, the Bose condensation occurs exactly at saturation density.  Saturation density is roughly the lowest density at which condensation can be tolerated due to  the existence of nuclei.  Therefore, due to the saturation of the pressure once Bose condensation occurs, $0.21~M_\odot$ is roughly the maximum NS mass that can be reached for a non-interacting S with mass less than 1941 MeV, unless a transition to new degrees of freedom replaces hadrons.  Such a low maximum mass is in clear contradiction with the observation of pulsars as massive as $2~M_\odot$  like PSR J0740+6620 \cite{Fonseca:2021wxt} or PSR J0348+0432 \cite{Antoniadis:2013pzd} and many, many lower-mass neutron stars.

The problem with S Bose condensation is alleviated by the plausible assumption that the effective mass of the S is medium-dependent and increases with density, similar to the behavior of the other baryon masses at supersaturation densities.
However, as we will show in this work, simply a medium-dependent mass for the S is insufficient to allow for neutron star masses consistent with
the observational constraint $M_{\rm max} \gtrsim 2~M_\odot$. 
This constitutes the ``sexaquark dilemma" to which this work is devoted. The same is true for hyperons -- the so-called hyperon puzzle.  It is known that the hyperon puzzle can be solved by quark deconfinement;  see, e.g., \cite{Shahrbaf:2019vtf,Shahrbaf:2020uau}
and references therein.
Therefore, we explore in the present paper various scenarios for the sexaquark dilemma and the viability of solution by quark deconfinement. 

In a recent work, McDermott et al. \cite{McDermott:2018ofd} argued that a deeply bound S -- in the mass range  considered here -- would be incompatible with the delayed neutrino signal from supernova 1987A.  However the inter-conversion amplitude between S and two $\Lambda$'s is naturally  small, and the value needed to not impact SN1987a cooling is comfortably in the expected range~\cite{fw22a,Farrar:2022mih}, so SN1987A cooling is not the show-stopper for sexaquark dark matter envisaged in  \cite{McDermott:2018ofd}.\footnote{Similarly the conclusions of of~\cite{Kolb:2018bxv} do not apply to sexaquark dark matter because the suppressed interconversion between sexaquark and two baryons inhibits the destruction of sexaquarks in the high-temperature hadronic phase.}
While the conditions relevant to the analysis of ~\cite{McDermott:2018ofd} are unlike the steady state conditions we consider, we note that their assumed hyperon-rich environment and medium independence of the masses, and neglect of the possibility of quark deconfinement would not be applicable 
for our solution of the S dilemma.

The structure of the present paper is as follows. In Sec. \ref{sec:Had} we present the formalism of calculation the equation of state (EoS) of hadronic matter based on the density dependent (DD) relativistic mean field (RMF) model in neutron star matter. In Sec.~\ref{sec:QM} the EoS of quark matter is obtained and the theory of two different approaches for constructing the hybrid stars are given. In Sec.~\ref{sec:res_GRDF} we present our results for the properties of purely hadronic stars including S particles and discuss their dilemmas. The deconfinement solution for this dilemma is discussed in Sec.~\ref{sec:results}. Finally, the summary and conclusion are given in Sec.~\ref{sec:conclusion}.

\section{A relativistic density functional approach to hypernuclear matter with sexaquark}
\label{sec:Had}

The study of nuclear matter based on relativistic approaches with a Lagrangian density including baryons and mesons as degrees of freedom has a long history \cite{Walecka:1974qa, Serot:1984ey, Reinhard:1989zi, Ring:1996qi}.
The original treatment as full-fledged field theories was subsequently replaced by the interpretation as effective field theories \cite{Rusnak:1997dj} and more recently as density functional theories \cite{dreizler2012density}. 
In this work the EoS of hadronic matter is obtained from a generalized relativistic density functional (GRDF) with baryon-meson couplings that depend on the total baryon density of the system. The original density functional for nucleonic matter considers the isoscalar $\sigma$ and $\omega$ mesons and the isovector $\rho$ meson as exchange particles that describe the effective in-medium interaction. The density dependence of the couplings is adjusted to describe properties of atomic nuclei \cite{Typel:1999yq,Typel:2005ba}. It has been confirmed that such GRDFs are successful in reproducing the properties of nuclear matter around nuclear saturation \cite{Klahn:2006ir}.

The explicit introduction of mesons in 
relativistic density functionals is not necessary for the description of nuclear matter since the theoretical formulation can be based on the baryonic degrees of freedom and their densities only. This is realized, e.g., in relativistic point-coupling models, see \cite{Sun:2019plt} and references therein. 
However, the use of mesons is convenient from a practical point of view as they represent a simple means to describe the relevant components of the effective interaction.

GRDFs with different functional forms of the density dependent couplings have been studied, e.g., in \cite{Typel:2018cap},
and various parameterizations are available in the literature \cite{Dutra:2014qga}.
In the present work the parameterization DD2 \cite{Typel:2009sy} is used for the $\sigma$, $\omega$, and $\rho$ couplings. 
It was obtained by fitting properties of finite nuclei: binding energies, charge and diffraction radii, surface thicknesses and spin-orbit splittings.
It predicts characteristic nuclear matter parameters that are consistent with recent constraints \cite{Oertel:2016bki}. 
In particular, a saturation density of $0.149065$~fm${}^{-3}$, a binding energy per nucleon of $16.02$~MeV, and an incompressibility of $K=242.7$~MeV in symmetric nuclear matter are found. The isospin dependence can be characterized by a symmetry energy of $J = 31.67$~MeV at saturation with a slope parameter of $L = 55.04$~MeV. Because no constraints were imposed at supra-saturation densities in determining the DD2 parameters, the EoS constitutes a pure extrapolation in this density range.
The DD2 model leads to a rather stiff EoS at high baryon densities with a maximum neutron star mass of $2.4$ solar masses in a pure nucleonic scenario of the strongly interacting system. This approach with nucleons has been extended to include light and heavy clusters as quasiparticles modified by medium effects \cite{Typel:2009sy,Pais:2016fdp} which are important in the finite-temperature EoS for astrophysical applications, e.g.,
the simulation of core-collapse supernovae or neutron-star mergers \cite{Oertel:2016bki}. The internal structure of these clusters is not taken in to account and they are treated as point-like particles.

In the description of neutron-star matter one has to consider that new baryonic degrees of freedom can become active with increasing density as the chemical potentials rise. Neglecting any interaction, a new species appears when the corresponding chemical potential crosses the particle mass.
This usually leads to a softening of the EoS as it is known
for hyperons - the so-called 'hyperon puzzle' which has been discussed in \cite{Shahrbaf:2019wex, Shahrbaf:2019vtf, Shahrbaf:2020uau} and the literature cited therein. However, if the interaction with an additional $\phi$ meson, which couples only to strangeness-carrying hyperons, is included in the GRDF, a sufficient
stiffening of the EoS can be achieved. 
In our present study, all particle in the octet of spin $1/2$ baryons are included in the model as degrees of freedom. Their interaction is described by the exchange of
$\sigma$, $\omega$, $\rho$, and $\phi$ mesons
with appropriately adjusted couplings to the baryons. The S particle is a further baryonic degree of freedom
in the GRDF with a density dependent mass shift that models the effect of the interaction with the medium.
For the description of neutron star matter electrons and muons are added in the GRDF to achieve charge neutrality. The electron and muon densities, assuming identical lepton chemical potentials, for given baryon density are determined by the condition of $\beta$ equilibrium. 

In the present work
with the application to neutron stars we can restrict ourselves to the case of matter at zero temperature. At densities below saturation there are no hyperons or sexaquarks and the unified crust EoS of the original GRDF-DD2 model with clusters is used. It contains the well-known sequence of nuclei in a body-centered cubic lattice with a uniform background of electrons and a neutron gas above the neutron drip line. The transition to homogeneous matter just below the nuclear saturation density is described consistently within the same approach. The main modification is to include the new degrees of freedom at supersaturation densities in the GRDF model. This will be described in the following subsections. All equations
follow the traditional convention of $\hbar = c = 1$ of nuclear physics.

\subsection{NS matter EoS with hyperons and sexaquark}

The hadronic part of the core in a neutron star is assumed to consist of homogeneous matter composed of strongly interacting baryons and charged, non-interacting leptons in full thermodynamic equilibrium. Explicitly, these degrees of freedom are protons ($p$), neutrons ($n$), hyperons ($\Lambda$, $\Sigma^{+}$, $\Sigma^{0}$, $\Sigma^{-}$, $\Xi^{0}$, $\Xi^{-}$), the sexaquark (S), electrons ($e$) and muons ($\mu$).  
All information on the thermodynamic properties
of the system can be obtained from a grand canonical thermodynamic potential density $\Omega(\{\mu_{i}\})$
that depends only on the chemical potentials $\mu_{i}$ of the individual particles at zero temperature. These can be expressed as
\begin{equation}
  \mu_{i} = B_{i} \mu_{b} + Q_{i} \mu_{q} + S_{i} \mu_{s} + L_{i} \mu_{l}
\end{equation}
with the individual baryon ($B_{i}$), charge ($Q_{i}$), strangeness ($S_{i}$) and lepton ($L_{i}$) numbers. Independent quantities are the baryon ($\mu_{b}$), charge ($\mu_{q}$),
strangeness ($\mu_{s}$) and lepton ($\mu_{l}$)
chemical potentials (assuming identical electron lepton and muon lepton chemical potentials). 
The condition of full equilibrium with respect to strangeness-changing reactions corresponds to $\mu_{s}=0$
and in case of $\beta$ equilibrium we have $\mu_{l}=0$. Thus only two independent chemical potentials, $\mu_{b}$ and $\mu_{q}$, remain. These have to be determined
for a given baryon density
\begin{eqnarray}
  \label{eq:n_B}
  n_{b} &=& \sum_{i} B_{i} n_{i}^{(v)} \nonumber\\
  &=& n_{p}^{(v)}+n_{n}^{(v)} +n_{\Lambda}^{(v)}+n_{\Sigma^{+}}^{(v)}+n_{\Sigma^{0}}^{(v)} +n_{\Sigma^{-}}^{(v)}\nonumber\\
  && +n_{\Xi^{0}}^{(v)}+n_{\Xi^{-}}^{(v)}
  +2n_{S}^{(v)} 
\end{eqnarray}
and total charge density 
\begin{eqnarray}
  n_{Q} & = & \sum_{i} Q_{i} n_{i}^{(v)}
  \\ \nonumber & = &
  n_{p}^{(v)}+n_{\Sigma^{+}}^{(v)} -n_{\Sigma^{-}}^{(v)}-n_{\Xi^{-}}^{(v)}
  -n_{e}^{(v)}-n_{\mu}^{(v)}
\end{eqnarray}
with the particle number (or vector) densities $n_{i}^{(v)}$, see below.
Requiring local charge neutrality corresponds to $n{q}=0$ and thus
the charge chemical potential $\mu_{q }$ will be fixed too.
Finally, the EoS of neutron star matter depends only on the baryon chemical potential $\mu_{b}$.

All constituent particles with vacuum rest masses $m_{i}$ are considered as quasiparticles in the medium
with effective masses $m_{i}^{\ast}=m_{i}-S_{i}$ and effective chemical potentials
$\mu_{i}^{\ast}=\mu_{i}-V_{i}$. These contain scalar and vector potentials, $S_{i}$ and $V_{i}$, that
describe the interaction of the particles in the medium. They arise from the coupling of the
meson fields $\sigma$, $\omega$, $\rho$, and $\phi$ with coupling strengths 
$\Gamma_{i\sigma}$, $\Gamma_{i\omega}$, $\Gamma_{i\rho}$,
and $\Gamma_{i\phi}$ to nucleons and hyperons
or they are modeled effectively by a mass shift $\Delta m_{S}$ for the sexaquark.
The couplings are assumed to depend on the total density of nucleons and hyperons 
\begin{equation}
 n_{cpl} = n_{b} - 2 n_{S}
\end{equation}
and the mass shift depends on the baryon density (\ref{eq:n_B}).
Using the meson field names also for the field strengths, the potentials for nucleons and hyperons have the form
\begin{equation}
  \label{eq:S_i}
  S_{i} = \Gamma_{i\sigma} \sigma
\end{equation}
and 
\begin{equation}
  \label{eq:V_i}
  V_{i} =  \Gamma_{i\sigma} \sigma
  + \Gamma_{i\omega} \omega
  +  \Gamma_{i\rho} \rho
  +  \Gamma_{i\phi} \phi
  + B_{i} V^{(r)} + W_{i}^{(r)}
\end{equation}
whereas they are given by 
\begin{equation}
  \label{eq:S_S}
  S_{S} =   -\Delta m_{S}
\end{equation}
and 
\begin{equation}
  \label{eq:V_S}
  V_{S} =  
  W_{S}^{(r)}
\end{equation}
for the sexaquark.
The rearrangement contributions $V^{(r)}$, $W_{i}^{(r)}$, see below, are required for thermodynamic consistency.
The coupling of a particle $i$ to a meson $m$ can be written as
\begin{equation}
\label{eq:Gamma_im}
    \Gamma_{im} = g_{im} \Gamma_{m}(n_{cpl})
\end{equation}
with prefactors $g_{im}$ 
and density dependent coupling functions $\Gamma_{m}(n_{cpl})$. 
The dependence of these functions 
on the baryon density $n_{b}$ is given by the functional forms as introduced in \cite{Typel:1999yq}
and used for the DD2 parameterization, i.e.,
\begin{equation}
    \Gamma_{m}(n_{cpl}) = \Gamma_{m}(n_{\rm sat}^{(v)}) f_{m}(x)
\end{equation}
with $x=n_{cpl}/n_{\rm sat}^{(v)}$, the couplings at saturation $\Gamma_{m}(n_{\rm sat}^{(v)})$ and functions
\begin{equation}
\label{eq:f_rat}
    f_{m}(x) = a_{m} \frac{1+b_{m}(x+d_{m})^{2}}{1+c_{m}(x+d_{m})^{2}}
\end{equation}
for $m=\sigma,\omega$ and
\begin{equation}
\label{eq:f_exp}
    f_{\rho}(x) = \exp[-a_{\rho}(x-1)]
\end{equation}
for the $\rho$ meson.
See \cite{Typel:2009sy} for the actual parameters and masses of the mesons. The coupling $\Gamma_{\phi}$ of the $\phi$ meson 
is assumed to be constant in the present model
with $\Gamma_{\phi}=\Gamma_{\omega}(n_{\rm sat}^{(v)})$. In the following, also
derivatives $\Gamma_{m}^{\prime}=d\Gamma_{m}/dn_{cpl}$ of the couplings will appear.

The coupling factors $g_{im}$ in eq.\ (\ref{eq:Gamma_im}) of nucleons and
hyperons to the vector mesons can be expressed with help of the quantum numbers $B_{i}$, $Q_{i}$, and $S_{i}$ of a particle $i$ as
\begin{eqnarray}
    g_{i\omega} &=& B_{i}+\frac{S_{i}}{3} \\
    g_{i\rho} &=& 2Q_{i}-B_{i}-S_{i} \\ 
    g_{i\phi} &=& \frac{\sqrt{2}}{3} S_{i}
\end{eqnarray}
following the usual SU(6) coupling scheme, see, e.g., \cite{Weissenborn:2011ut}.
The coupling factors for the $\sigma$ meson to the nucleons are given by $g_{p\sigma}=g_{n\sigma}=1$. For hyperons, they are determined by fixing their in-medium potential $U_{Y}$ with $Y=\Lambda,\Sigma,\Xi$ in symmetric nuclear matter at saturation density $n_{\rm sat}^{(v)}$ of the DD2 parameterization. Explicitly they are given by
\begin{eqnarray}
\label{eq:g_Ysigma}
    g_{Y\sigma} &=& \left[ \left( g_{Y\omega} \Gamma_{\omega}(n_{\rm sat}^{(v)}) 
    + \Gamma_{\omega}^{\prime}(n_{\rm sat}^{(v)})n_{\rm sat}^{(v)}\right)
    \omega_{\rm sat} \right.\nonumber\\
    && \left. -  \Gamma_{\sigma}^{\prime}(n_{\rm sat}^{(v)})n_{\rm sat}^{(s)} -U_{Y}  \right] \left( \Gamma_{\sigma}(n_{\rm sat}^{(v)}) \sigma \right)^{-1} 
\end{eqnarray}
with the vector density $n_{\rm sat}^{(v)}$ and the scalar density $n_{\rm sat}^{(s)}$ at saturation of the DD2 parameterization.

In principle, the interaction of the S particle with the medium can be described by a coupling to the mesons as for the other baryonic degrees of freedom. However, this requires to introduce several unknown meson-sexaquark couplings. 
Since the sexaquark will exist as a condensate at zero temperature, only the difference $V_{S}-S_{S}$ of the vector and scalar potentials appears in the equations. 
This effective potential rises, to lowest order, proportional to the baryon density. 
Thus it is sufficient to describe the interaction with the medium simply by a mass shift
\begin{equation}
\label{eq:Deltam_S}
    \Delta m_{S} = m_{S} x_{S} \frac{n_{b}}{n_{0}},
\end{equation}
giving the effective mass
\begin{equation}
\label{eq:m_S}
 m^*_{S} = m_{S} - S_{S} = m_{S}+ \Delta m_{S}.
\end{equation}
For simplicity in a first exploration, we describe this with a single, adjustable parameter $x_{S}$ and fixed constants $m_{S}$ and $n_{0}$. The mass shift (\ref{eq:Deltam_S}) rises linearly with the baryon density and represents an effective repulsive interaction with the medium for positive $x_{S}$. Due to this density dependence, 
there appears a rearrangement contribution in the vector potential (\ref{eq:V_S}) of the S particle. It depends linearly on the scalar density of the sexaquark and is thus considerably smaller than the mass shift as long as this density is small compared to the baryon density.

Obviously, electrons and muons are treated as free particles since they don't participate in the strong interaction and there is no electric potential in uniform matter.

The meson fields themselves are found from the meson field equations
\begin{eqnarray}
 m_{\sigma}^{2} \sigma  &=&  \Gamma_{\sigma} n_{\sigma} \\
 m_{\omega}^{2} \omega  &=&  \Gamma_{\omega} n_{\omega} \\
 m_{\rho}^{2} \rho  &=&  \Gamma_{\rho} n_{\rho} \\
 m_{\phi}^{2} \phi  &=&  \Gamma_{\phi} n_{\phi} 
\end{eqnarray}
with the source densities
\begin{eqnarray}
 n_{\sigma}  &=&  \sum_{i} g_{i\sigma} n_{i}^{(s)} \\
 n_{\omega}  &=&  \sum_{i} g_{i\omega} n_{i}^{(v)} \\
 n_{\rho}  &=&  \sum_{i} g_{i\rho} n_{i}^{(v)} \\
 n_{\phi}  &=&  \sum_{i} g_{i\phi} n_{i}^{(v)} ~.
\end{eqnarray}
The rearrangement contributions 
\begin{equation}
    V^{(r)} = \Gamma_{\omega}^{\prime}n_{\omega}\omega 
    + \Gamma_{\rho}^{\prime}n_{\rho}\rho 
    - \Gamma_{\sigma}^{\prime}n_{\sigma}\sigma 
\end{equation}
and
\begin{equation}
  W_{i}^{(r)} = n_{S}^{(s)} \frac{\partial\Delta m_{S}}{\partial n_{i}^{(v)}}
\end{equation}
in eqs.\ (\ref{eq:V_i}) and (\ref{eq:V_S}) 
arise due to the dependence of the couplings and mass shifts, respectively,
on the densities.

\subsection{Particle densities and relativistic density functional}

The densities of the particles depend on their effective chemical potentials $\mu_{i}^{\ast}$ and effective masses $m_{i}^{\ast}$. In a relativistic model, vector and scalar densities have to be distinguished.
The vector density of a fermion is given by
\begin{eqnarray}
  n_{i}^{(v)} &=& g_{i} \int \frac{d^{3}p}{(2\pi)^{3}} \: \theta(p_{i}-p) 
  = \frac{  g_{i}}{6\pi^{2}} p_{i}^{3}
\end{eqnarray}
with degeneracy factor $g_{i}$ and an integration up to the Fermi momentum
\begin{equation}
  p_{i} = \left\{ \begin{array}{lll}
    \sqrt{[\mu_{i}^{\ast}]^{2}-[m_{i}^{\ast}]^{2}} & \mbox{if} & \mu_{i}^{\ast} > m_{i}^{\ast}\\
    0 & \mbox{if} & \mu_{i}^{\ast} \leq m_{i}^{\ast}
  \end{array} \right. 
\end{equation}
at zero temperature. The expression for the scalar density of a fermion
\begin{eqnarray}
  n_{i}^{(s)} &=& g_{i} \int \frac{d^{3}p}{(2\pi)^{3}} \frac{m_{i}^{\ast}}{E(p,m_{i}^{\ast})}
   \: \theta(p_{i}-p) \nonumber \\
  &=& \frac{g_{i}}{4\pi^{2}} \left[\mu_{i}^{\ast} p_{i}- (m_{i}^{\ast})^{2}
    \ln \frac{\mu_{i}^{\ast}+p_{i}}{m_{i}^{\ast}} \right]
\end{eqnarray}
contains an additional factor $m_{i}^{\ast}/E$ in the integral as compared to the vector density with the energy 
\begin{equation}
  E(p,m_{i}^{\ast}) = \sqrt{p^{2}+[m_{i}^{\ast}]^{2}}
\end{equation}
depending on the momentum $p$, thus
$\mu_{i}^{\ast} = E(p_{i},m_{i}^{\ast})$ with the Fermi momentum $p_{i}$.
For the bosonic sexaquark (S), there is only a finite density at zero temperature
when the condition $\mu_{S}^{\ast}=m_{S}^{\ast}$ is satisfied and we have Bose-Einstein condensation. In this case the vector and scalar densities are identical and can be written as
\begin{equation}
  n_{S}^{(v)} = n_{S}^{(s)} = \left\{ \begin{array}{lll}
    g_{S}\xi_{S} & \mbox{if} & \mu_{S}^{\ast}=m_{S}^{\ast} \\
    0 & \mbox{else} &
  \end{array} \right.
\end{equation}
with degeneracy $g_{S}=1$ and a factor $\xi_{S}$. The latter can be determined
from the given baryon density $n_{b}$ as
\begin{eqnarray}
    \xi_{S}&=&\left(n_{b}-n_{n}^{(v)}-n_{p}^{(v)}
    -n_{\Lambda}^{(v)}-n_{\Sigma^{+}}^{(v)}-n_{\Sigma^{0}}^{(v)} -n_{\Sigma^{-}}^{(v)}\right.
    \nonumber \\
    &&\left. -n_{\Xi^{0}}^{(v)}-n_{\Xi^{-}}^{(v)}
    \right)/g_{S}
\end{eqnarray}
at the condensation point.

The relativistic density functional assumes the form of a
grand canonical thermodynamic potential density in the present approach.
It is given explicitly by the expression
\begin{equation}
\label{eq:Omega}
  \Omega(\{\mu_{i}\}) = \sum_{i \in \mathcal{F}} \Omega_{i} + \Omega_{S} + \Omega_{\rm meson}
  - \Omega_{\rm meson}^{(r)} - \Omega_{\rm mass}^{(r)}
\end{equation}
with a contribution 
\begin{equation}
  \Omega_{i} = -\frac{1}{4} \left[ \mu_{i}^{\ast}  n_{i}^{(v)} - m_{i}^{\ast}n_{i}^{(s)}\right] \: ,
\end{equation}
of the fermionic quasi particles $i \in \mathcal{F} =
\left\{ p, n, \Lambda, \Sigma^{+}, \Sigma^{0}, \Sigma^{-}, \Xi^{0}, \Xi^{-}\right\}$.
The condensate contribution of the bosonic sexaquarks is formally given by
\begin{equation}
  \Omega_{S} = g_{S}\xi_{S}\left(m_{S}^{\ast}-\mu_{S}^{\ast}\right) 
\end{equation}
and the meson contribution has the form
\begin{eqnarray}
  \lefteqn{\Omega_{\rm meson}}
  \\ \nonumber & = & -\frac{1}{2} \left(
  \Gamma_{\omega}n_{\omega}\omega 
    + \Gamma_{\rho}n_{\rho}\rho 
     + \Gamma_{\rho}n_{\rho}\rho 
    - \Gamma_{\sigma}n_{\sigma}\sigma 
  \right) \: .
\end{eqnarray}
Finally, the meson rearrangement contribution
\begin{equation}
  \Omega_{\rm meson}^{(r)} = V^{(r)}n_{cpl} 
\end{equation}
and the mass shift rearrangement contribution
\begin{equation}
  \Omega_{\rm mass}^{(r)} = \sum_{i=n,p,S} n_{i}^{(v)} W_{i}^{(r)}
\end{equation}
appear in eq.\ (\ref{eq:Omega}).
With the above definition of $\Omega$ the standard relations
\begin{equation}
  n_{i}^{(v)} = - \left. \frac{\partial \Omega}{\partial \mu_{i}} \right|_{\{\mu_{j}\}_{j\neq i}}
  \qquad
  n_{i}^{(s)} =  \left. \frac{\partial \Omega}{\partial m_{i}} \right|_{\{\mu_{j}\}}
\end{equation}
are valid as required for thermodynamic consistency.
It is easily noted that $\Omega_{S}=0$ because either $\xi_{S} = 0$ for $m_{S}^{\ast}\neq\mu_{S}^{\ast}$
or $m_{S}^{\ast}=\mu_{S}^{\ast}$ with $\xi_{S}>0$ if condensation occurs.
Generally, the condition $\mu_{S}^{\ast}\leq m_{S}^{\ast}$ applies to the model
with sexaquarks. Since S is uncharged and does not couple to the mesons, this corresponds to
\begin{equation}
  B_{S} \mu_{b}
  \leq m_{S}-S_{S}+W_{S}^{(r)}
  = m_{S}+\Delta m_{S}+n_{S}^{(s)} \frac{\partial\Delta m_{S}}{\partial n_{S}^{(v)}}
\end{equation}
with baryon number $B_{S}=2$. Thus there is a condition on the baryon chemical potential $\mu_{b}$ that is limited from above. If this maximum is reached, the baryon density can only rise by an increase of the condensate density of the sexaquark. With an increase of its effective mass due to the mass shift, the condensation point moves to higher baryon chemical potentials.

\subsection{Parameters and thermodynamic properties}
The rest masses $m_{i}$ of nucleons, hyperons and mesons used
in the GRDF are given in tables \ref{tab:masses_baryons} and \ref{tab:masses_mesons}.
\begin{table}[h]
    \centering
    \caption{Rest masses of nucleons and hyperons in the generalized relativistic density functional. The rest mass of the sexaquark depends on the scenario, see text.}
    
    \begin{tabular}{cc|cc}
    \hline 
    \hline 
      particle $i$   &  $m_{i}$ [MeV] &
      particle $i$   &  $m_{i}$ [MeV] \\
      \hline
      p & 938.272081 & $\Sigma^{0}$ & 1192.642\\
      n & 939.565413 & $\Sigma^{-}$ & 1197.449\\
      $\Lambda$ & 1115.683 & $\Xi^{0}$ &1314.86 \\
      $\Sigma^{+}$ & 1189.37 & $\Xi^{-}$ & 1321.71 \\
      \hline 
    \hline
    \end{tabular}
    \label{tab:masses_baryons}
\end{table}

\begin{table}[h]
    \centering
    \caption{Rest masses of mesons in the generalized relativistic density functional.}
    \begin{tabular}{ccccccccc}
    \hline 
      meson $i$   &  $\sigma$ & $\omega$ & $\rho$ & $\phi$ \\
      \hline 
    $m_{i}$ [MeV]    & 546.212459 & 783 & 763 & 1019.461 \\
    \hline
    \end{tabular}
    \label{tab:masses_mesons}
\end{table}
The baryon masses follow the recommendations
of the Particle Data Group \cite{ParticleDataGroup:2020ssz} except for the sexaquark that depends on the considered scenario, see section \ref{sec:res_GRDF}.
Here we assume the lowest mass of S that can be considered is determined by the value that avoids a condensation at baryon densities below the nuclear saturation density in symmetric nuclear matter in the DD2 model. The masses of the $\omega$ and $\rho$ mesons are standard values used in relativistic density functionals and $m_{\sigma}$ is a fit parameter of the DD2 interaction. 
The mass of the $\phi$ meson as well as those of the electron and muon are taken from \cite{ParticleDataGroup:2020ssz}. Values of the baryon-meson couplings $\Gamma_{m}$ at the reference density $n_{\rm sat}^{(v)}$ and the coefficients in the functions $f_{m}$, cf.\ eqs.\ (\ref{eq:f_rat}) and (\ref{eq:f_exp}), can be found in \cite{Typel:2009sy} and we recall that  
$\Gamma_{\phi}=\Gamma_{\omega}(n_{\rm sat}^{(v)})$. The reference density $n_{\rm sat}^{(v)}=0.149065$~fm$^{-3}$ is the nuclear saturation density of the DD2 parameterization corresponding to a scalar density of $n_{\rm sat}^{(s)}=0.139650$~fm$^{-3}$. The values $U_{\Lambda}=-28$~MeV,
$U_{\Sigma}=30$~MeV, and $U_{\Xi}=-14$~MeV from \cite{Raduta:2020fdn}
are adopted for the hyperon potentials at saturation in symmetric nuclear matter
to calculate the coupling factors (\ref{eq:g_Ysigma}).
The parameter $x_{S}$ in the mass shift
(\ref{eq:Deltam_S}) is varied in the model to explore changes of the EoS caused by different onsets for the condensation of sexaquarks.

All thermodynamic properties of the system can be derived from the grand canonical thermodynamic potential density (\ref{eq:Omega}), e.g., 
the pressure $P$ is simply given by $P=-\Omega$. The free energy density $\varepsilon$ is equal to the internal energy density $f$ and can be calculated easily as
\begin{equation}
    f = \varepsilon = \Omega+\sum_{i}\mu_{i}n_{i}^{(v)} \: .
\end{equation}
The speed of sound in unit of the speed of light is found from
\begin{equation}
\label{eq:c_s}
 c_{s}^2 = \frac{dP}{d\varepsilon}
\end{equation}
and should not exceed one to be physically permitted.

\section{High density equation of state}
\label{sec:QM}
\subsection{Quark matter EoS}
\label{ssec:QM}
At low temperatures and very high densities, it has been proven that the favorite state of three flavor color-superconducting quark matter would be the color-flavor-locking (CFL) phase \cite{Alford:1998mk}. Then one could ask what will happen to the quark matter state at lower densities,
in particular in the region of the quark-hadron phase transition.
In order to answer this question, one has to solve self-consistently the coupled gap equations for quark masses and diquark pairing gaps in the competing channels. 
Such a task has been attacked within the Nambu--Jona-Lasinio (NJL) model for dense quark matter \cite{Buballa:2003qv}. 
The solution which for the case of three flavors and three colors was presented for the first time in Refs. \cite{Ruester:2005jc,Blaschke:2005uj,Abuki:2005ms} shows that there is a corridor for the 2-flavor color superconducting (2SC) phase between hadronic matter and the CFL phase.

We consider here the scenario that at high enough density S particles are dissociated into a 2SC phase of quark matter which is microscopically described by non-local NJL (nlNJL) model while its formulation has been fitted \cite{Antic:2021zbn} to the simple parameterization of constant speed of sound (CSS) \cite{Zdunik:2012dj} at higher densities and zero temperature.
The microscopic nlNJL model is a covariant extension of the NJL model in which the quark fields interact via momentum dependent vertices \cite{Schmidt:1994di,GomezDumm:2005hy, Blaschke:2007ri}. It has been shown in \cite{Antic:2021zbn} that the EoS obtained from this method is fitted very well (regarding the $\chi^2$ value) to CSS parameterization for quark matter EoS. Indeed, a mapping from the 2-dimensional space of nlNJL model to the 3-dimensional space of CSS model has been performed in this paper using a simple functional. With this functional that is introduced in \cite{Antic:2021zbn}, the EoS of color superconducting quark matter can be obtained for each value of vector meson coupling ($\eta_V$) and diquark coupling ($\eta_D$). The resulting EoS which has a microscopic justification, is fitted to the following parameterization for CSS quark matter
  
\begin{equation}
\label{eq:CSS}
P_{QM}(\mu)= A(\mu/\mu_0)^{(1+1/c_s^2)} - B,
\end{equation}
where $\mu_0 = 1000$ MeV defines a scale for the chemical potential. In \eqref{eq:CSS}, there are three free parameters which have to be defined. The squared speed of sound $c_s^2=dP/d\varepsilon$ has to be large enough 
so that the maximum mass $M_{\rm max}$ of the corresponding neutron star sequence exceeds the observational lower bound on 
it, which presently amounts to $2.01~M_{\odot}$  at 68.3\% credibility \cite{Fonseca:2021wxt}. 
The prefactor $A$ could change the slope of the $P-\mu$ line and the effective bag pressure $B$ imposes the confinement effect at low densities in quark matter EoS and makes the pressure to have negative value in this region. Once the pressure as a function of chemical potential is obtained, one could calculate the baryon density ($n_b=dP/d\mu$) as well as the energy density ($\varepsilon=n_b\mu-P$) for the quark matter.
For the relation between the nlNJL and CSS models of quark matter, see also the recent work \cite{Contrera:2022tqh}.

In \cite{Antic:2021zbn}, for nlNJL parameters $0.70<\eta_D<0.80$ and $0.11<\eta_V<0.18$, the parameter mapping to the space of the CSS model resulted in the following range of parameters: $0.449<c_s^2<0.541$, $91.484<A[{\rm MeV/fm}^3]<101.116$ and $82.437<B[{\rm MeV/fm}^3]<92.290$.
In the current work, we use several different sets of these parameters in order to find out which of them is modeling the hybrid star with S in the outer core in such a way to fulfill all observational constraints. 

We consider four different scenarios for solving the S dilemma in compact stars denoted as follows, where the subscript 
$i$ labels two mass choices for the S ($i=1885, 1941$), in MeV:
\begin{itemize}
  \item Scenario DS$_i$Y: An early deconfinement (D) before sexaquark (S$_i$) and hyperon (Y) onset,
\item Scenario S$_i$DY: deconfinement after sexaquark but before hyperon onset,
\item Scenario S$_i$YD: deconfinement after both sexaquark and hyperon onset when S is prior,
\item Scenario YS$_i$D: like previous case but hyperon is prior.
 \end{itemize}
In this work, the value of $\eta_D$ has been taken to be $0.75$
in accordance to the Fierz transformation of a one-gluon exchange interaction \cite{Buballa:2003qv}. 

For the present hadronic matter EoS and the quark matter EoS, the transition from hadronic matter to deconfined quark matter occurs at low chemical potential. Moreover, because of the softening of the hadronic matter EoS at higher chemical potential after the appearance of hyperons, we encounter the reconfinement problem \cite{Zdunik:2012dj} which will be ignored in applying the Maxwell construction following the argument \cite{Shahrbaf:2019vtf} that once the deconfinement occurs at a certain critical density, the hadronic matter EoS is not valid anymore beyond that point.

For moving the transition point to higher chemical potentials, we may apply an extra bag pressure to the quark matter EoS which could be either a constant one ($B_0$) or a $\mu-$dependent one. The $\mu-$dependent bag pressure 
$B_{\rm eff}(\mu)$ is efficient for making 
a stronger phase transition with a big jump in density (a bigger difference in the slope of hadronic matter and quark matter EoS at the transition point) at the transition point.

The larger the density jump, the stronger the effect of softening the EoS which leads to more compact hybrid stars at the transition. This may help to fulfill the demanding constraint on the tidal deformability, $70 < \Lambda_{1.4} < 580$ at $1.4~M_\odot$, deduced from the gravitational wave signal of the binary neutron star merger GW170817 \cite{LIGOScientific:2018cki}.
For a sufficiently large jump in energy density 
($\Delta \varepsilon \gtrsim \varepsilon_c$, see 
\cite{Alford:2013aca} for details), 
a gravitational instability may occur which, after stability is recovered for sufficiently stiff quark matter at higher densities, leads to the formation of an alternate branch of hybrid stars in the $M-R$ diagram \cite{Gerlach:1968zz}. 
This potentially observable feature of a strong phase transition is accompanied with the phenomenon of mass-twin stars \cite{Glendenning:1998ag}. 
Due to the constraint of causality on the speed of sound, however, a parameterization of $B_{\rm eff}(\mu)$ with a sufficiently large value of $\Delta \varepsilon$
may not be admissible. 

The $\mu$-dependence of the bag pressure has been introduced in \cite{Alvarez-Castillo:2018pve} in the following form
\begin{equation}
\label{eq:bag}
B_{\rm eff}(\mu)=B_1f_<(\mu)~,
\end{equation}
with the switching functions
\begin{equation}
\label{eq:f<}
f_{<}(\mu)=\frac{1}{2}\left[1-\tanh\left(\frac{\mu-\mu_<}{\Gamma_<}\right)\right]~.\\
\end{equation}
While in  \cite{Alvarez-Castillo:2018pve} the parameterization of $B_{\rm eff}(\mu)$ resulted in mass-twins, in the present work (as also in Ref.~\cite{Contrera:2022tqh}), it will suffice to accommodate the tidal deformability constraint but not lead to mass-twins.
It is worth mentioning that in \eqref{eq:CSS}, the value of $B$ reads
\begin{equation}
\label{eq:totalbag}
B = B_0+B_{\rm eff}(\mu) 
\end{equation}
in which $B_0$ is a constant value and $B_{\rm eff}$ is defined according to \eqref{eq:bag}.
We have changed the parameters in above definition as well as the parameters of quark matter EoS several times to find the best sets (to the best of our knowledge) of parameters which not only fulfills all observational constraints but also results in an appropriate transition point to be a proper solution for one of the considered scenarios in this work.

\subsection{Hybrid EoS}

Within the two-phase approaches to hybrid neutron star matter, the most common phase transition constructions are the Maxwell construction (MC) and the recently developed crossover interpolations, see \cite{Baym:2017whm} for a recent review. 
While the MC has been in use already since the early days of discussing quark deconfinement in neutron stars \cite{Baym:1976yu} with varying but physically equivalent formulations, for the crossover interpolations a word of caution may be in order.

The idea of the interpolation construction pioneered in 
\cite{Masuda:2012ed,Masuda:2012kf} is to facilitate a thermodynamically consistent transition from a relatively soft hadronic EoS that could be trusted up to and slightly beyond saturation density, to a stiff quark matter EoS with a region of validity above 3-5 times saturation density.
Strictly speaking, in the crossover transition region neither of the input EoS are trustworthy. Therefore, no Maxwell construction would be applicable and alternatives have been developed to interpolate between soft hadronic and stiff quark matter EoS in the crossover region. 
The first construction \cite{Masuda:2012kf} was not formulated in natural variables of the thermodynamic potential, which was corrected in the second version \cite{Masuda:2012ed}, but both interpolations were defined by a ``mixing" of the input EoS, even in the crossover region where they are strictly speaking not applicable. 
Furthermore, the weight functions for the mixing were chosen as Fermi functions with the unphysical implication of a nonzero probability for quark matter at low and hadronic matter at high densities.
An alternative to facilitate treatment of the hadron-to-quark matter crossover is the replacement interpolation construction (RIC) developed in \cite{Ayriyan:2017nby,Abgaryan:2018gqp} to describe the situation of a mixed phase due to pasta structures in the hadron-to-quark matter transition. The RIC treatment was applied to the stiff-soft transition case in the context of solving the hyperon puzzle \cite{Shahrbaf:2019vtf}
without introducing an unphysical high-density hadronic component. A more general two-zone interpolation method that followed the intentions discussed in \cite{Baym:2017whm} has recently been developed in \cite{Ayriyan:2021prr} and used within a Bayesian analysis of neutron star constraints on the EoS.

Here we employ both the MC as well as RIC for the first order phase transition from DD2Y-T+S EoS to CSS quark matter EoS, to investigate 
the appearance of S particles in hybrid stars.
To do this, we try to find a stable hybrid star with the 2SC phase of quark matter in the core surrounded by a layer of nuclear matter or hypernuclear matter which includes S. 
\subsubsection{Maxwell construction}
In the MC, the Gibbs conditions for phase equilibrium have to be fulfilled globally, i.e.                            
 \begin{eqnarray}
 \mu_H&=&\mu_Q=\mu_c~,\\
 T_H&=&T_Q=T_c~,\\
 P_H(\mu_b,\mu_e)&=&P_Q(\mu_b,\mu_e)=P_c~.
 \end{eqnarray}
The above conditions guarantee the chemical, thermal and mechanical phase equilibrium at the transition point between hadronic phase and quark matter phase for which the critical values of the thermodynamic variables are denoted with the subscript $c$. We apply the conditions for the zero temperature case in the present work.
 
Within the MC scheme, the reconfinement phenomenon \cite{Zdunik:2012dj} can occur which consists in a second (unphysical) crossing of the hadronic and quark matter EoS.
It can be dealt with in the MC by ignoring the unphysical crossing. 
The reconfinement situation can also be removed by applying a RIC with negative value of $\Delta_p$, as it has been discussed in \cite{Shahrbaf:2020uau}. 

\subsubsection{Replacement interpolation construction}
Within the RIC, we assume that neither the hadronic matter nor the quark matter EoS  are reliable in a corridor around their unphysical crossing, e.g., in the reconfinement region. Therefore, two boundary points, i. e., $\mu_{H}$ before the reconfinement and $\mu_{Q}$ after the reconfinement, are defined so that the interpolated pressure could be described by a parabolic form between them,
\begin{equation}
 P_M(\mu)~=~\alpha_2(\mu-\mu_c)^2~+~\alpha_1(\mu-\mu_c)~+~(1+\Delta_P)P_c,
  \label{eq:pppressure}
\end{equation}
where the $\mu_c$ and $p_c$ corresponds to the critical point at which the wrong Maxwell transition from quark matter to hadronic matter has occurred.
The $\alpha_1$, $\alpha_2$ as well as $\mu_{H}$ and $\mu_{Q}$ could be obtained from the continuity conditions at the borders of the mixed-phase
\begin{eqnarray}
P_H(\mu_{H}) &=&  P_M(\mu_{H})~,\\
P_Q(\mu_{Q}) &=&  P_M(\mu_{Q})~,\\
\frac{\partial}{\partial\mu}P_H(\mu_{H}) &=&  \frac{\partial}{\partial\mu}P_M(\mu_{H})~,\\
\frac{\partial}{\partial\mu}P_Q(\mu_{Q}) &=&  \frac{\partial}{\partial\mu}P_M(\mu_{Q})~.
\label{eq:continuity}
\end{eqnarray}
The RIC allows us to have a solution for which the quark core of NS is surrounded by a mixture of usual nuclear matter, sexaquark and even hyperons if the upper boundary of the mixed phase goes beyond the hyperon onset. Therefore, we can have solutions for all scenarios in this way. Moreover, changing the parameters of the RIC enable us to have a larger part of the sequence in the sexaquark-nuclear phase than in the MC case.

\section{Purely hadronic stars and their problems}
\label{sec:res_GRDF}

Since numerous abbreviations have been used in this work, we collect them all together for the convenience of the readers in 
Appendix \ref{app:A}. 

With the generalized relativistic density functional (GRDF) for hadronic matter at hand, it is possible to explore the EoS and corresponding properties of neutron stars. There are various scenarios to be distinguished in the following. The simplest case with only nucleons and leptons corresponds to the original DD2 model as presented in \cite{Typel:2009sy}. Adding hyperons the model is called DD2Y-T  to distinguish it from the similar model DD2Y introduced in \cite{Marques:2017zju}. The DD2Y-T predictions were compared already to other EoS models with hyperons in \cite{Stone:2019blq}. Finally, after including also the sexaquark, there is the full model which will be denoted DD2Y-T+S in the following. 

To start, the mass of the sexaquark is taken to be constant. 
When the mass of the sexaquark is equal to $2054$ MeV, the onset of the S (sexaquark) occurs at about 0.25 fm$^{-3}$ (more than $1.5$ times the saturation density) with an immediate appearance of the BEC and the neutron star will collapse with increasing central density because the pressure remains constant.
 
The obvious suggestion is including the repulsive interaction of S by considering an increasing density-dependent mass for S particles, as done for hyperons.
Considering the S particle as a composite system of diquarks embedded in the baryonic medium suggests that a dissociation of the bound state into a many-body correlation in the continuum could be studied in a microscopic model.
But this requires intricate calculation that are not in the scope of the present work and we leave it for the future. 
In this work, we only consider the S particle as a point-like boson with a medium dependent mass. As it was mentioned in subsection A, we suggest that all possible substructures can lead to a medium dependent mass of S particles.

Hereafter, we consider a linear density dependence of the mass of the S in which the mass shift is positive so that the instability problem of NS is prevented since the corresponding increase of the chemical potential following the condensation criterion allows a rising pressure.
At zero density, the effective mass of S particles, i.e., $m_S^{\ast}$ is given by 
the vacuum mass of S particle, $m_S$, and the rate of increase that is given by a factor $x_{S}$ times the baryon density $n_{b}$ in the unit of the reference density $n_0= 0.15$~fm$^{-3}$,
\begin{equation}
\label{eq:m*}
m^*_S(n_{b},x_{S})=m_S\left(1+x_{S} \frac{n_{b}}{n_{0}}\right) \: .
\end{equation}
This assumption results in an increase of the onset density of S formation and subsequent condensation, so that there is still an increase of the pressure at higher densities.

An estimate of reasonable values for $x$ can be derived from the change of the effective potential $U_{i}=V_{i}-S_{i}$ of the octet baryons with the baryon density $n_{b}$ as 
\begin{equation}
    x_{i} = \frac{n_{0}}{m_{i}}\frac{dU_{i}}{dn_{b}} \: .
\end{equation}

\begin{figure}[!h]
	\includegraphics[width=0.48\textwidth]{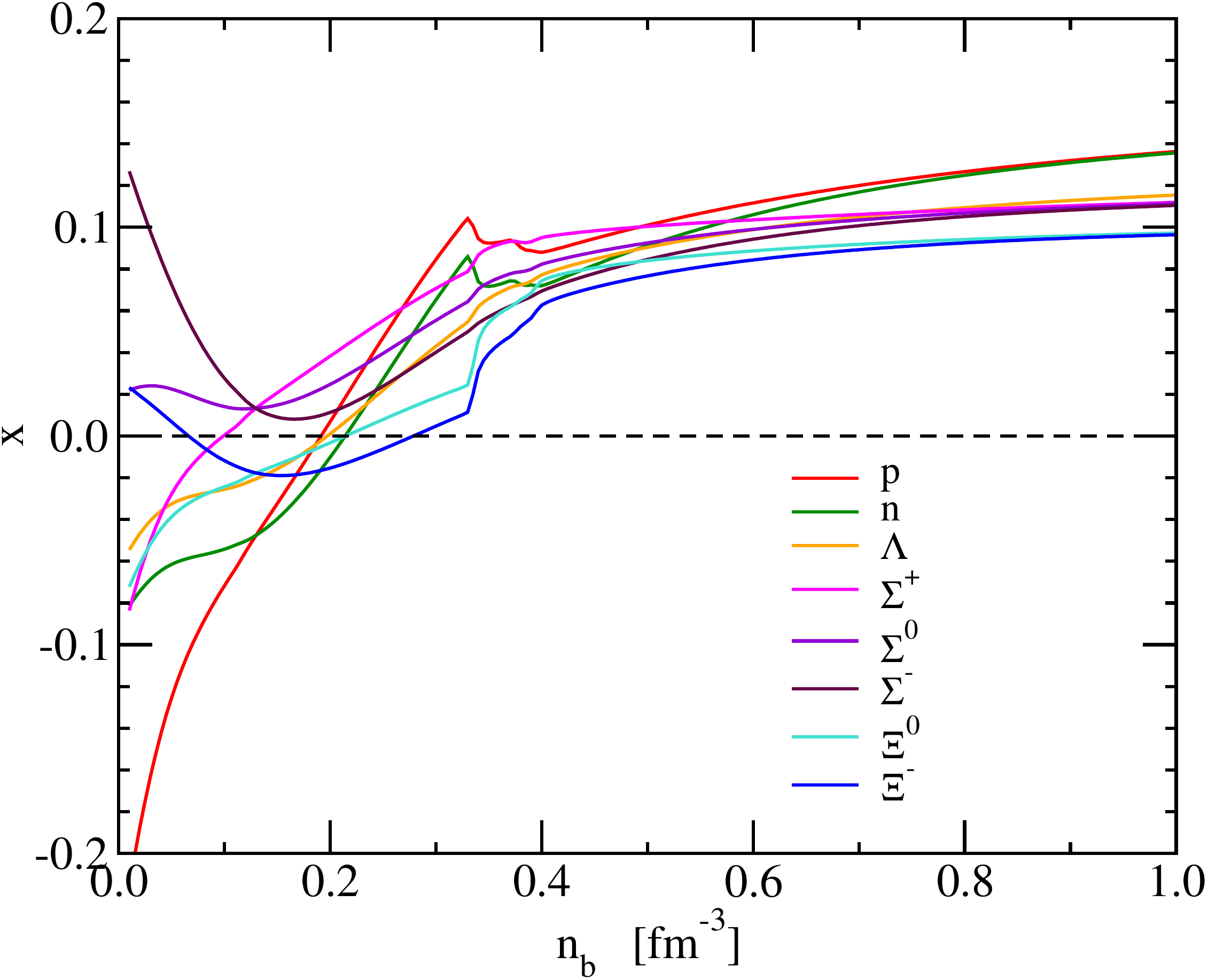}
	\caption{The effective slope of the mass shift for octet baryons calculated based on the effective potentials the in DD2Y-T model.
		\label{fig:effectivex}
	}
	\end{figure}

These effective slopes calculated in the DD2Y-T model are depicted in Fig.~\ref{fig:effectivex}.
The slopes for the various hyperons are typically large and positive near nuclear saturation density and transition to being slowly increasing beyond $n_b \approx 0.4 \, {\rm fm}^{-3}$.  As we shall see, without a transition to a deconfined phase the observed range of neutron star masses cannot be reproduced, so only a portion of the range shown in Fig.~\ref{fig:effectivex} will prove to be physically relevant.  It should also be noted that the constant-slope approximation we use for the S in the present work may be an over-simplification as it is only applicable for hyperons for relatively small ranges of $n_b$.

\begin{figure}[H]
	\includegraphics[width=0.48\textwidth]{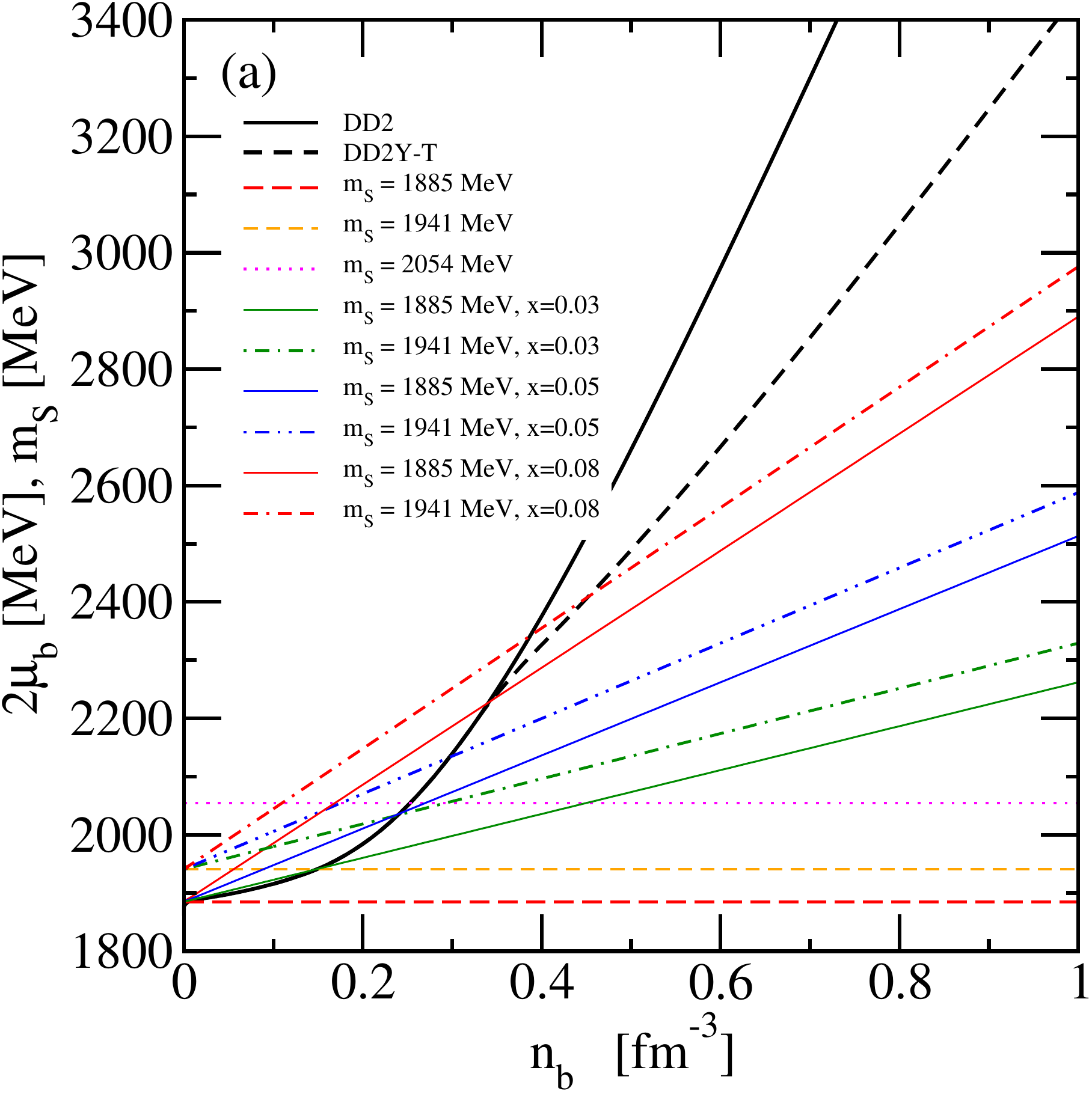}
	\includegraphics[width=0.48\textwidth]{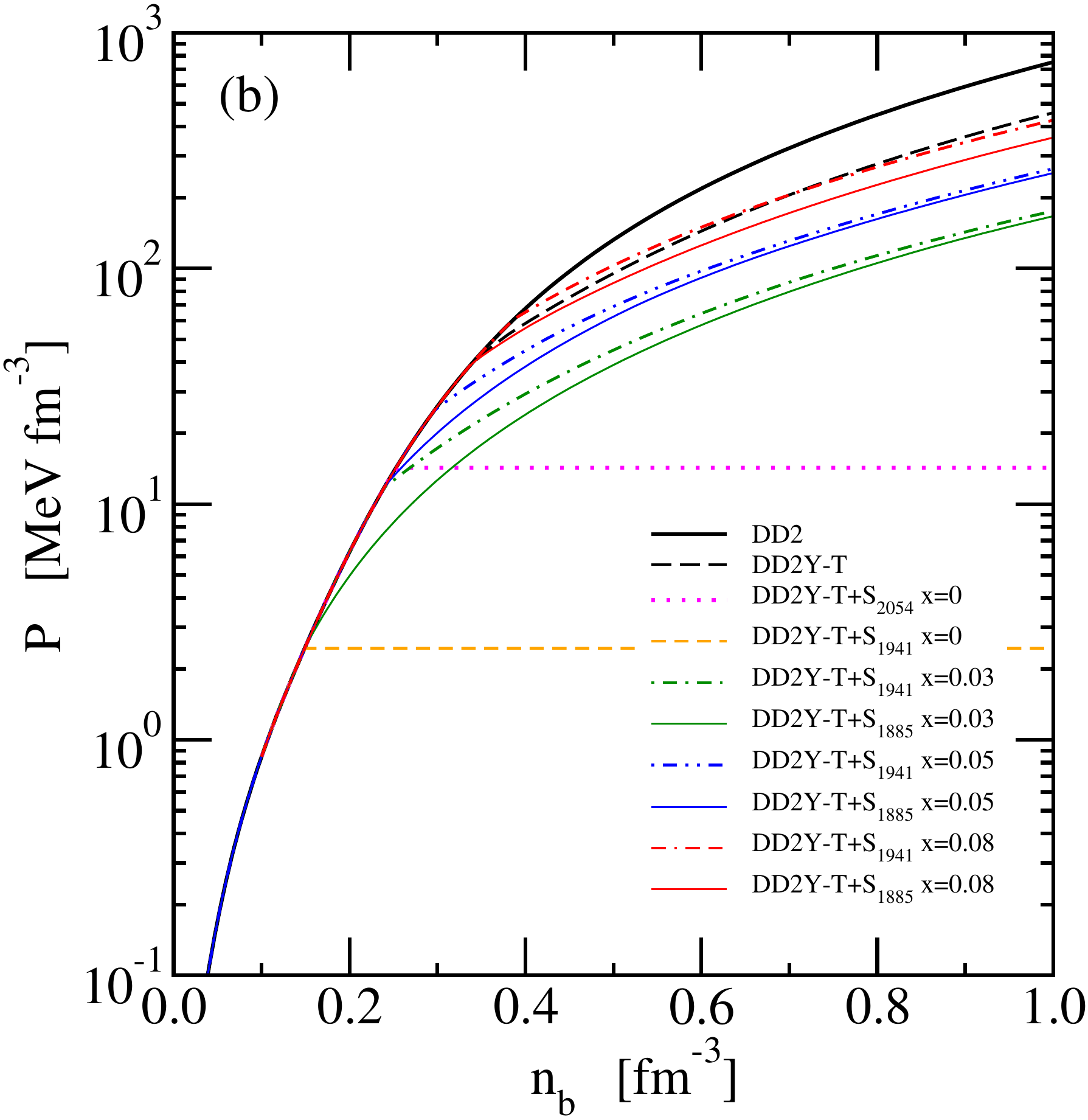}
	\caption{Panel (a) shows double baryonic chemical potential as a function of baryon density while panel (b) shows the pressure as a function of baryon density 
	for DD2, DD2Y-T and DD2Y-T+S for three different slopes of the mass shift of S particles.
		\label{fig:2mubn}
	}
	\end{figure}

The onset density for the condensation of S depends on its rest mass in vacuum and the slope parameter $x$. One can see this onset by plotting $2\mu_{b}$ as a function of density for different parameter choices, as shown in the panel (a) of Fig. \ref{fig:2mubn}. The onset density occurs as soon as the baryo-chemical potential of the hadronic matter fulfills $2\mu_b=m_S$.
The full and dashed black lines in figure \ref{fig:2mubn} show the value of $2\mu_{b}$
as a functions of $n_{b}$ for the DD2 and DD2Y-T models. 
The effective mass of S is also plotted in the same figure, for three values of the vacuum rest mass and several slope parameters. 
When the vacuum mass of S is $\leq 1885$ MeV, Bose condensation occurs for any density in the absence of repulsive interactions between S and nucleons and more careful modeling would be required.  Thus we do not consider lower masses for the S here.  

For $m_S = 1941$ MeV, condensation occurs exactly at nuclear saturation density if $x=0$, while for $S_{2054}$ the intersection point with the black lines occurs at a higher baryon density of about $0.25$~fm$^{-3}$. Obviously, the onset of condensation occurs at higher densities as $x$ increases.
Therefore, varying the value of $x$ results in different scenarios for NS regarding the priority of Y onset and S onset if the density-dependence of hyperon masses is taken as given.


Different behaviors of the effective mass of S are reflected in how the pressure changes with baryon density, as shown in panel (b) of Fig.~\ref{fig:2mubn} for illustrative cases. 
The pressure is constant beyond the onset density for Bose condensation  causing an instability of the neutron star calculated with the corresponding equations of state. For positive $x$, the increase of the effective mass causes a rise of the pressure; this could also be attributed to the effect of a repulsive potential at short distances.

\begin{figure}[h]
	\includegraphics[width=0.46\textwidth]{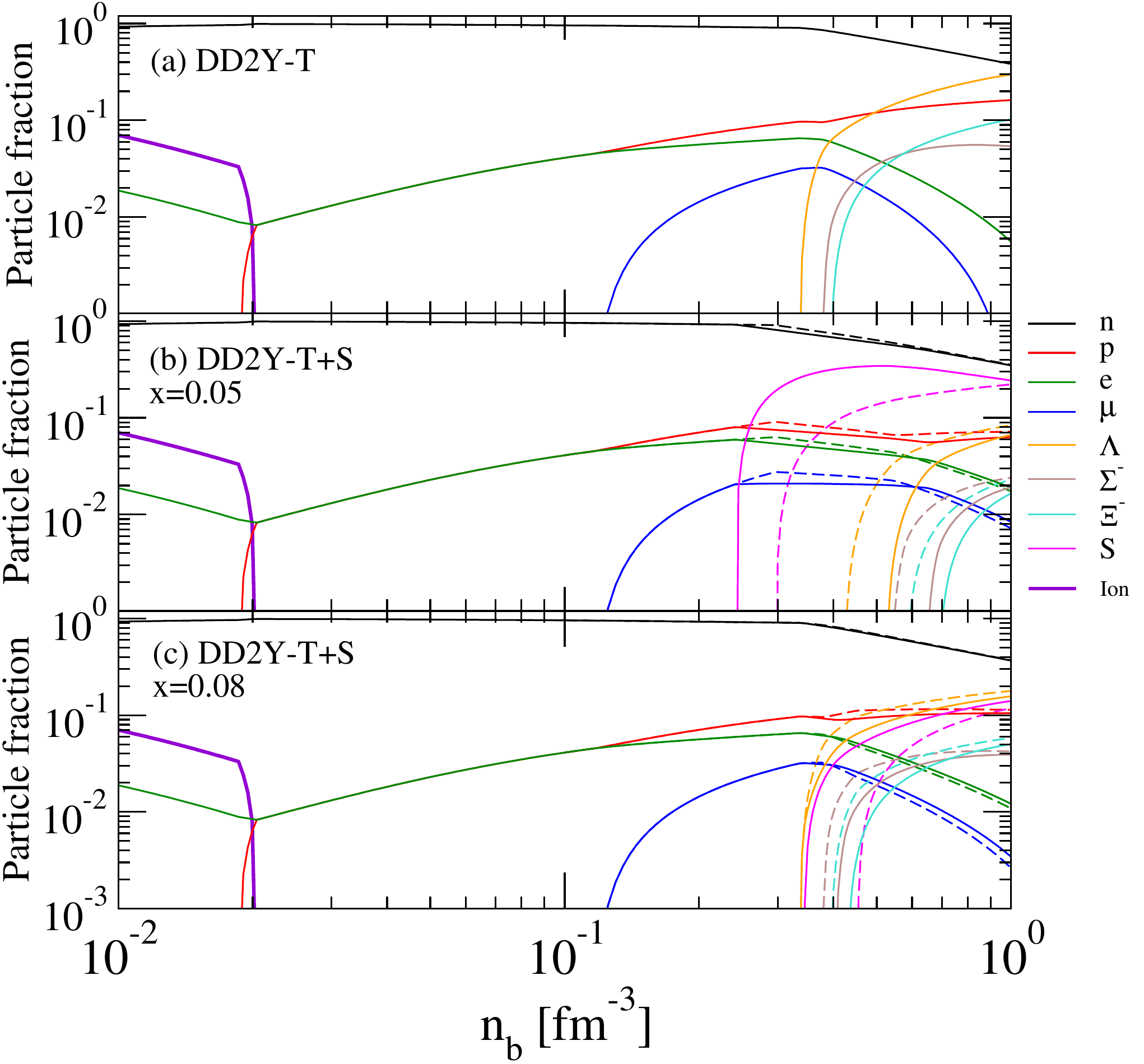}   
	\caption{Particle fraction of hadronic matter including an increasing S mass 
	in $\beta$-equilibrium for NS matter. The solid lines correspond to S$_{1885}$ while the dashed lines corresponds to S$_{1941}$. DD2Y-T, DD2Y-T+S with $x=0.05$ and DD2Y-T+S with $x=0.08$ are shown in panel (a), (b) and (c) respectively.
		\label{fig:eos2}
	}
\end{figure}

The particle fractions as a function of baryon density in $\beta$ equilibrated matter are shown in Fig.~\ref{fig:eos2} where S is taken to have an effective mass that increases with the density. The dashed lines show the results for the heavier S, i.e., S$_{1941}$ ($m_{S}=1941$ MeV) and the solid lines correspond to the lighter S, i.e., S$_{1885}$ ($m_{S}=1885$ MeV). As the figure shows, for $x=0.05$ the S particles appear before hyperons but for $x=0.08$, S onset occurs after the hyperon onset. However, for the lower mass of S, the hyperon onset is very close to the S onset and the only hyperon which appears before S, is the lightest one, i.e., $\Lambda$.
It is concluded from Fig.~\ref{fig:eos2} that the appearance of S affects the hyperon onset. When there is a delay in S$_{1941}$ onset because of the larger mass, hyperons appear at lower densities compared to the S$_{1885}$ case. 

The speed of sound $c_{s}$ is a quantity that can be derived directly from the pressure and energy density, see equation (\ref{eq:c_s}). It has to be smaller than the speed of light to have a causal EoS.
\begin{figure}[!bht]
	\includegraphics[width=0.44\textwidth]{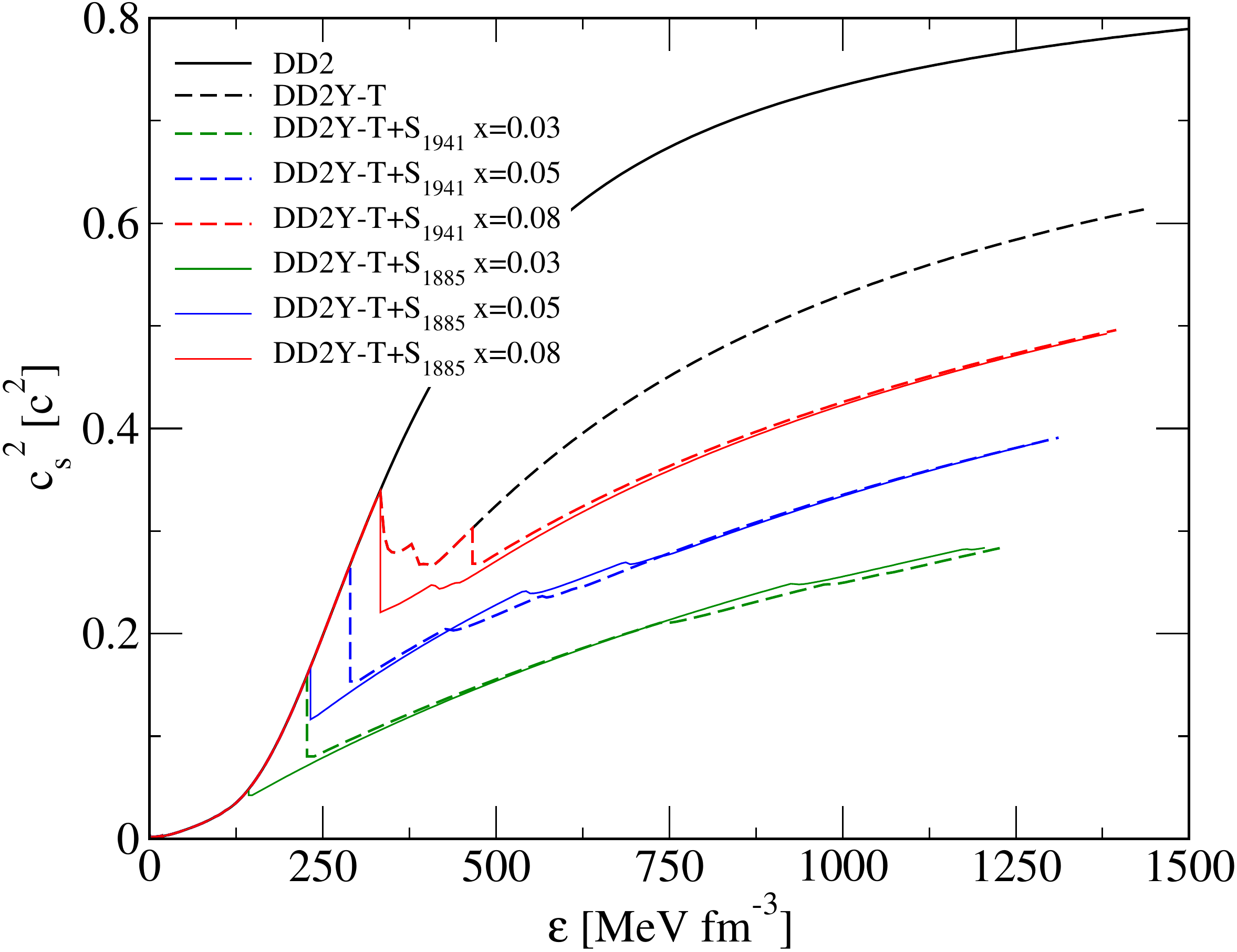}
	\caption{The square of speed of sound as a function of energy density for the purely hadronic EoS cases.
		\label{fig:cs2hadronic}
	}
\end{figure}

The square of $c_{s}$ for the hadronic equations of state is depicted in Fig. \ref{fig:cs2hadronic} showing that the causality condition is fulfilled in the whole range of energy densities for all combinations of S mass and slope parameter $x$.

\subsection{Sexaquark and hyperon dilemma in neutron stars}
\begin{figure}[H]
	\includegraphics[width=0.45\textwidth]{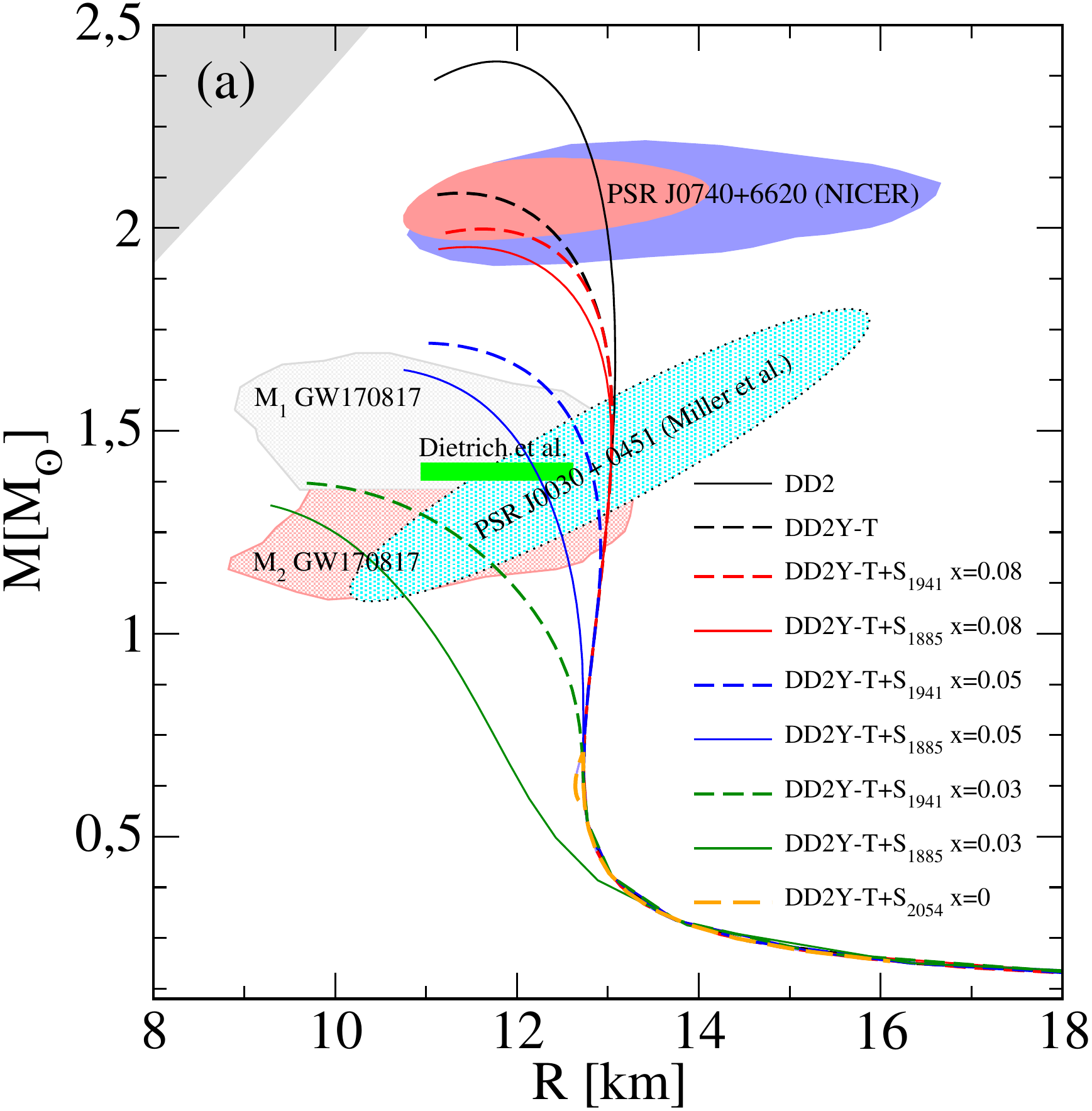}\\
	\includegraphics[width=0.44\textwidth]{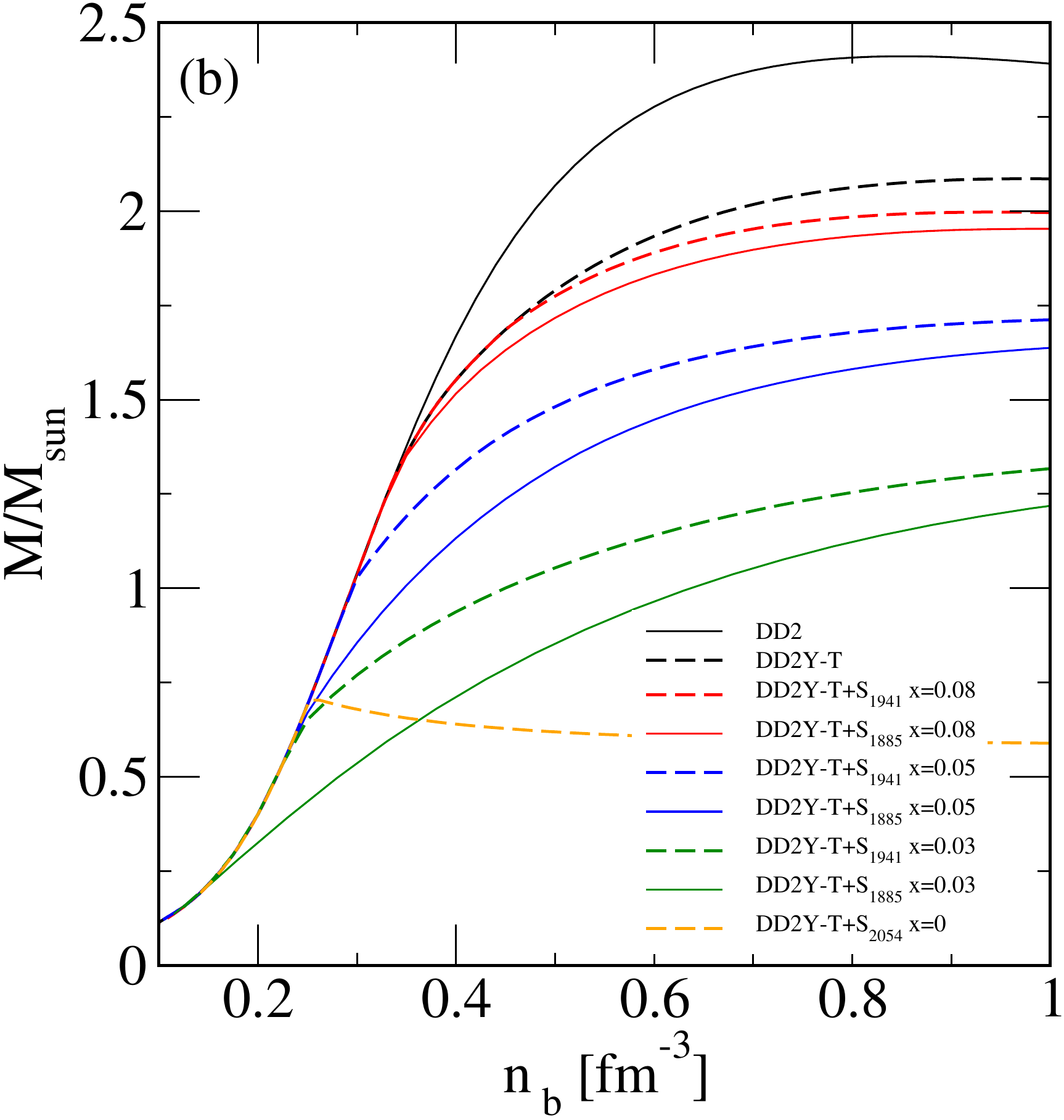}
\caption{Mass-radius relation (panel (a)) and mass versus central density (panel (b)) for compact stars in which the S particle is supposed to have an effective  mass $m_S^{\ast}(n_{b},x)$ given by equation (\ref{eq:m*}) with $m_S=1941$, and $1885$ MeV and $x=0.03, 0.05, 0.08$. The solid lines correspond to S$_{1885}$ while the dashed lines correspond to S$_{1941}$. For a comparison the new $1-\sigma$ mass-radius constraints from the NICER analysis of
observations of the massive pulsar PSR J0740+6620 \cite{Fonseca:2021wxt} are indicated in red \cite{Riley:2021pdl} and blue \cite{Miller:2021qha} regions. Additionally, the green bar marks the radius of a $1.4$ solar mass neutron star from a joint analysis of the gravitational-wave signal GW170817 with its electromagnetic counterparts at $90\%$ confidence \cite{Dietrich:2020efo}.  
The cyan region is from the NICER mass-radius measurement on PSR J0030+0451 \cite{Miller:2019cac}. The gray and light orange regions corresponds to the estimates of the components of the binary system labeled as $M_1$ and $M_2$ of the GW170817 merger \cite{abbott2018gw170817}. 
\label{fig:Mx} 
}
\end{figure}

Each combination of $m_{S}$ and $x$ leads to an equation of state of neutron star matter and corresponding mass-radius relation that can be compared to the predictions of the original DD2 and DD2Y-T models without sexaquark.
In Fig.~\ref{fig:Mx}, the mass-radius (panel (a)) and mass-density (panel (b)) relations for neutron stars obtained from DD2 and DD2Y-T as well as DD2Y-T+S are shown when the S particles are supposed to have a mass that increases with density.

The S dilemma for the DD2Y-T+S models, i.e., the decrease of the maximum neutron star masses due to the softening of the EoS with the appearance of S, is also seen. As the figure shows, for $x=0.03, 0.05$ the maximum mass of the neutron stars does not reach 2M$_\odot$ and more repulsive effects are needed in hadronic matter to reach the lower bound of the maximum mass of neutron stars, as exemplified by $x=0.08$. 

Another important constraint for an EoS is the tidal deformability of a neutron star, deduced by the LIGO and Virgo collaborations from GW170817 
\cite{LIGOScientific:2018cki}
and translated to a multi-messenger constraint on the radius at $1.4~M_\odot$ in \cite{Dietrich:2020efo}, shown as green bar in panel (a) of
Fig.~\ref{fig:Mx}.  
The only line which crosses it is the blue solid line corresponding to DD2Y-T+S$_{1885}$ with $x=0.05$.

\begin{figure}[!htb]
 	\includegraphics[width=0.45\textwidth]{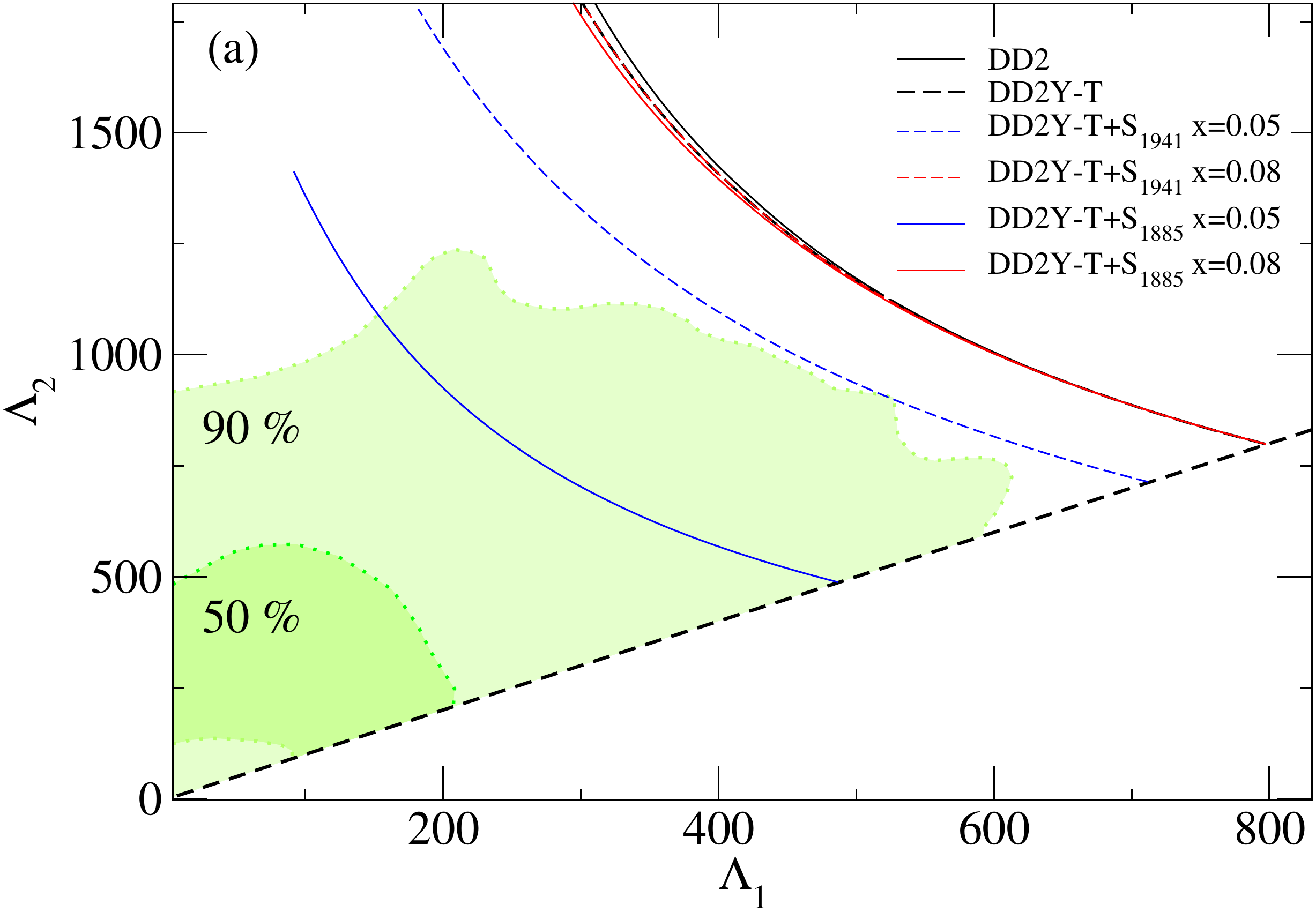}\\
 	\includegraphics[width=0.45\textwidth]{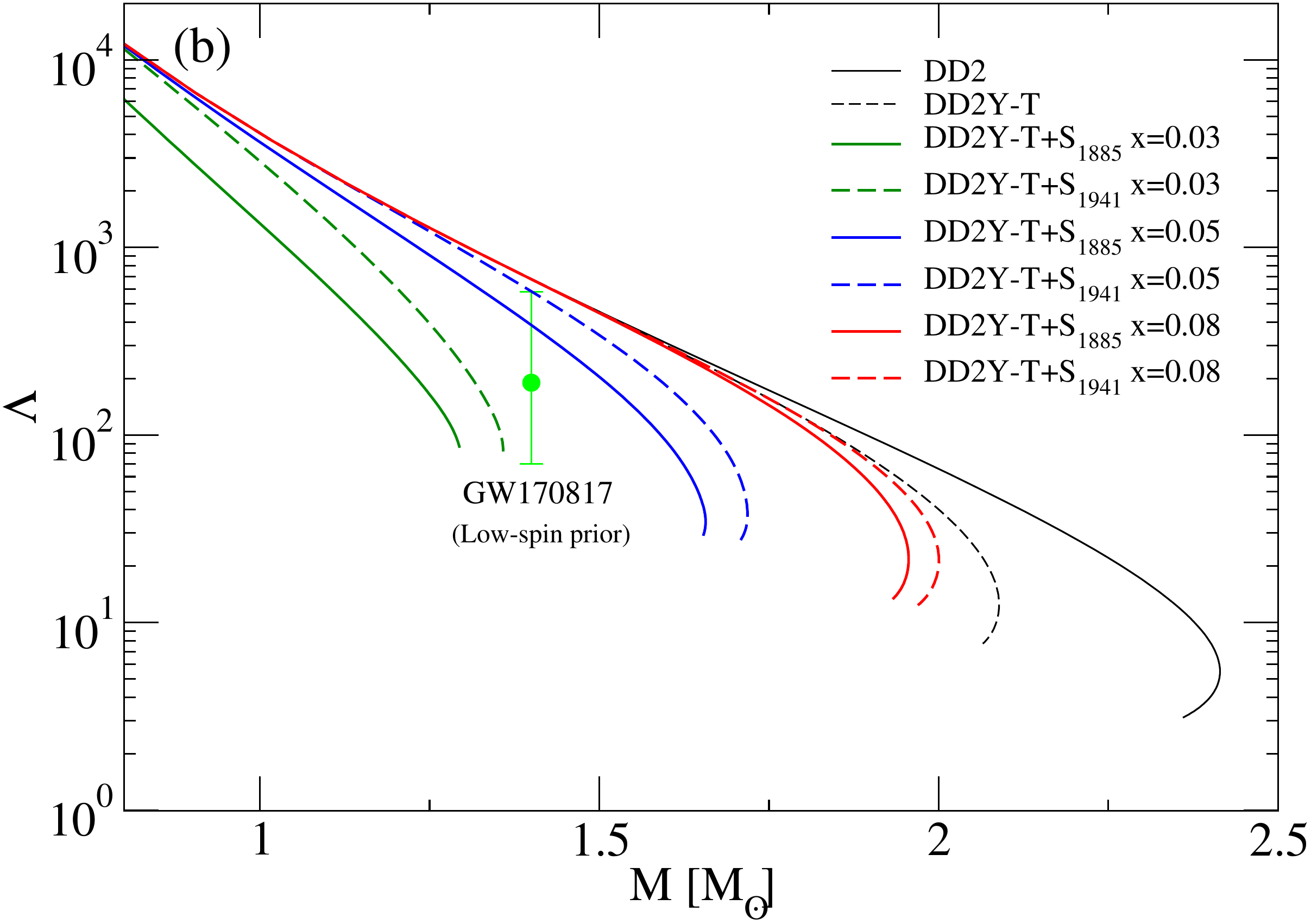}
 	\caption{(a): Tidal deformability parameters $\Lambda_1$ and $\Lambda_2$ of the high- and low-mass components of the binary merger. The results have been shown for $m_S=1941, 1885$ MeV and $x=0.050, 0.08$. The solid lines correspond to S$_{1885}$ while the dashed lines correspond to S$_{1941}$. The green region is the placed constraints on the tidal effects by LIGO and Virgo collaboration from GW170817 
 	\cite{LIGOScientific:2018cki}.
 	(b): Dimensionless tidal deformability $\Lambda$ as a function of the star mass. The green line
shows the $\Lambda_{1.4}$ constraint from the low-spin prior analysis of GW170817 \cite{LIGOScientific:2018cki}.
 		\label{fig:tidal11}
 	}
 \end{figure}

The tidal deformability parameters $\Lambda_1$ and $\Lambda_2$ of the high- and low-mass components of the binary merger event have been constructed from the $\Lambda(M)$ relation 
\cite{LIGOScientific:2018cki}.
We show in Fig.~\ref{fig:tidal11} the observational constraints for the tidal deformability parameters along with predictions for the various equations of state. 

As can be seen, the only EoS which lies in the green credibility region, is DD2Y-T+S when $m_S=1885$~MeV and $x=0.05$.  However, this model does not fulfill the constraint on the  maximum mass, $M_{\rm max}\ge 2~M_\odot$ \cite{Fonseca:2021wxt}, see Figs.~\ref{fig:Mx} and \ref{fig:tidal11} panel (b). 
It should also be stressed that the standard hadronic matter model, DD2Y-T, also fails to satisfy 
the tidal deformability constraint.  

Therefore we conclude that within our setting, where the stiff nucleonic RMF parameterization "DD2" (that violates the tidal deformability constraint) sets the hadronic matter baseline, a purely hadronic EoS is ruled out -- independent of the existence of a sexaquark. 

\begin{figure}[!thb]
        \includegraphics[width=0.55\textwidth]{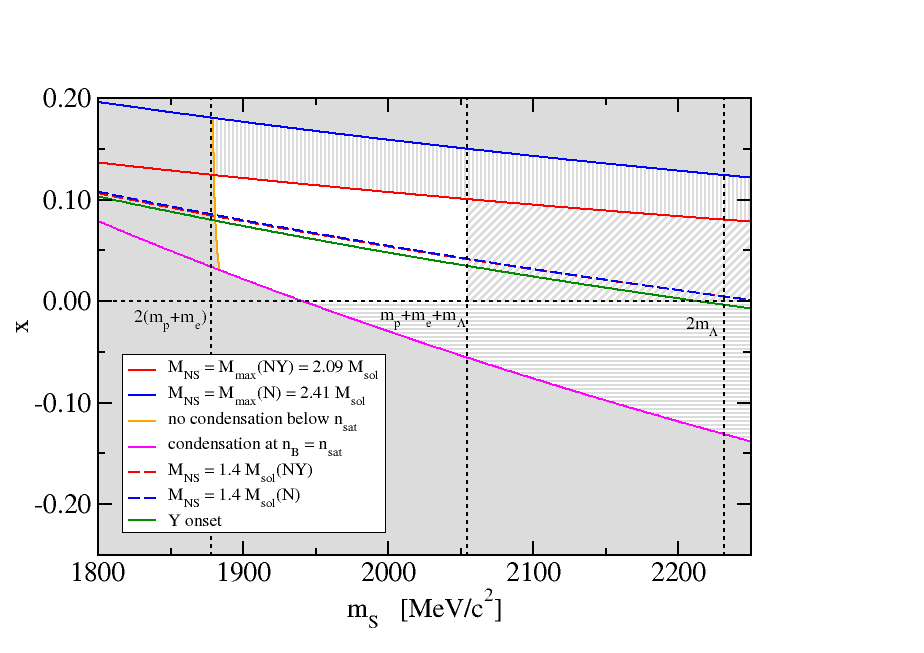}
        \caption{Allowed (white) and excluded (light and dark grey) regions in the plane of $m_{S}$ and $x$ values as defined by constraints on the sexaquark condensation. See text for details and explanation of the additional lines.
        The region of masses $m_S>m_p+m_e+m_\Lambda = 2054$ MeV is also excluded by the requirement of sexaquark stability on cosmological timescales, if S is the dark matter.
 For further details, see text.
        \label{fig:xmassS}
}
        \end{figure}

In Fig.~\ref{fig:Mx}, some special values for the slope factor $x$ and $m_S$ have been selected based on the range which fulfills the constraints of the combination of $m_S$ and $x$. These constraints for $x$ as a function of $m_S$ are shown in Fig. \ref{fig:xmassS}.
The boundaries are defined as follows:
\begin{enumerate}
    \item S appears below the maximum mass of NS in matter with
    only nucleons: region above blue solid line is excluded.
    \item S appears below the maximum mass of NS in matter with nucleons and hyperons: region above red solid line is excluded.
    \item Condensation of S occurs only above the saturation density: region below violet solid line is excluded.
     \item Condensation of S doesn't occur below the saturation density: region left of yellow solid line is excluded.
    \item No decreasing mass of S (i.e., constant pressure above density of
    condensation and below endpoint of neutron star $M$-$R$ relation): region below $x=0$ is excluded.
    \item Avoiding an unstable S that decays in $10^{-10}$s: region above $2054$ MeV is excluded ($m_S>m_p+m_e+m_\Lambda = 2054$ MeV) 
\end{enumerate}
Therefore, only the white region remains.
However, the border of the white region which lies on the line
$x=0$ is also strictly excluded  within the framework of our analysis because it leads to Bose-Einstein condensation (BEC) of a constant-mass S which limits the maximum mass of purely hadronic neutron stars to $M_{\rm max}\lesssim 0.7~M_\odot$.
A scenario according to which such a BEC of S  triggers  a catastrophic rearrangement of the neutron star structure (star quake) which entails the conversion to a hybrid neutron star with a color superconducting (CFL) quark matter core has recently been considered in \cite{Blaschke:2022knl}.

In addition to these boundaries, three dotted vertical lines for specific values of the sexaquark mass are shown. Furthermore the green line corresponds to the condition of S condensation at the onset of hyperons. The sexaquark appears before (after) hyperons for points below (above) this line. Dashed blue and red lines, which are almost on top of each other, show S condensation at the center of a neutron star, which includes nucleons and nucleons-hyperons, respectively, with a mass of $M = 1.4$~M$_\odot$.

In the next sections we discuss hybrid EoS which include a deconfined phase of matter to understand if they favor, exclude or constrain the existence of a sexaquark.


\section{Deconfinement solution for the hyperon and sexaquark dilemmas}
\label{sec:results}
It was mentioned in section \ref{sec:QM} that two different types of construction are used for dealing with hybrid stars. We compare their results regarding the maximum mass, radius, tidal deformability and the sequence of particles onset for constructed hybrid stars in this section.

 The resultant parameters for the EoS which have been obtained as a hybrid solution for different scenarios using the MC and RIC are listed in \tableautorefname~\ref{tab:parameters} and \tableautorefname~\ref{tab:parameters3}, respectively. The quark matter parameters have been selected and tuned in such a way to describe a desirable critical point for the transition from hadronic phase to quark phase. (The corresponding hadronic matter and quark matter EoS for each scenario which includes S$_{1941}$ are given in Appendix B in Fig.~\ref{fig:eos5} in the P-$\mu$ plane while the ones which include S$_{1885}$ are shown in Fig.~\ref{fig:eos6}.) As we already mentioned, the ``reconfinement" phenomenon at higher chemical potentials has been ignored when a Maxwell construction is employed.

 The effect of applying a RIC is to shift the deconfinement onset to higher densities than in a Maxwell construction which, therefore, allows S particles to exist in the core of higher mass neutron stars -- around the mass of the lightest observed NS. Around the reconfinement point, neither the hadronic EoS nor the quark matter one is reliable. Indeed, nature prefers the phase with the higher pressure (transition occurs from a stiff EoS to a softer one) but at the same time a transition from deconfined quark matter to a confined hadronic phase is nonphysical. In this situation, the RIC with a negative value of $\Delta_p$ in \eqref{eq:pppressure} is a mathematical approach that allows to describe a crossover transition from the softer hadronic phase to the stiffer quark phase. It is worth mentioning that RIC with a negative value of $\Delta_p$ results in an interpolated EoS which is stiffer than both the hadronic and quark matter EoS's. 
One should notice that for each set of hadronic EoS and quark matter EoS, there is always a minimum value of $\Delta_p$ for which causality is fulfilled, i.e., $c_s^2={(n/\mu)}/{(d^2P/d\mu^2)}<1$.  Therefore not only $d^2P/d\mu^2$ must be positive but also it has to be greater than $n/\mu$. The first value of $\Delta_p$ in \tableautorefname~\ref{tab:parameters3} for each set is the minimum allowed value.
 
\begin{figure*}[htb]
\includegraphics[width=1.0\textwidth,height=0.4\textwidth]{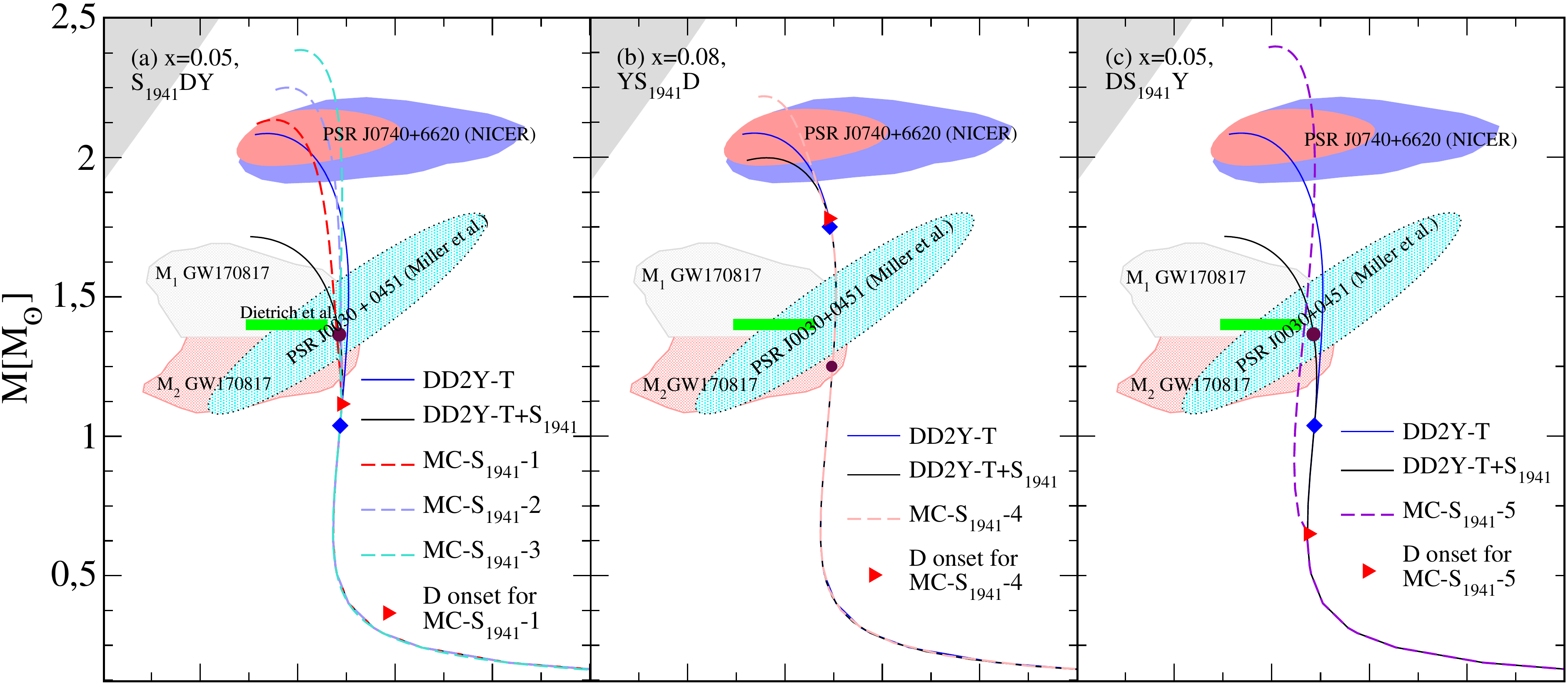}
\\
\includegraphics[width=1.0\textwidth,height=0.4\textwidth]{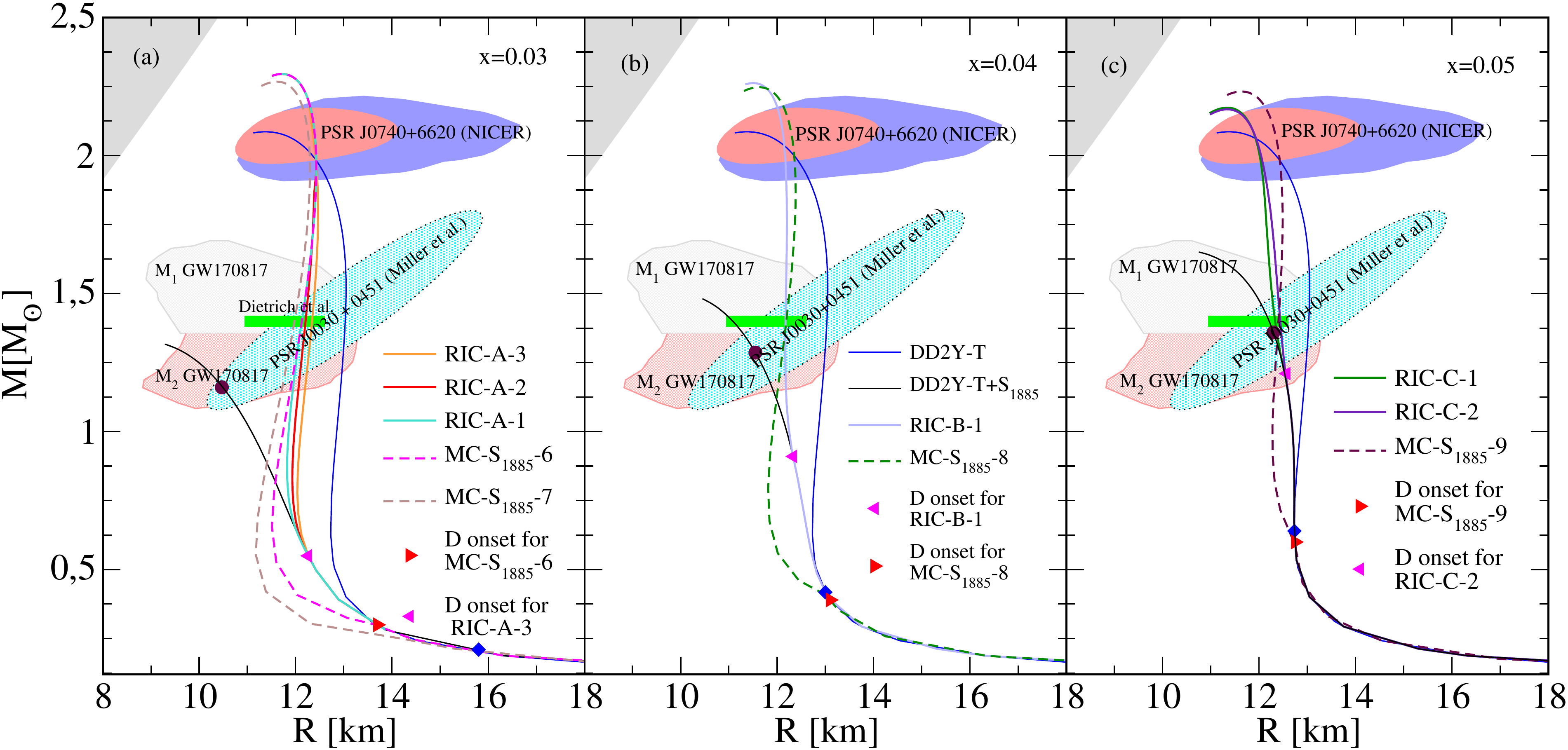}
\caption{The upper panels show the mass-radius
relation for the hybrid star in each scenario which includes S$_{1941}$ based on the \tableautorefname~\ref{tab:parameters}. Mass-radius
relation for the constructed hybrid stars based on the \tableautorefname~\ref{tab:parameters} when S$_{1885}$ is included in hadronic EoS and also based on \tableautorefname~\ref{tab:parameters3} are shown in the lower panels. The Blue diamond and maroon circle show the S onset and Y onset respectively in all panels. Moreover, DD2Y-T and DD2Y-T+S lines have been shown with blue and black solid lines in all panels.  Colored regions are described in the caption of Fig.~\ref{fig:Mx}. 
\label{fig:MRhybrid}
}
\end{figure*}

The mass-radius relations of the hybrid stars corresponding to the EoS plotted in Fig.~\ref{fig:eos5} are shown in Fig.~\ref{fig:MRhybrid}.
 
It can be seen that all obtained solutions fulfill all observational constraints regarding the mass-radius of neutron stars except for the recent radius constraint for the $1.4$ solar mass neutron star obtained by Dietrich et al. \cite{Dietrich:2020efo} which is not fulfilled by any of MC lines in upper panels. Only MC-S$_{1941}$-5 crosses it marginally while all hybrid solutions with S$_{1885}$ in lower panels cross this green bar. We expect this issue will result in remarkable difference in tidal deformability plots for upper and lower panels.


\begin{figure*}[htb]
 	\includegraphics[width=0.48\textwidth]{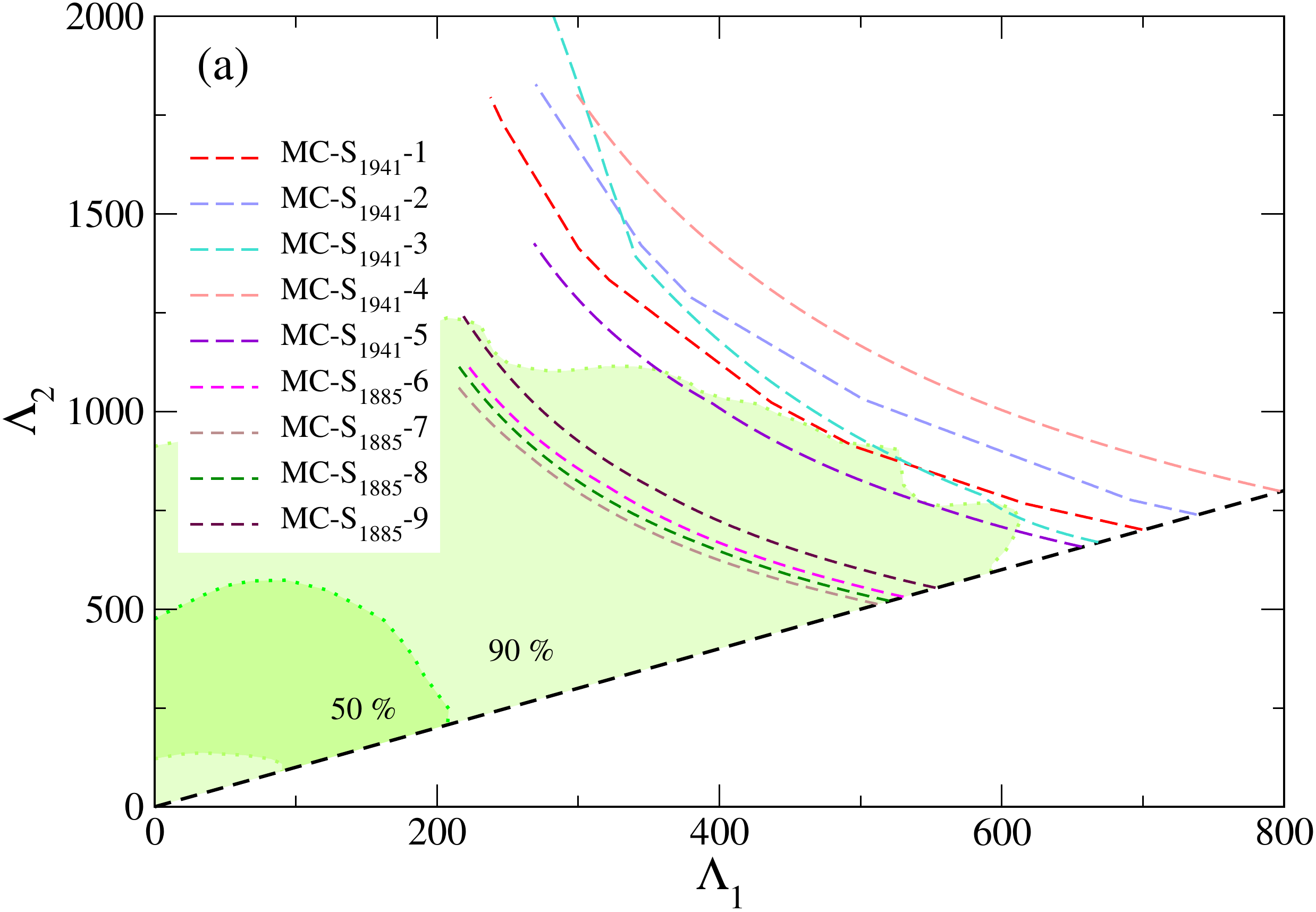}
 	\includegraphics[width=0.48\textwidth]{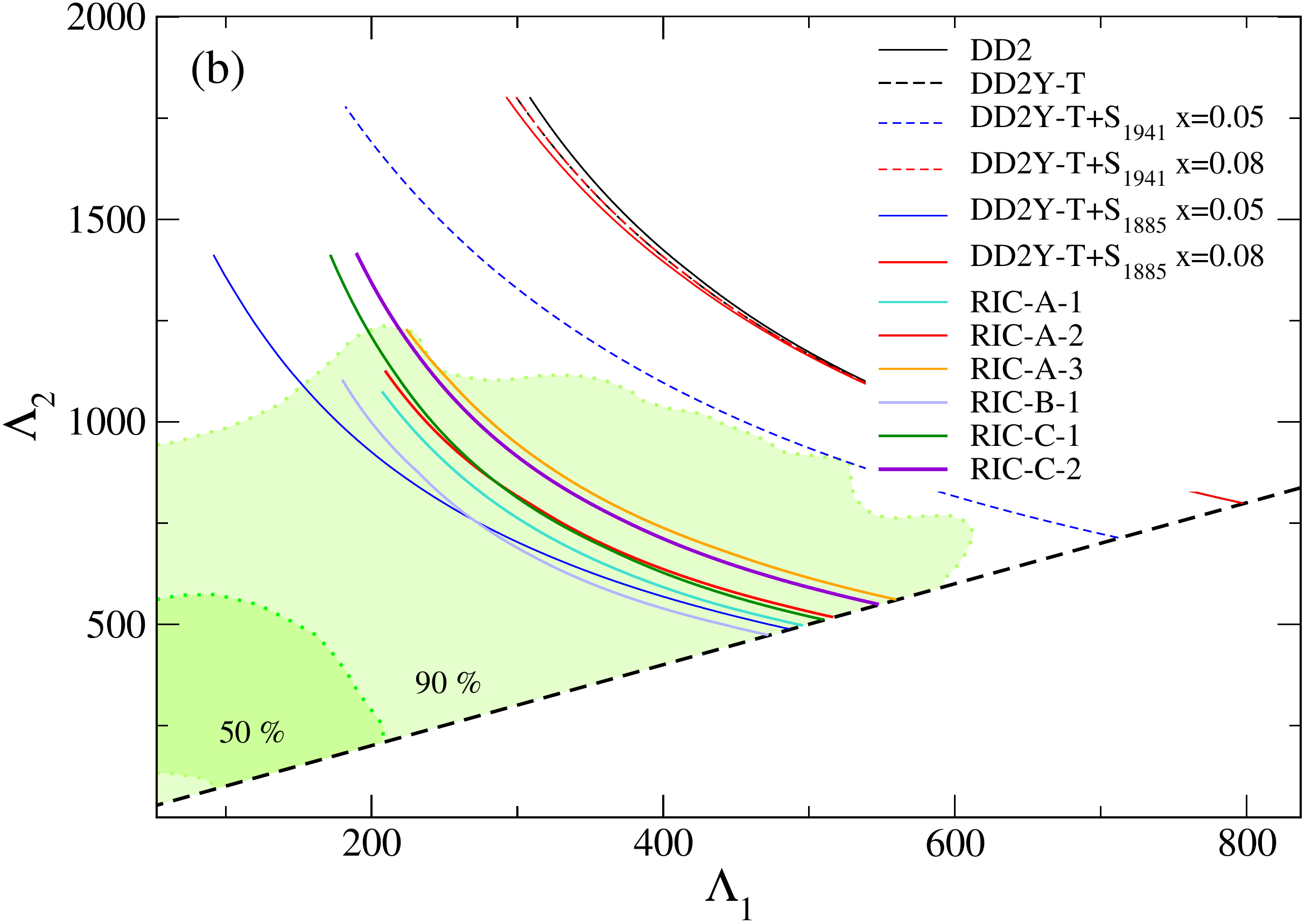}
 	\caption{Tidal deformability parameters $\Lambda_1$ and $\Lambda_2$ for all MC hybrid stars based on \tableautorefname~\ref{tab:parameters} are shown in panel (a). In panel (b) the tidal deformability parameters $\Lambda_1$ and $\Lambda_2$ correspond to all RIC hybrid stars based on \tableautorefname~\ref{tab:parameters3} are shown. In panel (b), the pure hadronic lines are also depicted for comparison. The only hadronic line that fulfills the $90\%$ credibility constraint is DD2Y-T+S$_{1885}$ with $x=0.05$ which can not reach the necessary maximum mass for fulfilling the mass constraint of neutron stars, See Fig.~\ref{fig:Mx}.
 		\label{fig:tidal22}
 	}
 \end{figure*}

 The maximum mass is obtained via MC for MC-S$_{1941}$-5 and MC-S$_{1941}$-3 because the stiffest quark matter EoS with $\eta_v=0.17$ which results in the maximum value of the speed of sound, i.e., $c_s^2/c^2=0.52$, has been used for constructing these two solutions. As the lower panels of Fig.~\ref{fig:MRhybrid} show, the sets that include S$_{1885}$ are mainly a solution for the early deconfinement scenario and the second crossing corresponding to reconfinement is also visible in the $M-R$ curves. If one neglects this early deconfinement and apply RIC to the second crossing, the appearance of S is also possible at interpolated region. Evidently, the  onset of deconfinement has been shifted to higher mass for RIC compared to MC in the lower panels of Fig.~\ref{fig:MRhybrid} and therefore, the constructed hybrid stars are a solution for S$_{1885}$DY scenario instead of DS$_{1885}$Y. Comparing the three lower panels of this figure, one sees that for fixed mass of S, increasing the slope of the mass shift, $x$, increases the reconfinement density and therefore allows the interpolated region to cover higher masses of hybrid stars, around $1.4 M_{\odot}$.
 
 The softer EoS has the smaller radius and therefore, around $1.4~M_\odot$, the softest EoS should have the best tidal deformability. As Fig.~\ref{fig:MRhybrid} shows, the $M-R$ curves with S$_{1885}$ which are shown in the lower panels, have the smaller radius. 
 The tidal deformability has also been calculated for all obtained hybrid stars and the results are plotted in Fig. \ref{fig:tidal22}. In this figure, panel (a) shows the results for MC while the results for RIC are plotted in panel (b).
 
It can be seen in Fig. \ref{fig:tidal22} that the tidal deformability results for MC solutions with S$_{1941}$ are either totally out of the green credibility region or marginally cross it, especially the YS$_{1941}$D scenario which is strongly disfavored by the results of tidal deformability.

However for the hybrid star solutions with S$_{1885}$, not only the M-R constraints are fulfilled but also the tidal deformability puzzle has been solved and the observational constraints from GW170817 with the $90\%$ credibility is fulfilled. Both MC and RIC with S$_{1885}$ result in hybrid stars compatible with both mass-radius constraints and tidal deformability constraints.   Since (neglecting repulsion between S and nucleons) condensation of S occurs at saturation density or below if $m_S \leq 1885$ MeV, and $m_S = 1885$ MeV is the mass for which the condensation occurs exactly at saturation density, $m_S $ near 1885 MeV may be most favorable for compatability with observational constraints of neutron stars~\cite{Dietrich:2020efo}.
  

\begin{figure}[!htb]
	\includegraphics[width=0.48\textwidth]{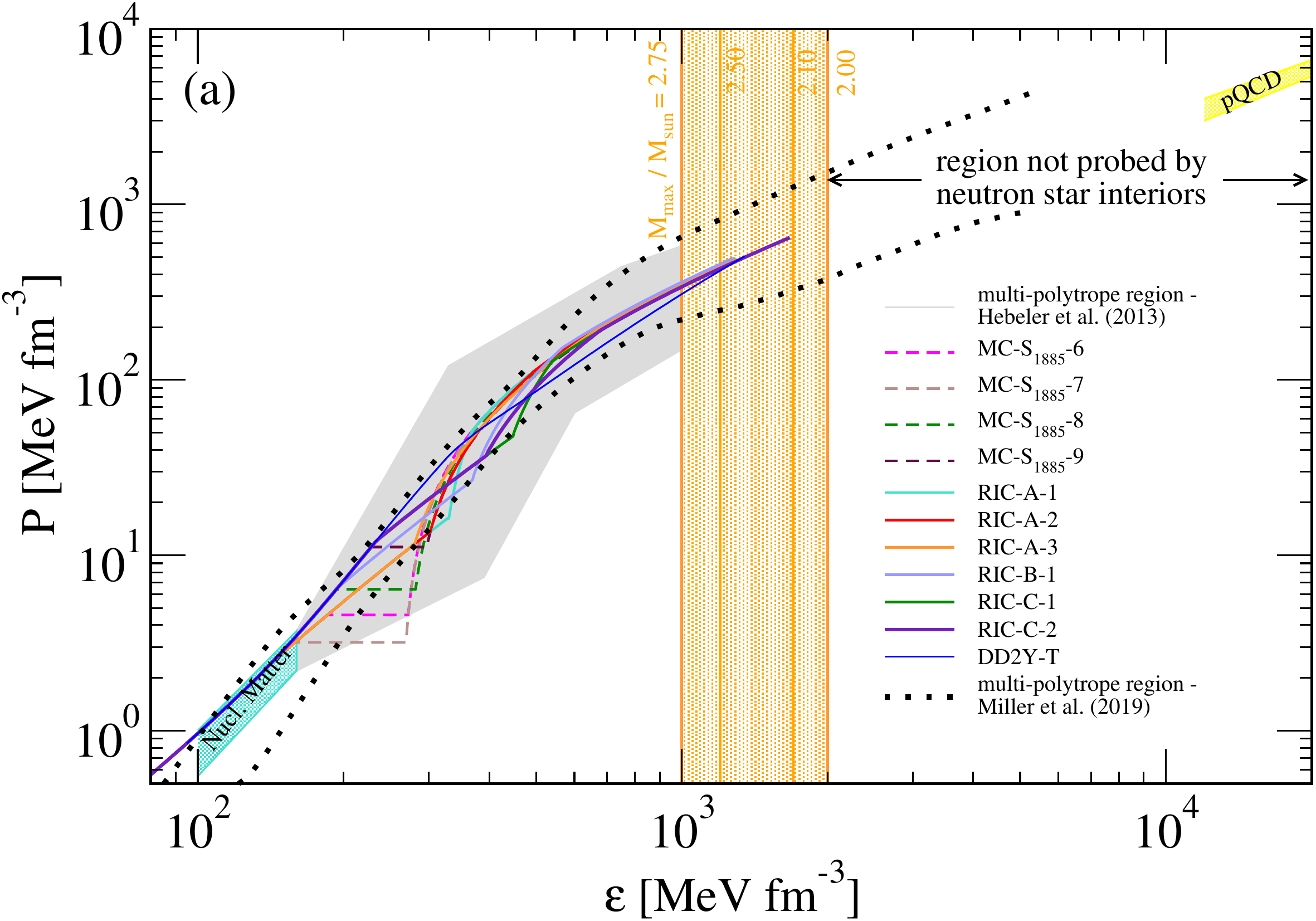}\\
	\includegraphics[width=0.48\textwidth,height=0.28\textwidth]{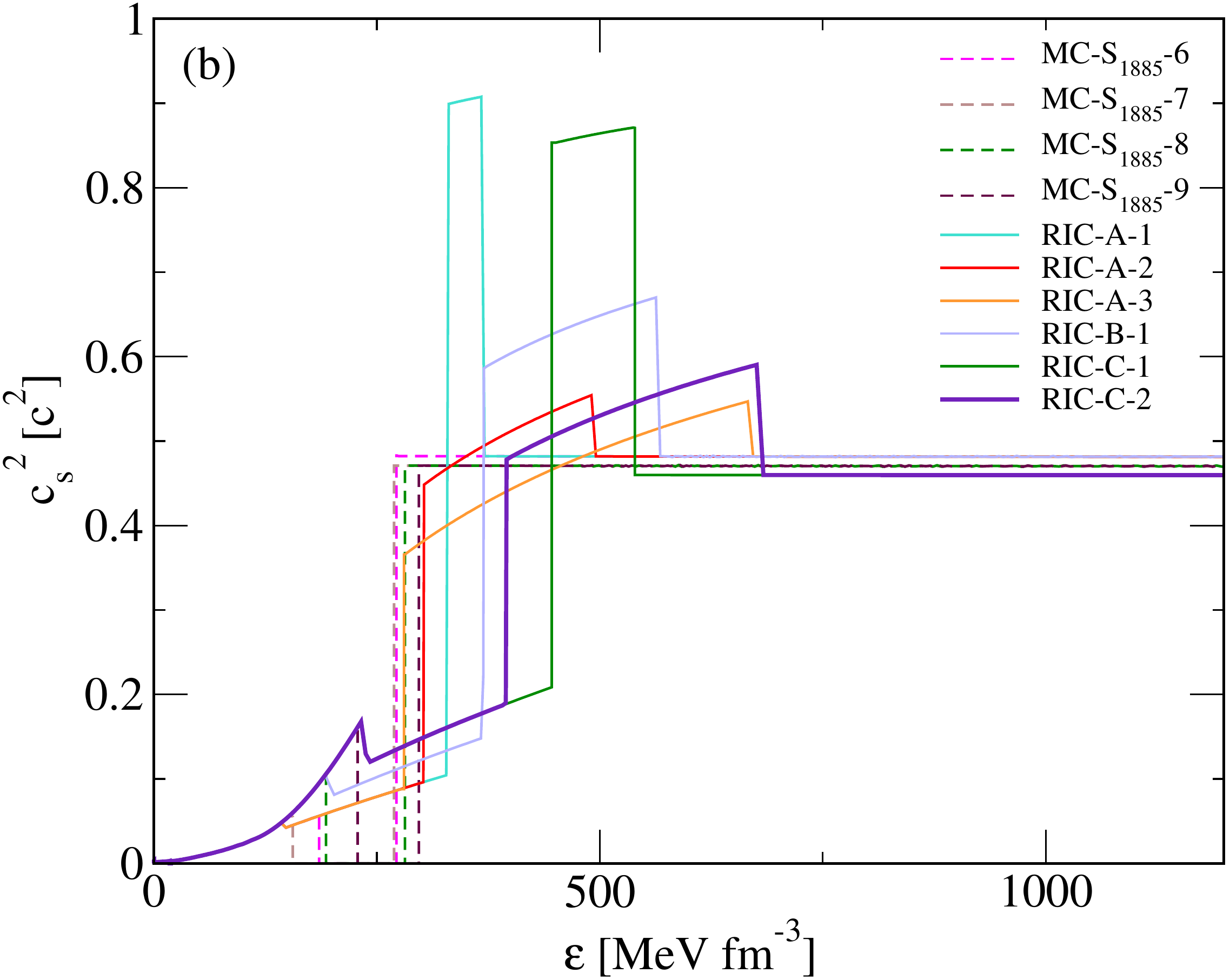}
	\caption{Panel (a): Pressure as a function of energy density for the obtained EoS. The horizontal lines show the Maxwell construction. The gray region corresponds to the EoS constraint from \cite{Hebeler:2013nza} while the narrow region between black dotted lines corresponds to the analysis of \cite{Miller:2021qha}.
	Panel (b): Squared sound speed for these EoS.
		\label{p-e hebeler}
	}
\end{figure}

Panel (a) of Fig.~\ref{p-e hebeler} shows that all obtained EoS are almost inside the gray region which has been introduced as the accepted region for $P-\varepsilon$ lines in \cite{Hebeler:2013nza}.
From panel (b) of this figure one concludes that
the causality constraint $c_s\le 1$ (in units of the speed of light) is fulfilled for the all solutions which include S$_{1885}$.

\begin{figure*}[!htb]
\includegraphics[width=0.43\textwidth]{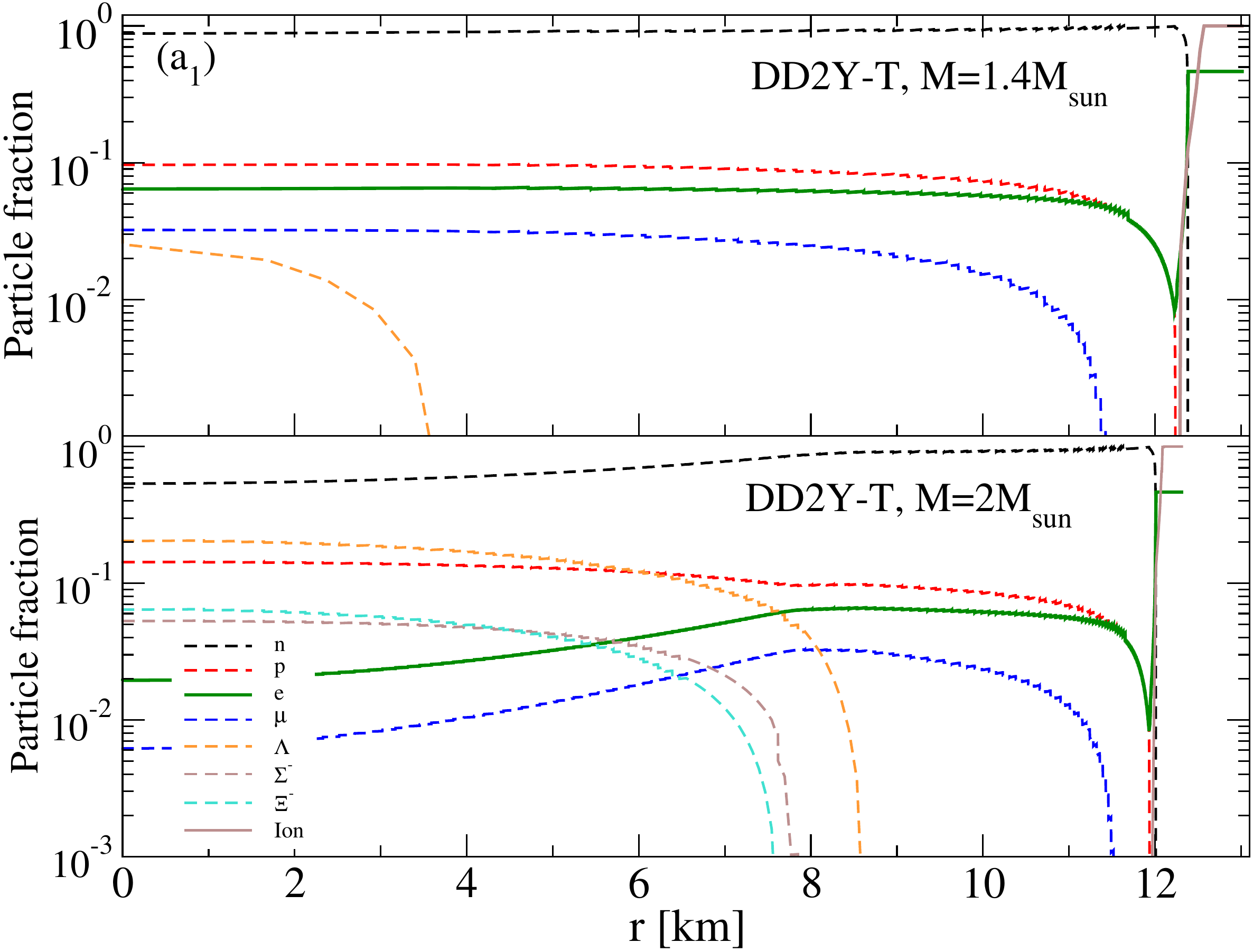}
\includegraphics[width=0.43\textwidth]{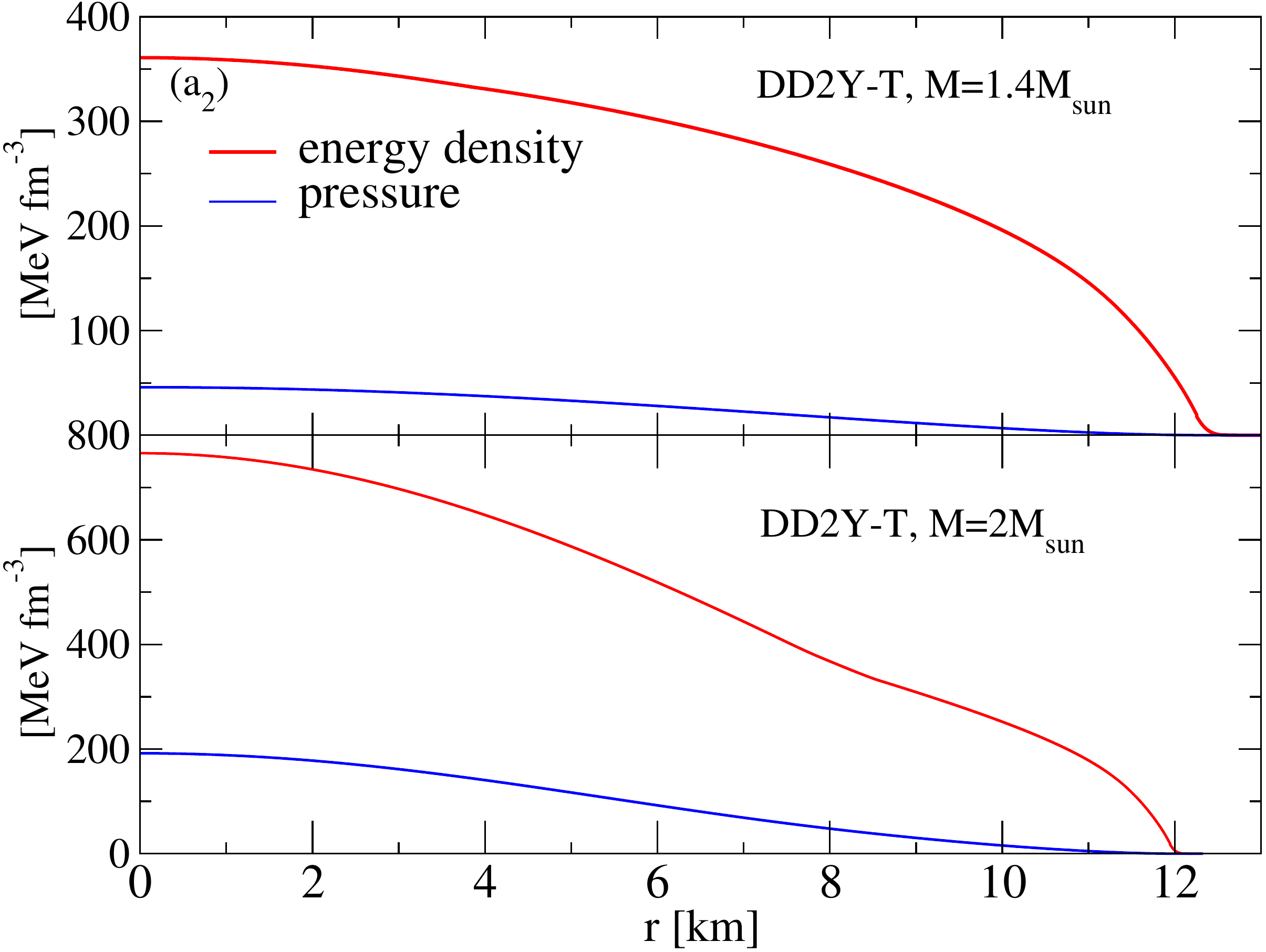}
\includegraphics[width=0.43\textwidth]{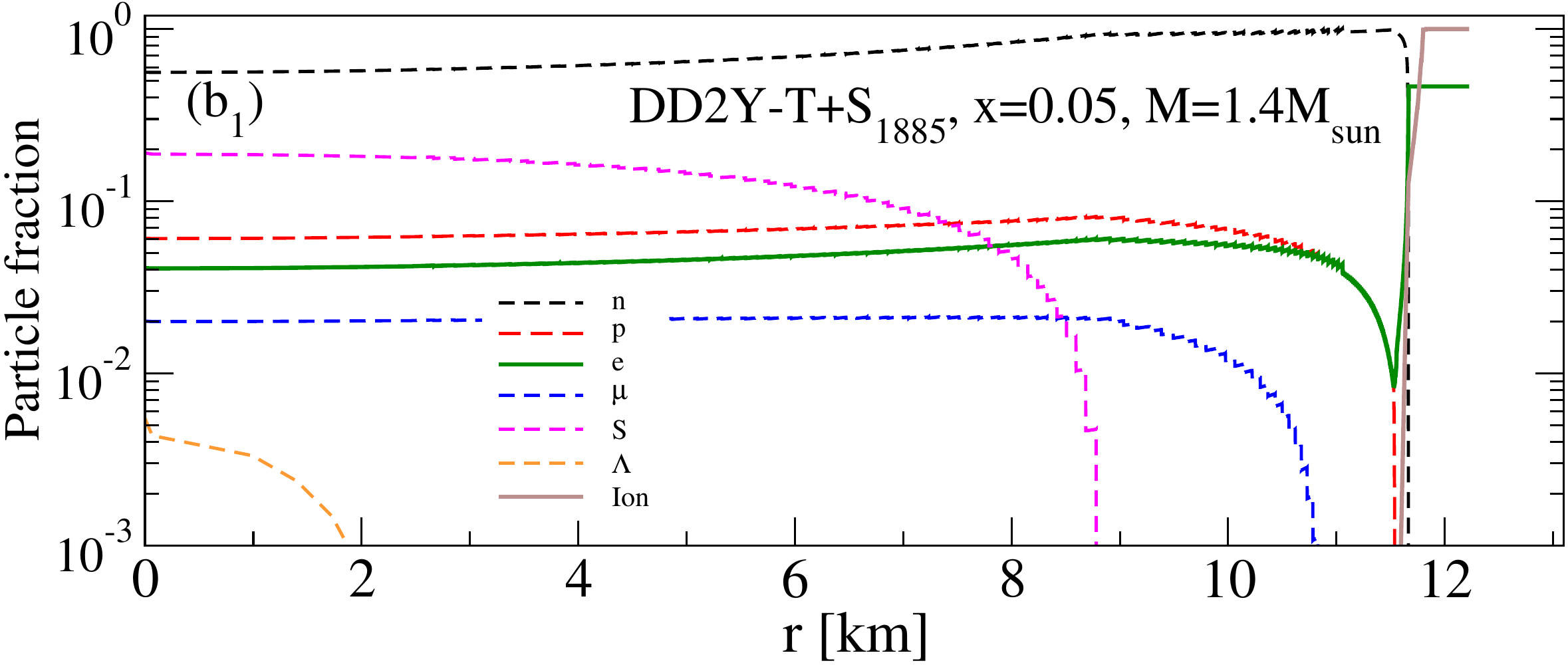}
\includegraphics[width=0.43\textwidth]{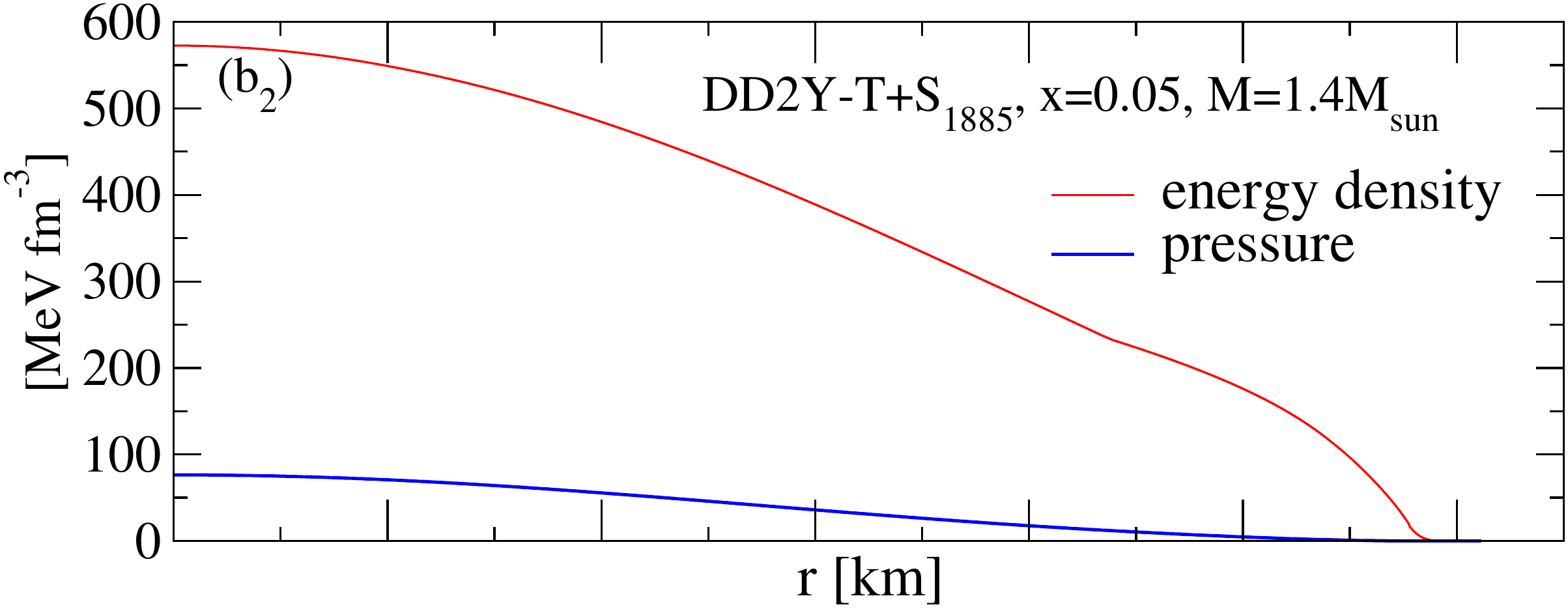}
\includegraphics[width=0.43\textwidth]{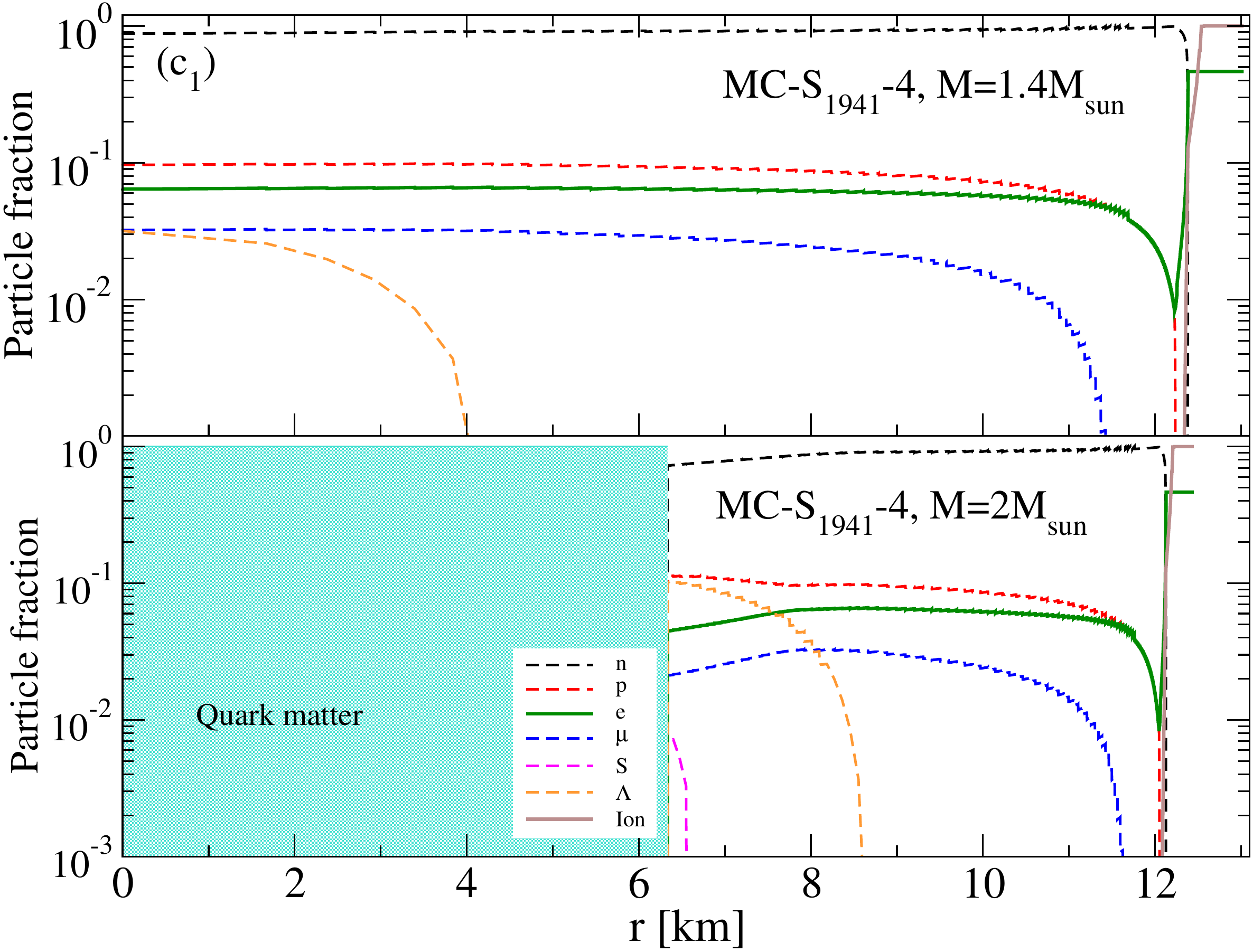}
\includegraphics[width=0.43\textwidth]{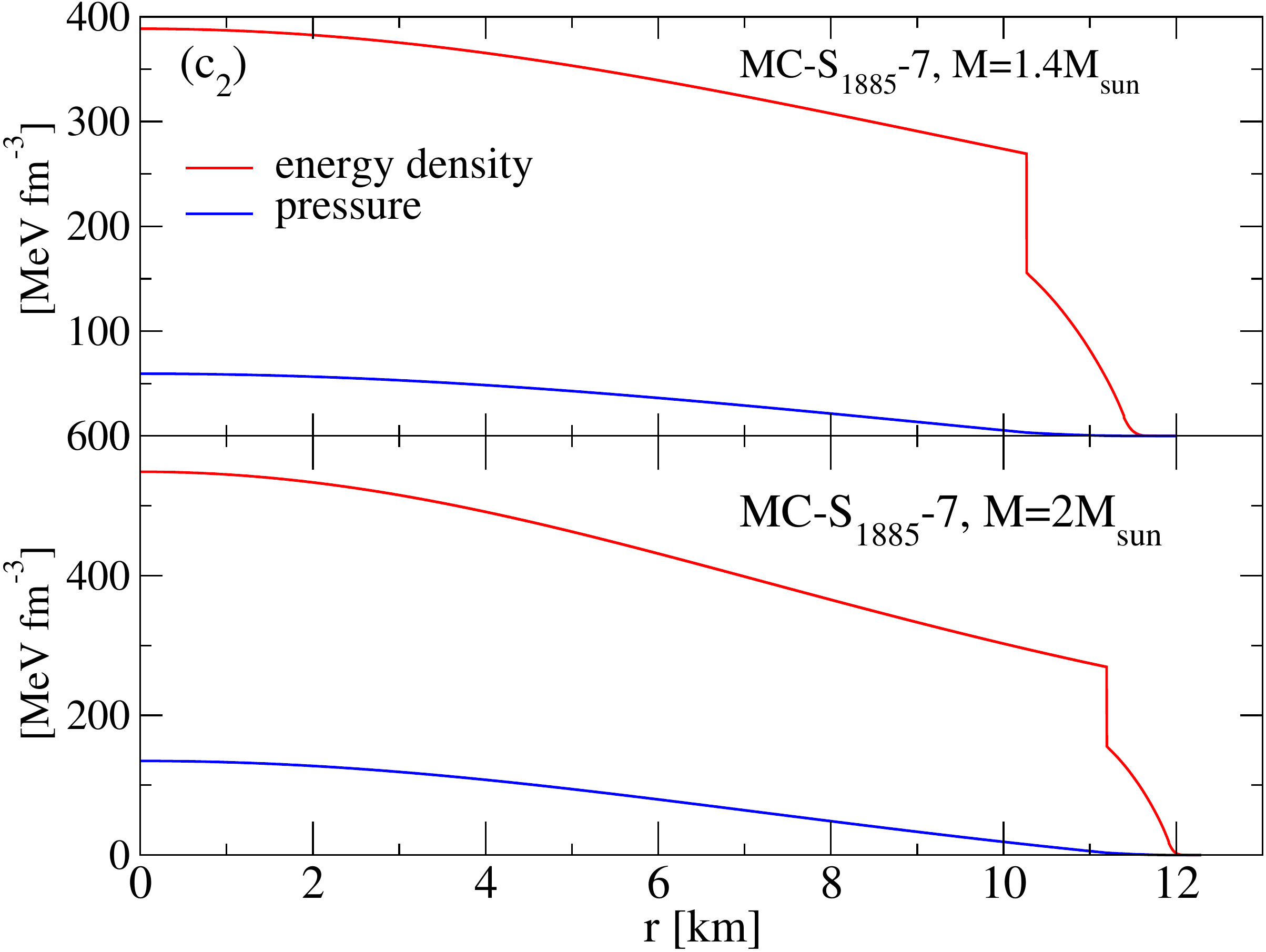}

\caption{The profile of the stars for DD2Y-T, DD2Y-T+S$_{1885}$ and MC-S$_{1941}$-4. The particle fraction as a function of the distance from the center of star is shown in panels $(a_1)$ DD2Y-T, $(b_1)$ DD2Y-T+S$_{1885}$ and $(c_1)$ MC-S$_{1941}$-4. The pressure and the energy density of each star are shown in panels $(a_2)$, $(b_2)$ and $(c_2)$. The profiles correspond to $M=2M_{\odot}$ and $M=1.4M_{\odot}$ for each star. The EoS for DD2Y-T+S$_{1885}$ when $x=0.05$ is a soft one which cannot reach to $2M_{\odot}$  None of these models satisfies both maximum mass and tidal deformability constraints.
\label{fig:profile1}
}
\end{figure*}

\begin{figure*}[!htb]
	\includegraphics[width=0.45\textwidth]{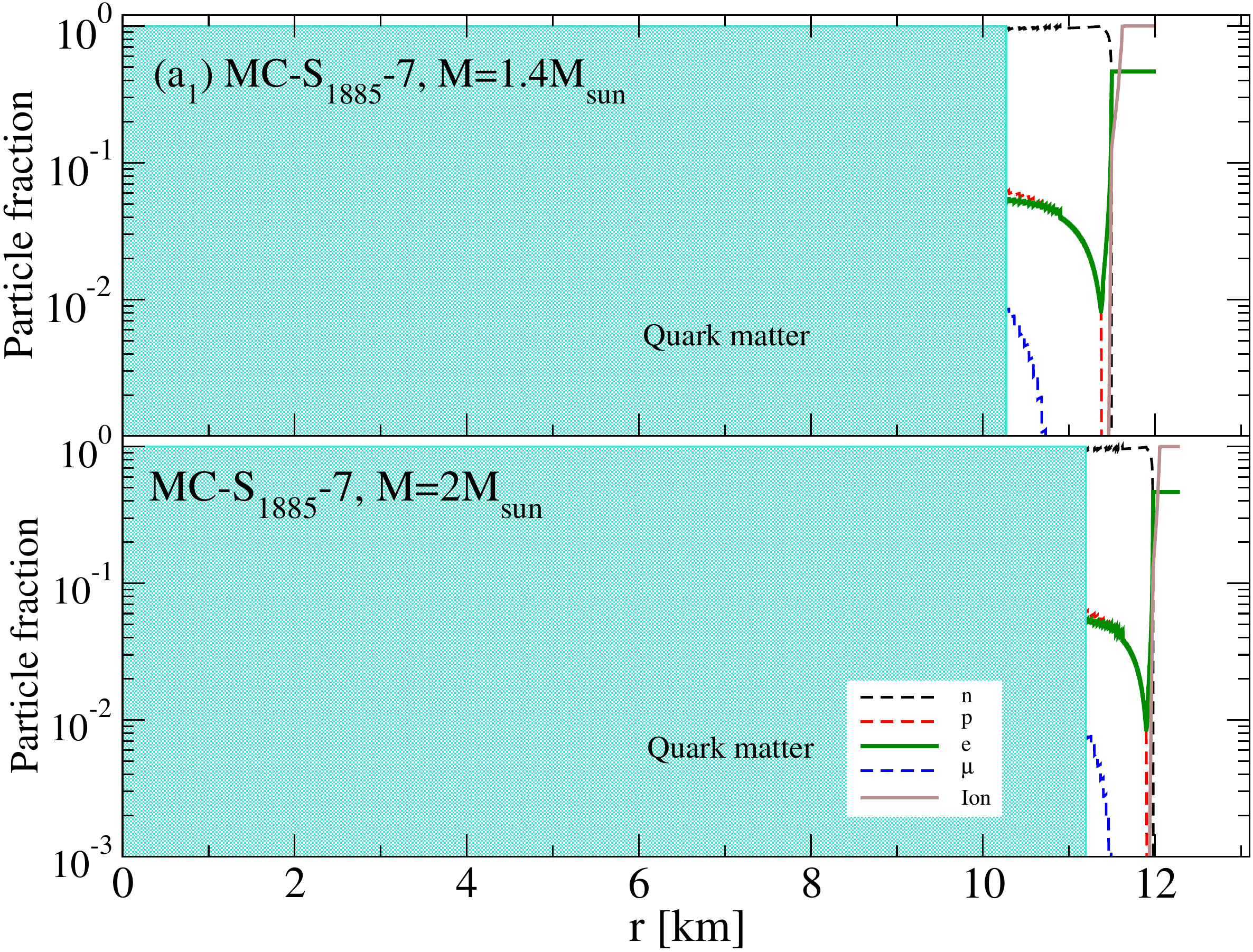}
	\includegraphics[width=0.45\textwidth]{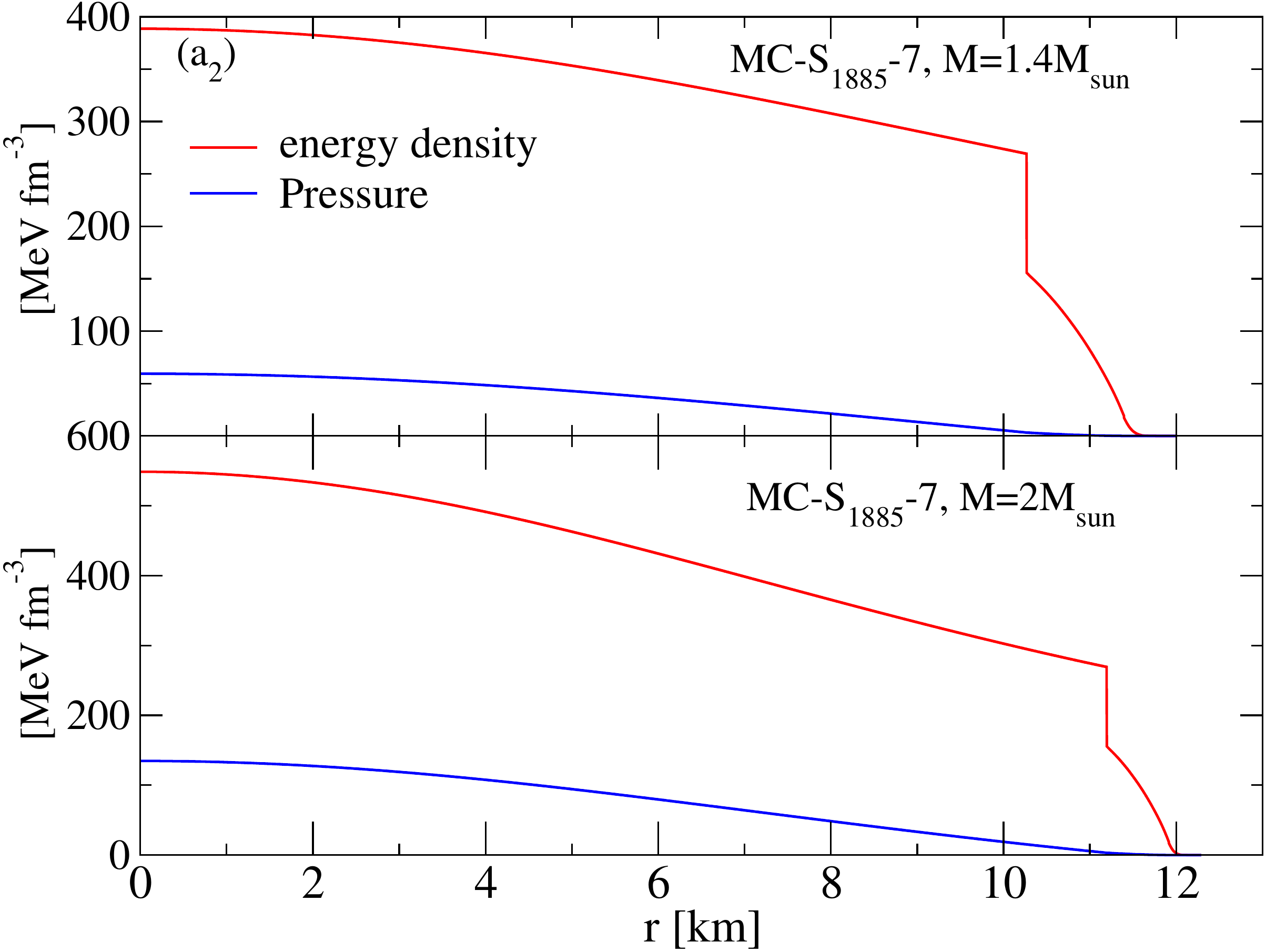}\\
	\includegraphics[width=0.45\textwidth]{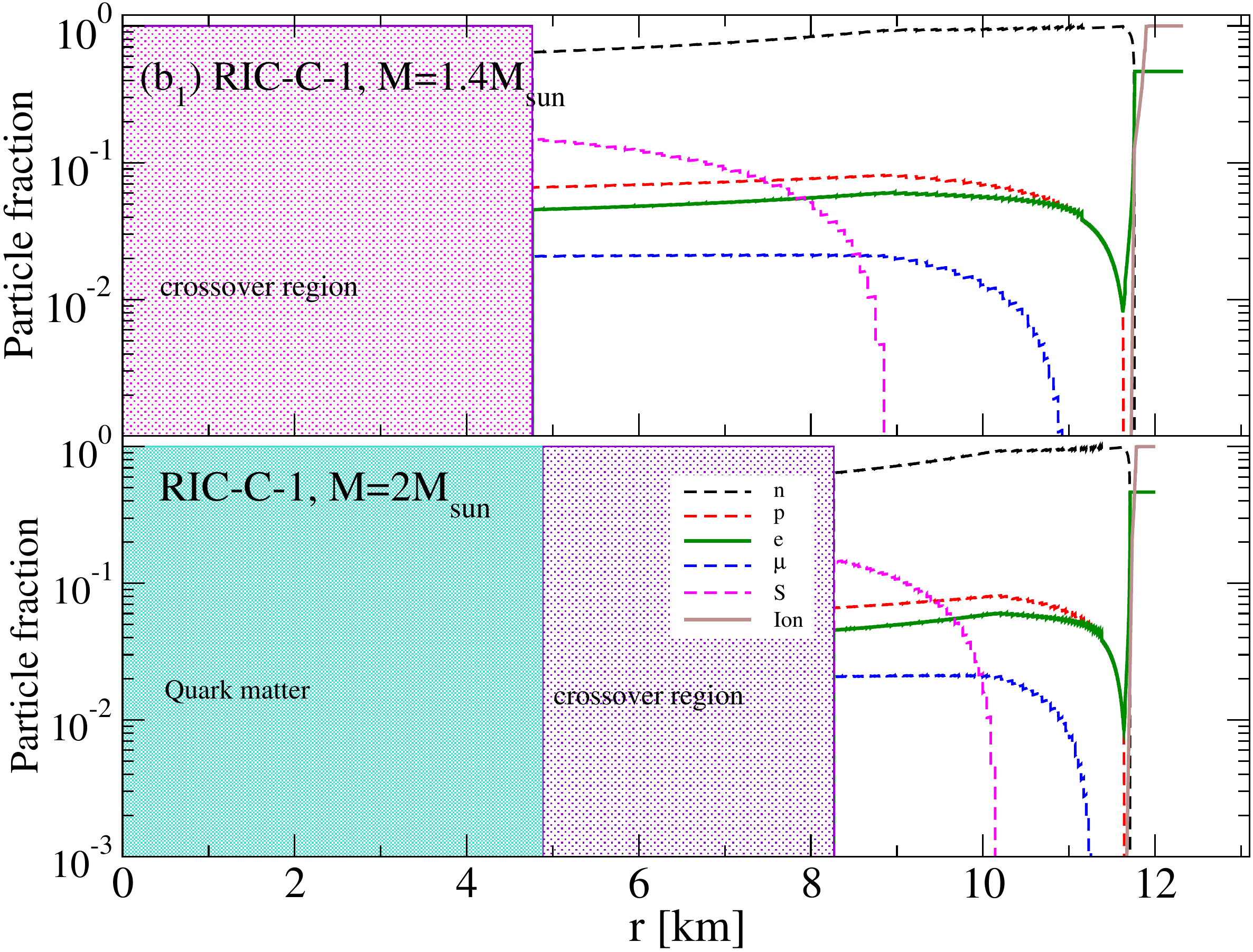}
	\includegraphics[width=0.45\textwidth]{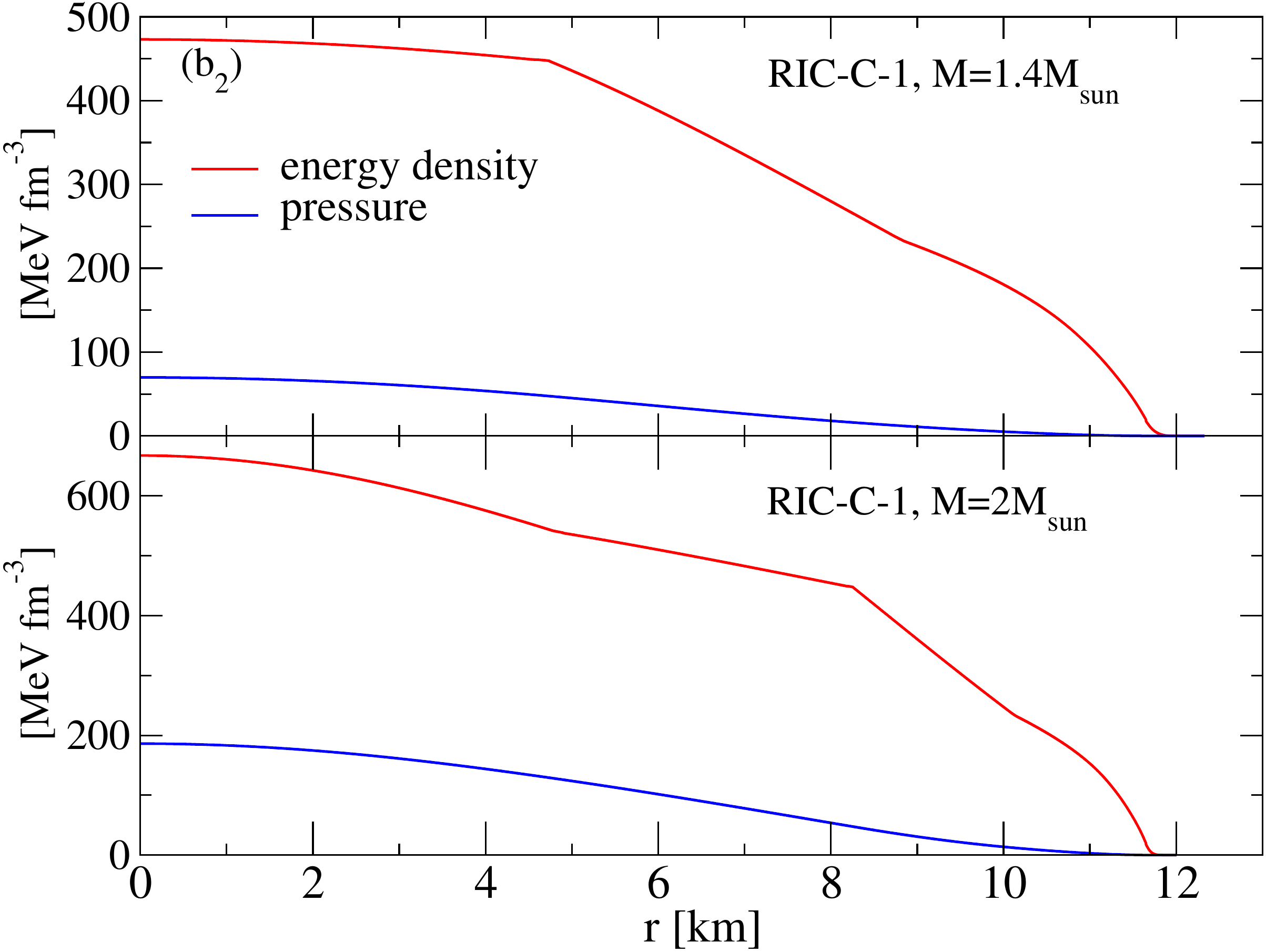}\\
	\caption{Same as Fig.~\ref{fig:profile1}, but for MC-S$_{1885}$-7 (panels $(a_1)$ and $(a_2)$) and RIC-C-1 (panels $(b_1)$ and $(b_2)$), both of which satisfy both the maximum mass and tidal deformability constraints. In the first model, MC-S$_{1885}$-7, outside the quark matter core there are only nucleons and leptons, which in the RIC-C-1 model there is a layer outside the quark matter core which has a significant sexaquark component.
		\label{fig:profile2}
	}
\end{figure*}

As can be seen in the lower panels of Figs.~\ref{fig:MRhybrid} and panels (a) and (b) of Fig.~\ref{fig:tidal22}, we have found two different types of stable hybrid stars including S, which fulfill all modern observational constraints for pulsars.  Taking $m_S=1885$ MeV with medium-dependent mass, and using both
the MC
as well as the 
RIC, we have a transition from DD2Y-T+S on the exterior to a 2SC phase quark matter on the interior.  We display the radial structure of a number of illustrative cases, with the composition in panels $(a_1)$, $(b_1)$ and $(c_1)$ and energy density and pressure in panels $(a_2)$, $(b_2)$ and $(c_2)$.  The stars in Fig.~\ref{fig:profile1} do not satisfy both the M-R and tidal deformability constraints, while those in Fig.~\ref{fig:profile2} satisfy all constraints.

In Fig.~\ref{fig:profile1}, panels $(a_1)$ and $(a_2)$ show the pure hadronic star from DD2,  panels $(b_1)$ and $(b_2)$ show the hyperonic star including S when $m_S=1885$ MeV and $x=0.05$, and panels $(c_1)$ and $(c_2)$ show the MC-S$_{1941}$-4.   Fig.~\ref{fig:profile2} shows MC-S$_{1885}$-7 and RIC-C-1 which are representative of the different types of successful hybrid solutions with S.  As expected there is a jump in energy density for the MC solutions but we don't see this jump in RIC which is a crossover transition.  Among the solutions for which the profile of the star is plotted, the only solution which has a quark matter core for $M=1.4M_{\odot}$ is MC-S$_{1885}$-7 because of the early deconfinement.

One can see the sequence of the onset of different particles at different radius in the profiles of the composition of each figure. The only star in which $\Lambda$ hyperon could appear using the Maxwell construction is MC-S$_{1941}$-4;  it gives a solution for YS$_{1941}$D scenario but it is not compatible with tidal deformability constraint. 
Employing RIC, a stable hybrid star with a typical radius of about 12 km can be obtained; it has an inner core of quark matter while a layer with S and neutrons connects it to a normal nuclear matter crust.  Not only does the sexaquark appear in this model but also all constraints from neutron stars are well fulfilled.

Thus we have shown that -- contrary to the expectation of \cite{McDermott:2018ofd} that a deeply bound dibaryon would be incompatible with neutron stars -- the observed NS properties are compatible with the hadron spectrum containing a stable sexaquark, for parameters similar to those commonly adopted for hyperons and  deconfined quark matter.  Whether the S appears in neutron stars or not depends on how early matter deconfines to quarks.

\section{Conclusions}
\label{sec:conclusion}

Motivated by a stable sexaquark\footnote{S=$uuddss$, with $m_S<2054$ MeV to assure a lifetime greater than the age of the Universe~\cite{Farrar:2022mih}.} being a good dark matter candidate, we have investigated whether the maximum mass and tidal deformability measurements of neutron stars are compatible with the existence of such a particle.  To address this question we extended a relativistic mean field model of hadrons to include sexaquarks and used a causal phenomenological EoS to describe the quark matter phase, restricting parameters to the range suggested by two flavor superconductivity.  A first-principles treatment of the cross-over region between hadronic and quark matter phases is not available for all EoS parameters of interest, so we investigate two different approaches.  When the conditions for applicability of the Maxwell construction are met  we use that, and otherwise we use a Replacement Interpolation construction.

We find that the existence of a stable sexaquark is well-compatible with both the maximum mass and the highly constraining tidal deformability of GW170817, given present knowledge of the properties of hadronic and quark matter.  
Whether or not a sexaquark exists, within our framework the most massive stars must have a quark matter core with stiff EoS to support their high mass. 
\footnote{There are purely hadronic models, some of them including hyperons, with a softer EoS at high densities than the DD2 parameterization  which can satisfy the maximum mass and tidal deformability constraints. However, this requires a modification of the high-density couplings with more parameters, e.g., DD2-F \cite{Alvarez-Castillo:2016oln}, or
specifically adjusted hyperon-meson couplings, e.g., \cite{Thapa:2021kfo},
or the introduction of additional coupling mechanisms, e.g., \cite{Grigorian:2018bvg}.}  
Among the successful hybrid star solutions which we develop, we find two general types:  those in which the quark matter core is surrounded by nucleons, and others in which the core and interpolated region is surrounded by a layer with a substantial sexaquark (but negligible hyperon) fraction. This potentially may lead to an observable signature of sexaquarks in the cooling curve or kilonova properties.  

Satisfying the tidal deformability constraints on neutron stars of $\approx 1.4~M_\odot$ while accommodating neutron star masses above $2~M_\odot$ is very challenging.  At low density the EoS must be quite soft to produce compact $\approx 1.4~M_\odot$ neutron stars, while at high density it must be stiff to support high mass stars.  Intriguingly, the sexaquark neatly solves this problem by naturally producing the needed softening.  However the softening at low density needed to fit current deformability constraints can also be achieved by an early onset of the quark matter phase.  Thus within our treatment we cannot judge whether neutron star properties may actually call for a sexaquark or not.  We also stress that our approach in this paper has been entirely phenomenological; theoretical work is needed to decide whether the parameters leading to the successful models with sexaquarks can emerge from fundamental QCD, and to accurately treat the cross-over between the quark and hadron phases.  (The same caveats also apply to models without sexaquarks.)

\section*{Acknowledgements}
We thank Sanjay Reddy for valuable discussions. We also acknowledge discussions with Kazem Azizi. This work was supported by the Polish National Science Centre (NCN) under grant number 2019/33/B/ST9/03059. 
D.B. received support from the Russian Fund for Basic Research (RFBR) under grant No. 18-02-40137 and from the Russian Federal Program "Priority-2030".
The research of G.R.F. is supported by NSF-PHY-2013199 and the Simons Foundation.
D. A-C. acknowledges support from the Bogoliubov-Infeld program for the collaboration between JINR Dubna and Polish Institutions.
This work is part of a project that has received funding from the European Union’s Horizon 2020 research and innovation program under the grant agreement STRONG – 2020 - No 824093.
We are grateful to the COST Actions CA15213 "THOR" and CA16214 "PHAROS" for their support of our networking activities.


\appendix

\section{Guide to terminology}
\label{app:A}

For the convenience of the readers, we collect in 
this Appendix the explanations of the numerous abbreviations that have been used in the main text 
of this work.

\begin{description}
\label{tab:abbreviations}
\item[DD2]{Parameterization of a GDRF for hadronic matter including only nucleons}
\item[DD2Y-T]{DD2 with hyperons by S. Typel}
\item[DD2Y-T+S]{DD2Y-T with sexaquark}
\item[GRDF]{Generalized relativistic density functional}
\item[MC]{Maxwell Construction}
\item[RIC]{Replacement Interpolation Construction}
\item[DS$_i$Y]{Deconfinement (D) before sexaquark (S$_i$) and hyperon (Y) onset, where
   i=1885, 1941, 2054 denotes to the mass of S in [MeV]}
\item[S$_i$DY]{Deconfinement after sexaquark but before hyperon onset}
\item[S$_i$YD]{Deconfinement after both sexaquark and
hyperon onset when S is prior}
\item[YS$_i$D]{Deconfinement after both sexaquark and
hyperon onset when hyperon is prior}
\item[CSS$_j$, j=1,2,...,9]{Constant speed of sound parameterization for quark matter EoS, where $j$ denotes to the number of the set in \tableautorefname~\ref{tab:parameters}}
\item[MC-S$_i$-j, j=1,2,...,9]{Hybrid solution which has been obtained from Maxwell construction where $j$ denotes the number of sets in \tableautorefname~\ref{tab:parameters}}
\item[CSS-A]{Constant speed of sound parameterization 
for quark matter EoS where A denotes 
the case with  x = 0.03 in \tableautorefname~\ref{tab:parameters3}}
\item[CSS-B]{Constant speed of sound parameterization for quark matter EoS where B denotes 
the case with  x = 0.04 in \tableautorefname~\ref{tab:parameters3}}
\item[CSS-C]{Constant speed of sound parameterization for quark matter EoS where C denotes 
the case with  x = 0.05 in \tableautorefname~\ref{tab:parameters3}}
\item[RIC-A-k, $k=1,2,3$]{Hybrid solution obtained with RIC when S$_{1885}$ is in hadronic matter EoS. 
A is defined the same as in CSS-A and $k$ corresponds to the different sets for replacement interpolation construction in \tableautorefname~\ref{tab:parameters3}}
\item[RIC-B-k, $k=1,2$]{Same as RIC-A-k, for A $\leftrightarrow$ B, corresponding to CSS-B}
\item[RIC-C-k, $k=1$]{Same as RIC-A-k, for A  $\leftrightarrow$ C, corresponding to CSS-C}
\end{description}


\section{Hadronic and quark matter EoS in the P-$\mu$ plane}
\label{app:B}

In this appendix we illustrate the construction of the hybrid EoS via MC and RIC given the hadronic and quark matter EoS as input.

\begin{figure*}[!htb]
    \includegraphics[width=0.32\textwidth]{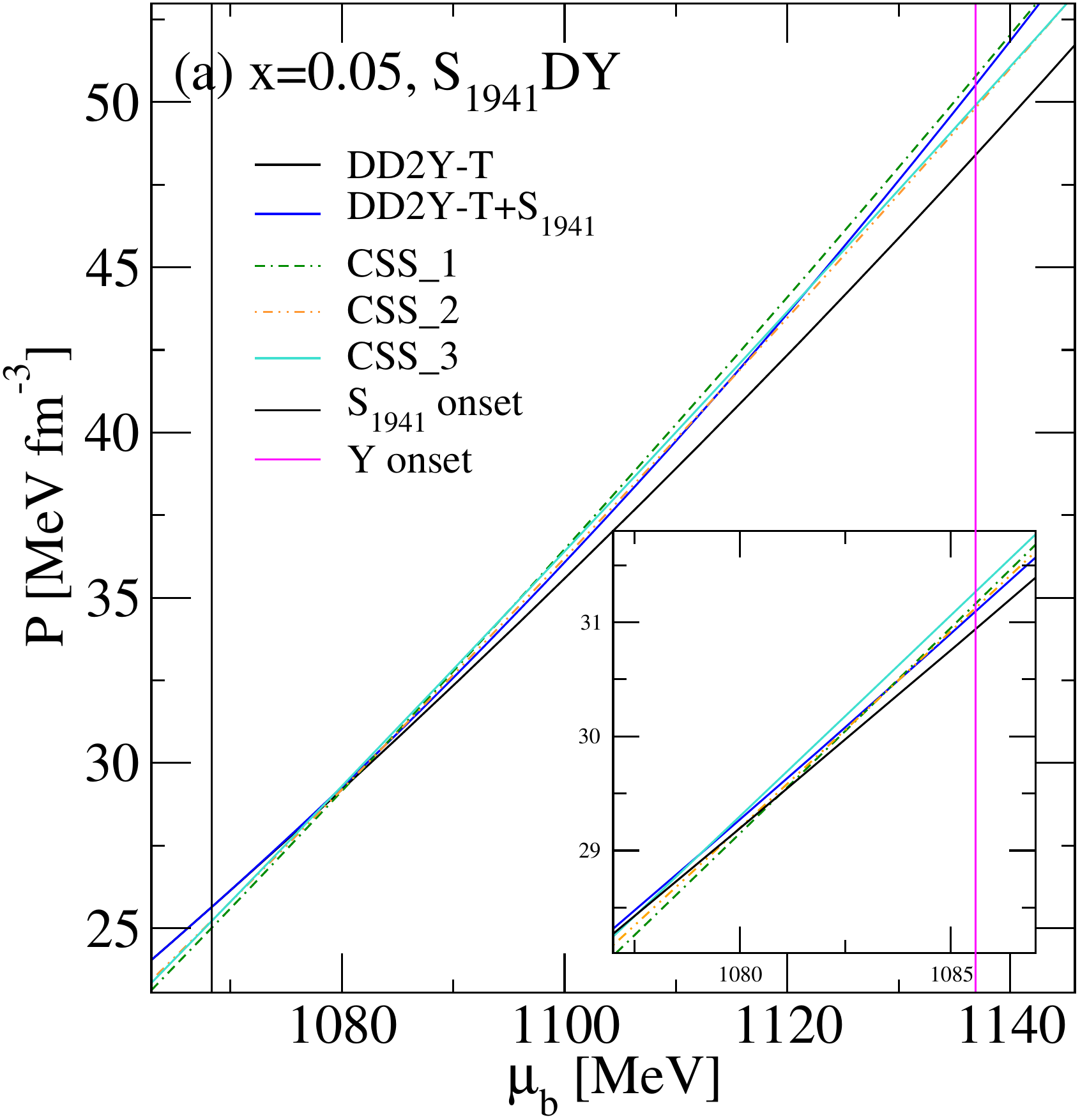}
	\includegraphics[width=0.34\textwidth]{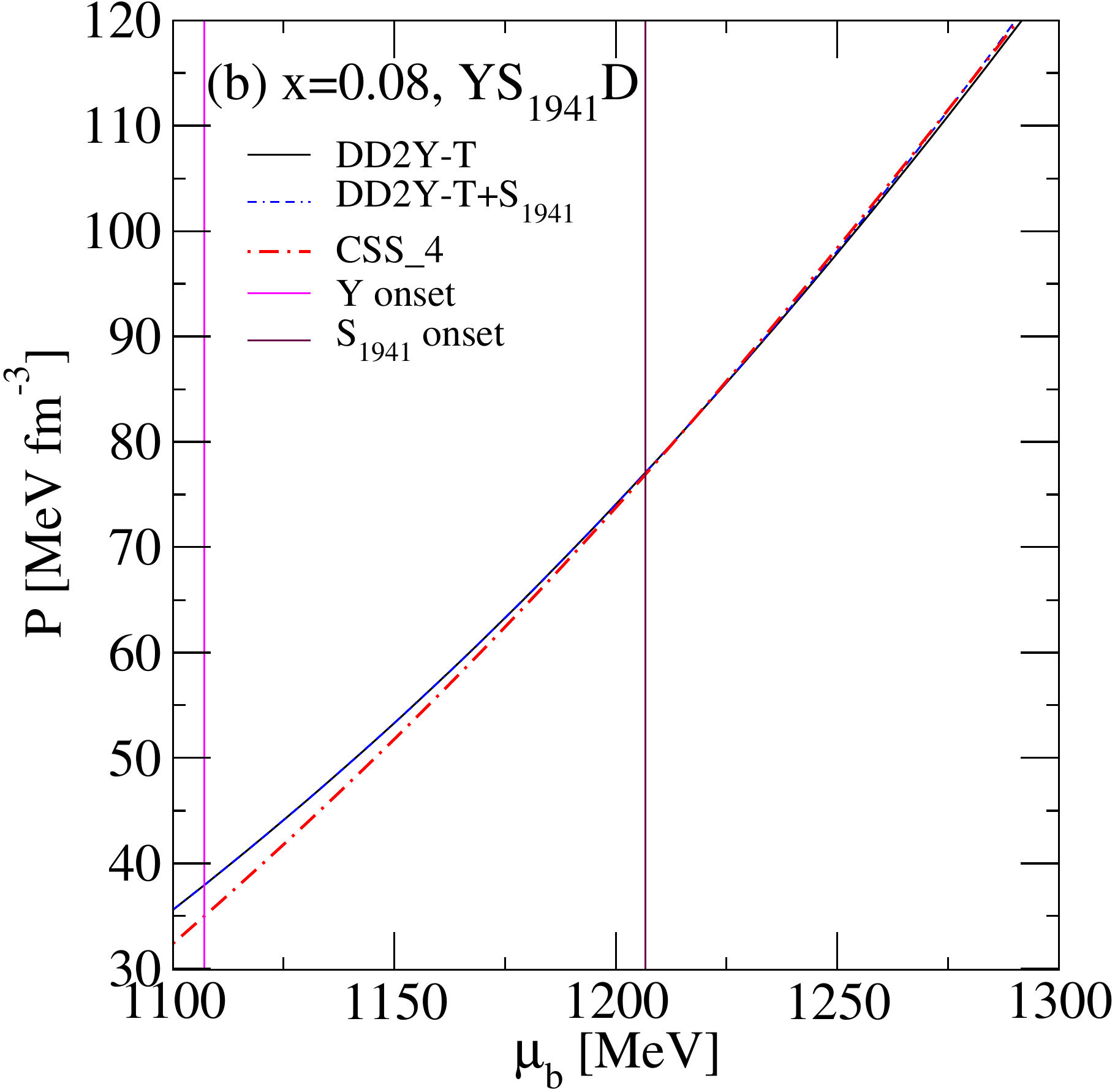}
	\includegraphics[width=0.32\textwidth]{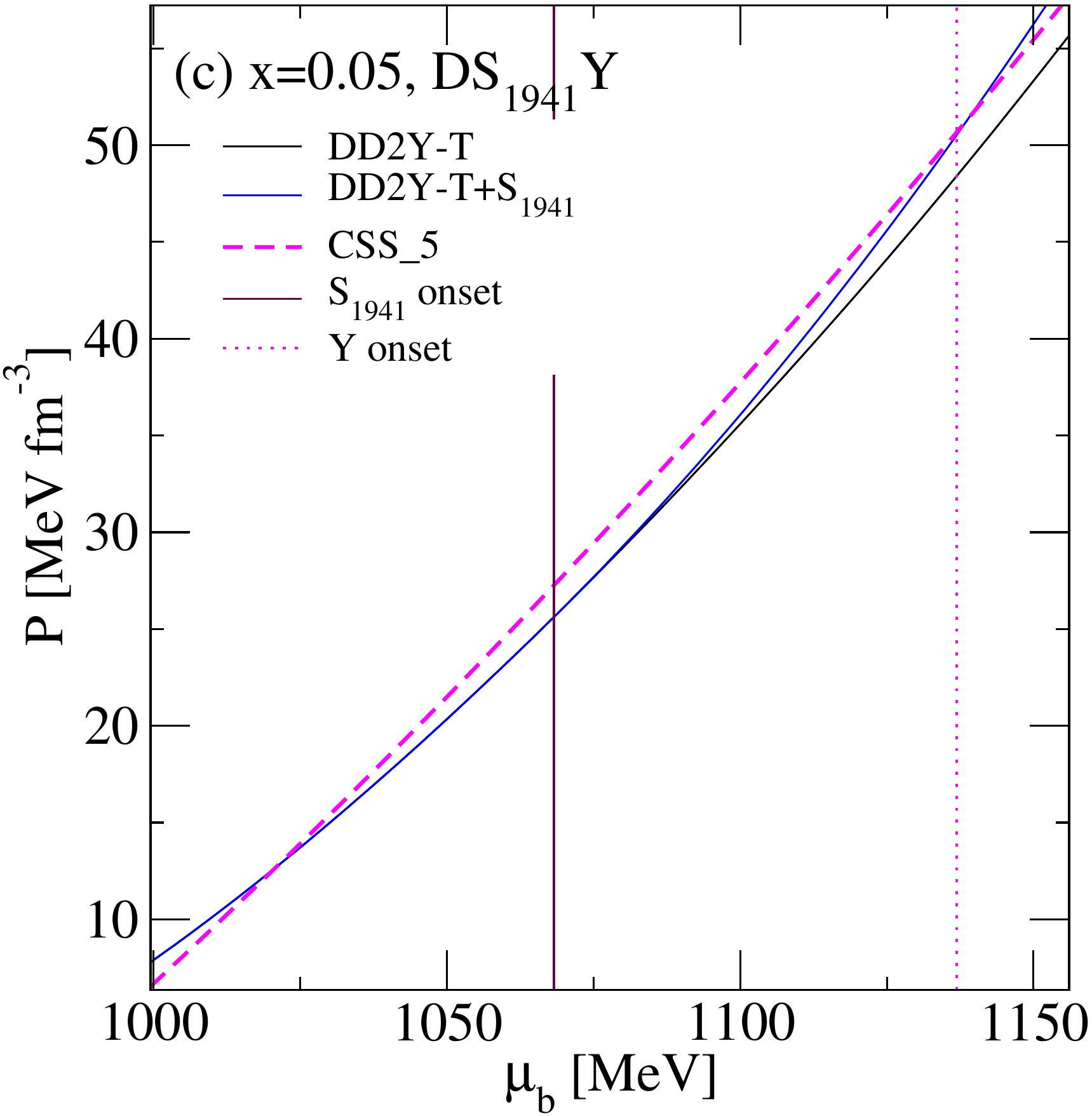}
	\caption{MC for different scenarios of hybrid stars. The solutions for the S$_{1941}$DY scenario when the deconfinement occurs after S$_{1941}$ onset and before Y onset has been shown in panel (a). Panel (b) shows the solution for the YS$_{1941}$D scenario when the deconfinement occurs after both S onset and Y onset and panel (c) shows the solution for the DS$_{1941}$Y scenario when the deconfinement occurs before S$_{1941}$ onset and Y onset. The curves labelled CSS$\_$1 to CSS$\_$5 correspond to the quark matter EoS which after MC give rise to the parameter sets MC-S$_{1941}$-1 to MC-S$_{1941}$-5 which are given in \tableautorefname~\ref{tab:parameters}.
\label{fig:eos5}
	}
\end{figure*}

In Fig.~\ref{fig:eos5}, the MC is performed for different scenarios of hadronic and quark matter EoS combinations for the case of $m_S=1941$ MeV. 
As was discussed in subsection \ref{ssec:QM}, the reconfinement phenomenon is ignored at higher chemical potentials when a Maxwell construction is employed. The CSS$\_$4 is almost masqueraded with the hadronic matter EoS around the transition point. However, a $\mu$-dependent bag pressure has been used for constructing CSS$\_$3 and CSS$\_$4 to produce a bigger change in the slope at the transition point and therefore a bigger jump at the density, but a bigger slope for the quark matter EoS is not compatible with fulfilling the causality condition.
 
\begin{figure*}[!thb]
    \includegraphics[width=0.48\textwidth]{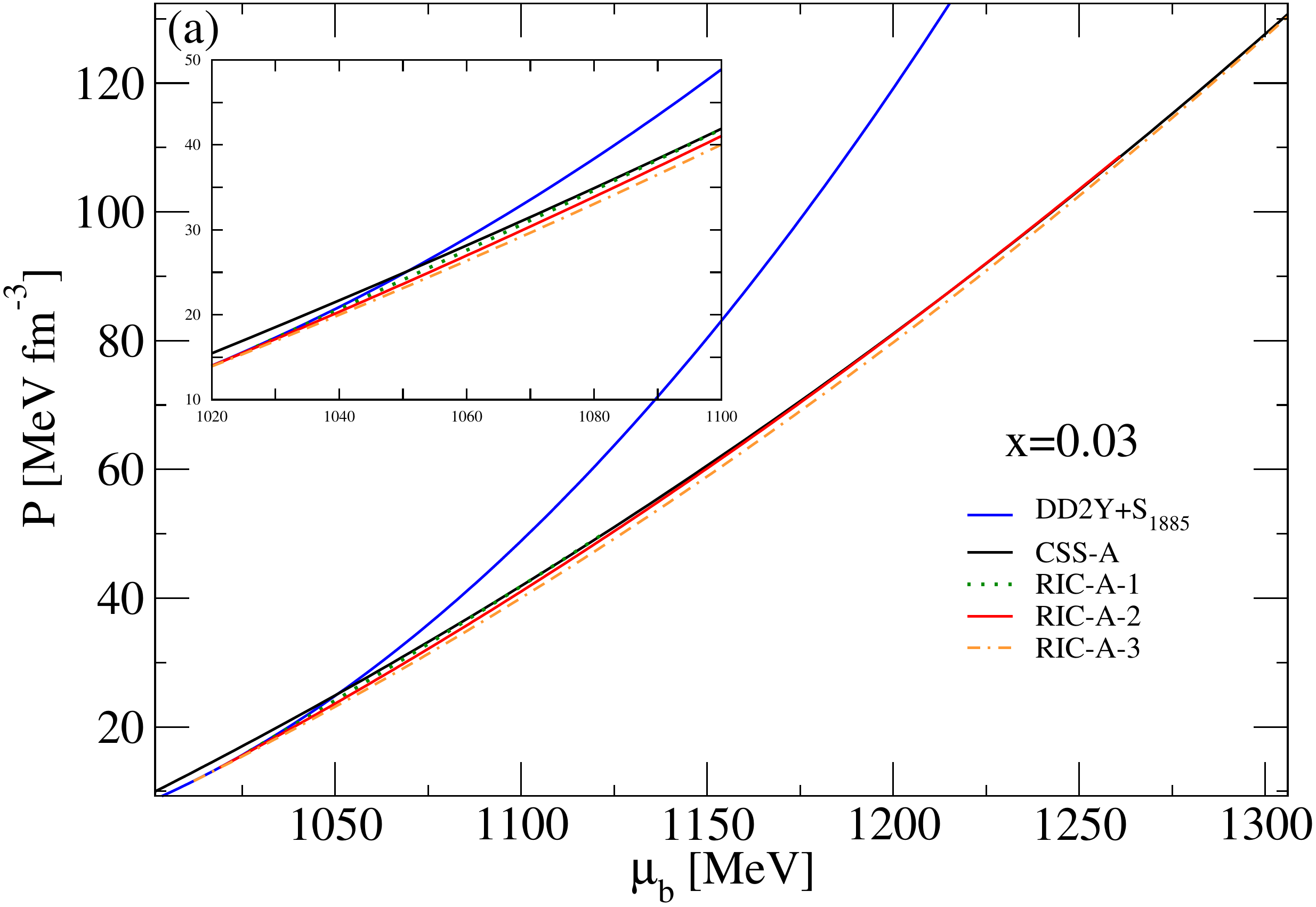}
	\includegraphics[width=0.47\textwidth]{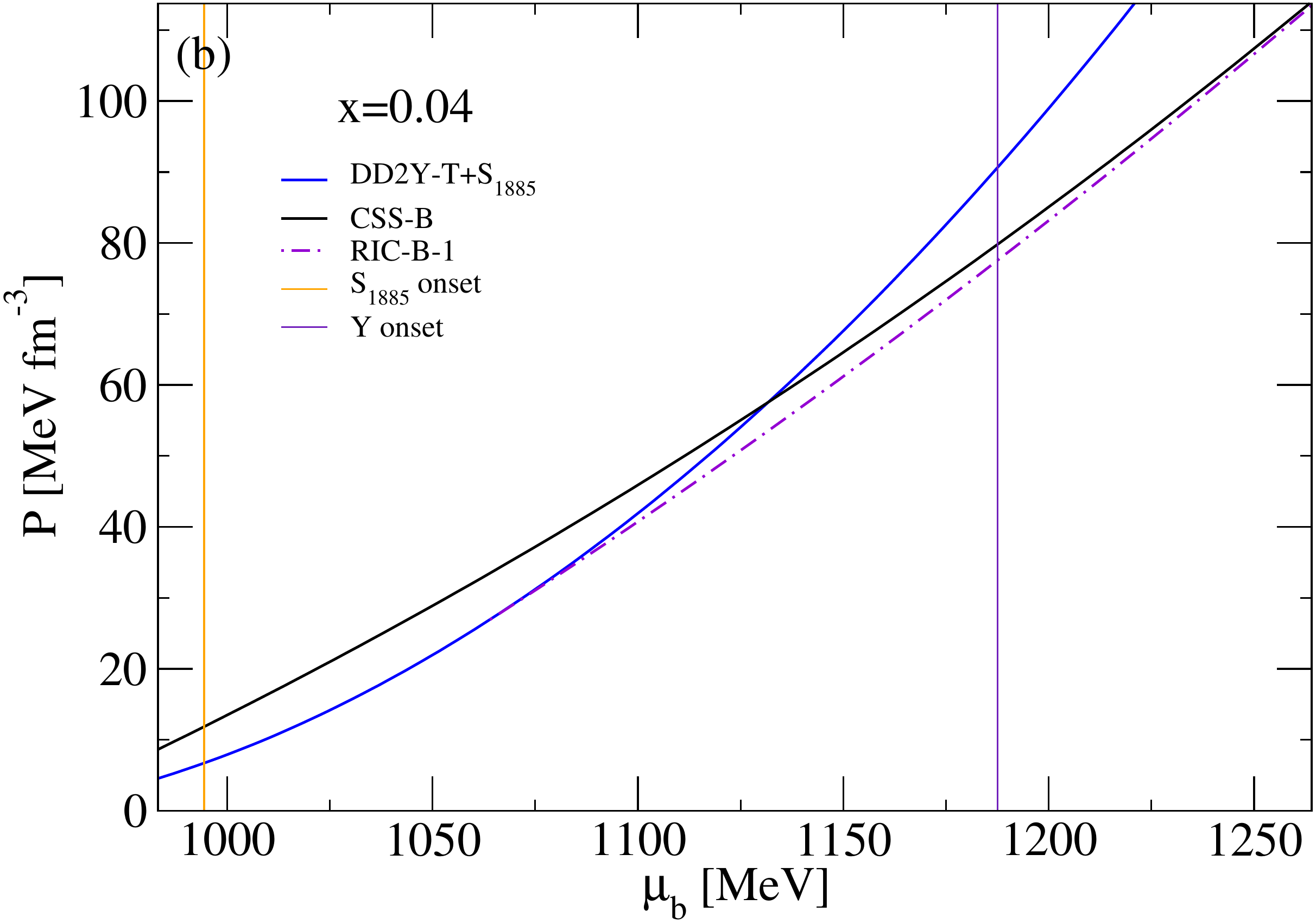}\\
	\includegraphics[width=0.48\textwidth]{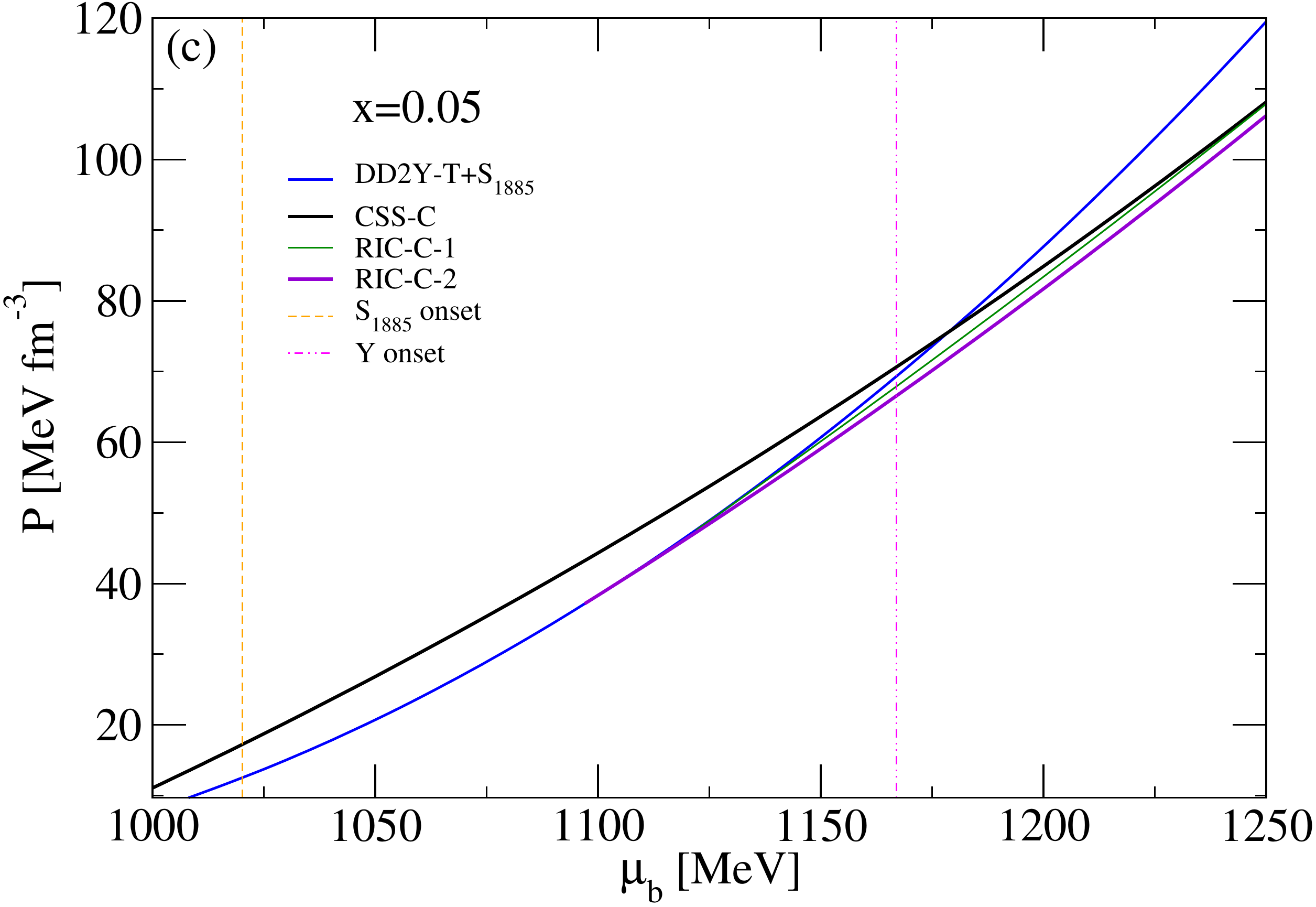}
	\includegraphics[width=0.47\textwidth]{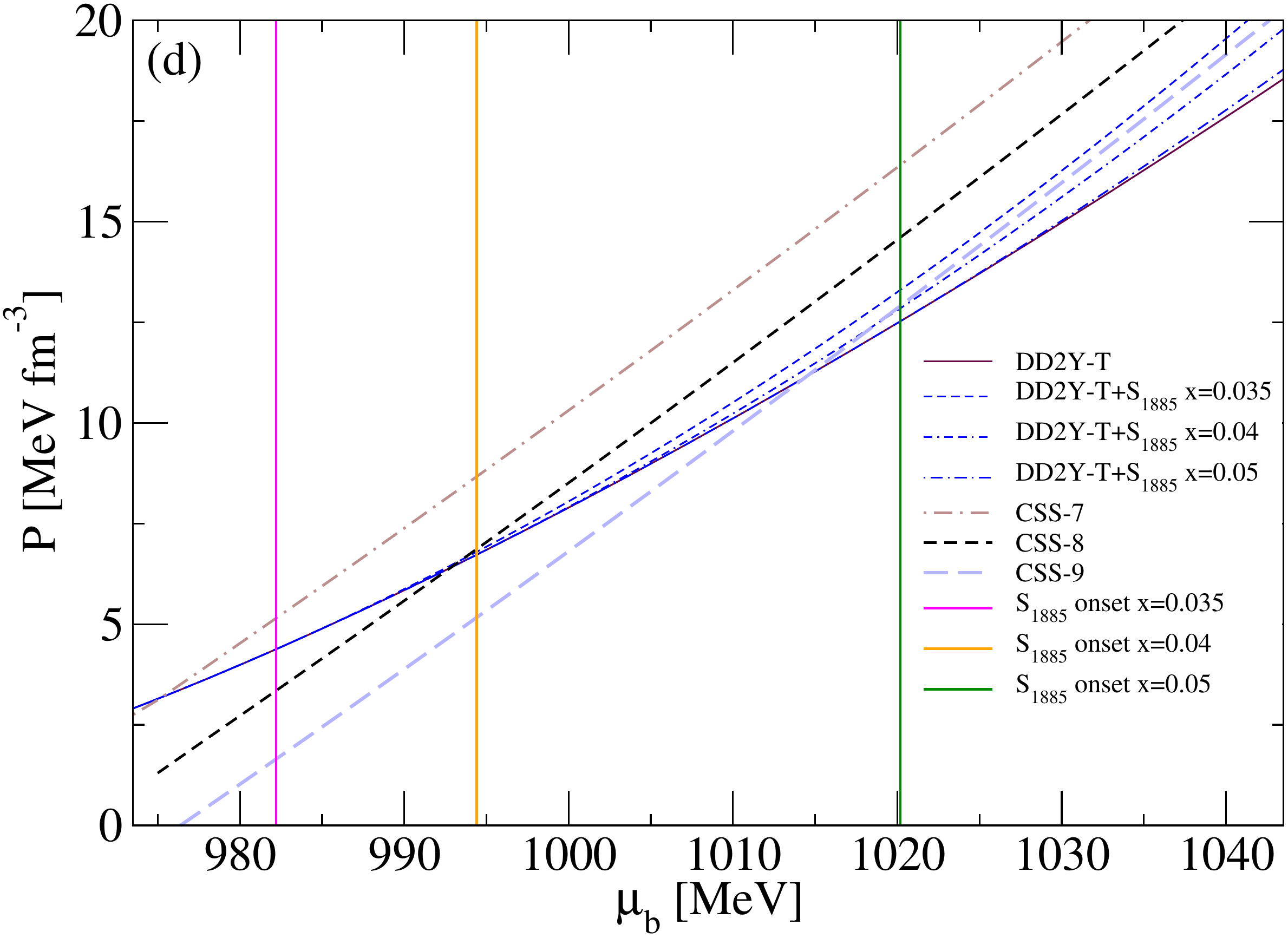}
	\caption{MC and RIC solutions when $S_{1885}$ is included in the hadronic matter EoS. The curves labelled CSS-A, CSS-B and CSS-C correspond to the quark matter EoS which after RIC give rise to the parameter sets RIC-A, RIC-B and RIC-C given in \tableautorefname~\ref{tab:parameters} while CSS-7, CSS-8 and CSS-9 correspond to the quark matter EoS 
	which after MC give rise to the sets MC-S$_{1885}$-7, MC-S$_{1885}$-8 and MC-S$_{1885}$-9 which are given in \tableautorefname~\ref{tab:parameters3}.
\label{fig:eos6}
	}
\end{figure*}

In Fig.~\ref{fig:eos6}, MC and RIC are performed for the case $m_S=1885$ MeV.
Panel (d) shows a MC when an early deconfinement happens while the other panels show RIC with the possible minimum value of $\Delta_p$ for which the mechanical stability is fulfilled. The described situation of the reconfinement and crossover transition by RIC is depicted in this figure.

\begin{table*}[!bht]
 \caption{\label{tab:parameters}
  Parameter sets characterizing hybrid EoS which have been investigated as a solution for MC when S$_{1941}$ or S$_{1885}$ are included in hadronic matter.
  }
 \begin{ruledtabular}
 \begin{tabular}{lccccccccccc}
 set & $m_S$ & $~~x~~$ & $~~\eta_V~~$ & \textrm{c}$_s^2$/\textrm{c}$^2$ & \textrm{A} & \textrm{B}  & \textrm{B}$_0$ & {B}$_1$ & \textrm{$\mu_<$}& \textrm{$\Gamma_<$} & scenario\\ 
  & [\textrm{MeV}] &  &  &    & [\textrm{MeVfm}$^{-3}$] & [\textrm{MeVfm}$^{-3}$]  & [\textrm{MeVfm}$^{-3}$]  & [\textrm{MeVfm}$^{-3}$]&  [\textrm{MeV}] & [\textrm{MeV}] &  \\
 \hline
MC-S$_{1941}$-1 & 1941 &  0.05 & 0.11 & 0.45 & 93.49 & 81.73 & 9.0 & 0 & 0 & 0 & S$_{1941}$DY\\
MC-S$_{1941}$-2 & 1941 &  0.05 & 0.14 & 0.48 & 94.99 & 85.50 & 5.7 & 0 & 0 & 0 & S$_{1941}$DY\\
MC-S$_{1941}$-3 & 1941 &  0.05 & 0.17 & 0.52 & 96.85 & 90.18 & 0 & 7.0 & 1020 & 110 & S$_{1941}$DY\\
MC-S$_{1941}$-4 & 1941 &  0.08 & 0.14 & 0.48 & 94.99 & 85.50 & 0 & 10.0 & 1250 & 100 & YS$_{1941}$D\\
MC-S$_{1941}$-5 & 1941 &  0.05 & 0.17 & 0.52 & 96.85 & 90.18 & 0 & 0 & 0 & 0 & DS$_{1941}$Y\\
 MC-S$_{1885}$-6  & 1885 & 0.030 & 0.14 & 0.48 & 94.99  & 85.50 & 0 & 0 & 0 & 0  &  S$_{1885}$DY\\
MC-S$_{1885}$-7 & 1885 & 0.035 & 0.13 & 0.47 & 94.45 & 84.14 & 0 & 0 & 0 & 0  &  DS$_{1885}$Y\\
MC-S$_{1885}$-8 & 1885 & 0.040 & 0.13 & 0.47 & 94.45 & 84.14 & 1.8 & 0 & 0 & 0 & DS$_{1885}$Y\\
MC-S$_{1885}$-9 & 1885 & 0.050 & 0.13 & 0.47 & 94.45 & 84.14 & 3.5 & 0 & 0 & 0  &  DS$_{1885}$Y\\
 %
 \end{tabular}
 \end{ruledtabular}
 \end{table*}
 
 \begin{table*}[!bht]
 \caption{\label{tab:parameters3}
  Parameter sets characterizing hybrid EoS which have been investigated as a solution for RIC when S$_{1885}$ is included in hadronic matter. In different sets, A corresponds to $x=0.03$, B corresponds to $x=0.04$ and C corresponds to $x=0.05$. }
 \begin{ruledtabular}
 \begin{tabular}{lccccccccccc}
 set & $m_S$ & $~~x~~$ & $~~\eta_V~~$ & \textrm{c}$_s^2$/\textrm{c}$^2$ & \textrm{A} & \textrm{B}  & \textrm{B}$_0$ & $\Delta_p$ & $\mu_H$ & $\mu_Q$ & scenario \\ 
  & [\textrm{MeV}] & &  &    & [\textrm{MeVfm}$^{-3}$] & [\textrm{MeVfm}$^{-3}$]  & [\textrm{MeVfm}$^{-3}$]  &  &  [\textrm{MeV} & \textrm{MeV}]  &  \\
 \hline
RIC-A-1 & 1885 & 0.03 & 0.14 & 0.48 & 94.99  & 85.50 & 0 & -3$\%$ & 1027.76 & 1121.92 & S$_{1885}$DY\\
RIC-A-2 & 1885 &   0.03 & 0.14 & 0.48 & 94.99  & 85.50 & 0 & -5$\%$ & 1019.32 & 1261.74 & S$_{1885}$YD \\
RIC-A-3  & 1885 &  0.03 & 0.14 & 0.48 & 94.99  & 85.50 & 0 & -7$\%$ & 1012.81 & 1418.92 & S$_{1885}$YD\\
RIC-B-1 & 1885 &   0.04 & 0.14 & 0.48 & 94.99  & 85.50 & -4 & -7$\%$ & 1064.10 & 1335.80 & S$_{1885}$YD\\
RIC-C-1 & 1885 & 0.05 & 0.12 & 0.46 & 93.95 & 82.88 & 0 & -3$\%$ & 1122.39 & 1286.74 & S$_{1885}$YD\\
RIC-C-2 & 1885 & 0.05 & 0.12 & 0.46 & 93.95 & 82.88 & 0 & -5$\%$ & 1097.33 & 1398.25 & S$_{1885}$YD\\
 %
 \end{tabular}
 \end{ruledtabular}
 \end{table*}
\pagebreak[4]

\bibliography{sexaquark}

\end{document}